\documentclass{jfm}
\usepackage{natbib}
\usepackage{upmath}
\usepackage[british]{babel}
\usepackage{amsmath,bm}
\usepackage{amssymb}
\usepackage{amsbsy}
\usepackage{amscd}
\usepackage{amstext}
\usepackage{tabularx}
\usepackage{float}
\usepackage{makeidx}
\usepackage{amsmath}
\usepackage{subfigure}
\usepackage{afterpage}
\usepackage[T1]{fontenc}
\usepackage[latin1]{inputenc}
\usepackage{multirow}
\usepackage{color}
\usepackage{xcolor}
\usepackage{makecell}
\usepackage[colorlinks,	
linkcolor   =blue,
anchorcolor =blue,
citecolor   =blue,
urlcolor    =blue]{hyperref}

\usepackage{graphicx} % not used later, figures must be supplied as hardcopies, use \vspace
\usepackage{graphics,epsfig}
\usepackage{amsfonts}
\usepackage{psfrag}
\usepackage{tikz}
\usepackage{pifont}

\ifCUPmtlplainloaded \else
  \checkfont{eurm10}
  \iffontfound
    \IfFileExists{upmath.sty}
      {\typeout{^^JFound AMS Euler Roman fonts on the system,
                   using the 'upmath' package.^^J}%
       \usepackage{upmath}}
      {\typeout{^^JFound AMS Euler Roman fonts on the system, but you
                   dont seem to have the}%
       \typeout{'upmath' package installed. JFM.cls can take advantage
                 of these fonts,^^Jif you use 'upmath' package.^^J}%
      }
  \else
  \fi
\fi

\ifCUPmtlplainloaded \else
  \checkfont{msam10}
  \iffontfound
    \IfFileExists{amssymb.sty}
      {\typeout{^^JFound AMS Symbol fonts on the system, using the
                'amssymb' package.^^J}%
       \usepackage{amssymb}%
       \let\le=\leqslant  \let\leq=\leqslant
         \let\geq=\geqslant
      }{}
  \fi
\fi

\ifCUPmtlplainloaded \else
  \IfFileExists{amsbsy.sty}
    {\typeout{^^JFound the 'amsbsy' package on the system, using it.^^J}%
     \usepackage{amsbsy}}
    {\providecommand\boldsymbol[1]{\mbox{\boldmath $##1$}}}
\fi

\providecommand\bcdot{\boldsymbol{\cdot}}

\newsavebox{\astrutbox}
\sbox{\astrutbox}{\rule[-5pt]{0pt}{20pt}}

\newcommand\p{\ensuremath{\partial}}

\newcommand\ie{i.e.}

\def\u{{\boldsymbol{u}}}

\def\U{{\boldsymbol{U}}}
\def\Ub{{\U_b}}

\def\up{{\boldsymbol{u}}^+}
\def\pp{p^+}

\def\n{{\boldsymbol{n}}}

\def\p{\partial}
\def\ar{\textsc{ar}}

\usepackage{xcolor}
\definecolor{myorange}{RGB}{255, 102, 0}

\title[Numerical analyses of the flow past a short rotating cylinder]
{Numerical analyses of the flow past a short rotating cylinder}

\author[Y. Yang, C. Wang, R. Guo, M. Zhang]%
{
Yongliang Yang$^{1,2}$, Chenglei Wang$^3$, Rui Guo$^1$ \and Mengqi Zhang$^2$\thanks{Email address for correspondence: {mpezmq@nus.edu.sg}}
}

\affiliation{
  $^1$ School of Mechanical Engineering, Nanjing University of Science and Technology, Nanjing 210094, PR China   \\
  $^2$ Department of Mechanical Engineering, National University of Singapore, 9 Engineering Drive 1, 117575 Singapore \\
  $^3$ Department of Mechanical Engineering, The Hong Kong Polytechnic University, Kowloon, Hong Kong SAR, People's Republic of China
  \\ [\affilskip]
}

\date{\today}

\graphicspath{{figures/}}
\newlength\savewidth

\begin{document}
\maketitle

\begin{abstract}

This work studies the three-dimensional flow dynamics around a rotating circular cylinder of finite length, whose axis is positioned perpendicular to the streamwise direction. Direct numerical simulations and global stability analyses are performed within a parameter range of Reynolds number $Re=DU_\infty/\nu<500$ (based on cylinder diameter $D$, uniform incoming flow velocity $U_\infty$), length-to-diameter ratio $\ar=L/D\leq2$ and dimensionless rotation rate $\alpha=D\Omega/2U_\infty\leq2$ (where $\Omega$ is rotation rate). By solving Nav\-ier--Sto\-kes equations, we investigated the wake patterns and explored the phase diagrams of the lift and drag coefficients. For a cylinder with $\ar=1$, we found that when the rotation effect is weak ($0\leq\alpha\lesssim0.3$), the wake pattern is similar to the unsteady wake past the non-rotating finite-length cylinder, but with a new linear unstable mode competing to dominate the saturation state of the wake. The flow becomes stable for $0.3\lesssim\alpha\lesssim0.9$ when $Re<360$. When the rotation effect is strong ($\alpha\gtrsim0.9$), new low-frequency wake patterns with stronger oscillations emerge. Generally, the rotation effect first slightly decreases and then sharply increases the $Re$ threshold of the flow instability when $\alpha$ is relatively small, but significantly decreases the threshold at high $\alpha$ ($0.9<\alpha\leq2$). Furthermore, the stability analyses based on the time-averaged flows and on the steady solutions demonstrate the existence of multiple unstable modes undergoing Hopf bifurcation, greatly influenced by the rotation effect. The shapes of these global eigenmodes are presented and compared, as well as their structural sensitivity, visualising the flow region important for the disturbance development with rotation. This research contributes to our understanding of the complex bluff-body wake dynamics past this critical configuration.

\end{abstract}

\section{Introduction}
The study of flows around rotating bluff bodies, including cylinders and spheres, constitutes a fundamental problem in fluid dynamics. It provides critical insights into vortex formation and wake dynamics, which are relevant to various natural phenomena and engineering applications. For instance, these flows are critical in circulation control on an airfoil \citep{Tennant1976}, heat exchangers \citep{Roslan2012}, laminar/turbulence separation \citep{Afroz2017} and design of guided rockets \citep{DECELIS2017204}. Besides, the use of rotating effects to control the wake flow past bluff bodies has attracted much attention \citep{Gad-el-Hak1991,modi1997moving}, including its applications such as Flettner rotors utilising the Magnus effect \citep{Seifert2012}. However, there is a research gap in the study of flow past a rotating cylinder with two free ends. Experimental studies and real-world applications typically involve finite-length rotating cylinders. In contrast, theoretical and numerical studies traditionally consider infinitely long rotating cylinders. This discrepancy highlights the need for further investigation on the effect of aspect ratio and free ends. To address this research gap, our study aims to examine the three-dimensional (3-D) flow past a short rotating cylinder, an area that previous researchers have not explored. By investigating this problem, we hope to contribute to the knowledge base on the flow around rotating bluff bodies and provide insights that are relevant to various engineering applications.

For a non-rotating infinitely long circular cylinder, it is well known that the wake first experiences a Hopf bifurcation at $Re\approx46.7$ and then a 3-D wake transition at $Re\approx190$ \citep{williamson1996three,Williamson1996}. The non-dimensional number $Re$ quantifies the ratio between inertia and viscosity. The rotation effect enriches the flow dynamics of the wake flow. \cite{kang1999laminar} conducted a two-dimensional (2-D) numerical study of the flow past a rotating circular cylinder and showed that rotation could effectively suppress the vortex shedding (mode I) found in the stationary cylinder at $\alpha > \alpha_c$ (where $ \alpha_c$ is the critical dimensionless rotation rate), and the relationship between the lift/drag and rotation rate in the range of $0\leq \alpha \leq2.5$ is significantly different from that predicted by potential flow theory.
Subsequently, the numerical studies of 2-D rotating cylinders by \cite{stojkovic2002effect,stojkovic2003new}, \cite{mittal2003flow} and \cite{mittal2004three} revealed that when the rotation rate $\alpha$ is relatively large, there is a secondary instability phenomenon (model II) characterized by low-frequency vortex shedding. Especially, \cite{mittal2003flow} brought to light this instability mechanism of 2-D perturbations by the global stability analysis, which will be extended in the current study, focusing on a rotating finite cylinder, to account for 3-D perturbations. Built upon the previous works, \cite{el2008three} extended the neutral stability curves for these wakes in the $Re$-$\alpha$ plane by direct numerical simulations (DNS) and Landau model. 
The experiments by \cite{kumar2011flow} provided evidence of the existence of mode II at $Re=200,300, 400$ and $0<\alpha<5$. The experimental study of \cite{LINH2011} also reported observations of the low-frequency mode II vortex. Their experimental Strouhal number and wake patterns agree well with numerical data of \cite{mittal2003flow}. Later, \cite{pralits2010instability,pralits2013three} conducted an extensive study of the linear global dynamics of the 2-D rotating cylindrical wake flow. The authors explored neutral stability curves on the $(Re, \alpha)$ plane, providing a comprehensive understanding of this phenomenon. They also observed multiple steady solutions at high $\alpha$, explaining the decay of the secondary shedding wake. More recently, \cite{sierra2020bifurcation} fully described the bifurcation, neutral curves and global instability modes in the parameter space $(Re, \alpha)\subset[0, 200]\times[0, 10]$, exploring the relations among Takens-Bogdanov bifurcations, cusps and generalized Hopf bifurcations when varying the parameters in the rotating cylinder wake flow. 
To sum up, for the infinitely-long rotating cylinders, mode I and mode II are fundamentally different flow phenomena. The mode I undergoes a supercritical Hopf bifurcation and becomes linearly unstable at $0\leq \alpha \leq 2$, which is related to the classical B\'{e}nard-von-K\'{a}rm\'{a}n vortex street, characterized by alternating vortices with opposite signs of spanwise vorticity. On the other hand, the physical mechanism of the linear instability mode II $(4.5\leq \alpha<6)$ is featured by low-frequency vortex shedding with the same vorticity sign. 

Further research has shown that rotation can lead to complex 3-D instabilities. Numerical investigations by \cite{rao_2013a,rao_2013b} demonstrated several 3-D modes becoming unstable to spanwise perturbations in the steady and unsteady regimes of $Re=400$ flows. Five 3-D modes were identified to be unstable in the mode I shedding regime, while four 3-D modes were observed in the steady flow regimes for $\alpha \geq 2$.  \cite{radi2013experimental} proved experimentally the existence of the above numerically predicted 3-D modes. They additionally showed a highly 3-D wake and the absence of 2-D periodic shedding at high $\alpha$ $(\rm{\ie}$$\ Re=200, \alpha=4.5)$ previously reported in \cite{mittal2003flow}.  

\cite{navrose_mittal_2015} conducted 3-D numerical studies and found that the span length of the rotating cylinder plays an important role in the evolution of the wake with $Re\in[200,350], \alpha\in[0,5]$. Specifically, only linear global modes with wavelengths that are integer multiples of the cylinder span are selected for growth in nonlinear DNS. It is thus necessary to consider 3-D configurations that take into account the spanwise length in studies of rotating bluff-body flows. 

Researchers have also studied the flow past a rotating bluff body of other forms. It is instructive to review relevant works on these flows as they will also be discussed in this study. The flow past a rotating sphere, along either the transverse axis or the streamwise axis, received considerable attention. \cite{citro2016} applied the global linear stability analysis (LSA), adjoint-based structural sensitivity analysis and weakly nonlinear analysis (WNL) to reveal the mechanism of flow instability around a rotating sphere around the transverse axis. They characterised the evolution processes of the first (at low $\alpha$) and second (at high $\alpha$) instability modes. \cite{fabre2017flow} further improved the WNL from than that in \cite{Fabre2012} to achieve a better comparison between the WNL expansion result and the DNS. Namely, the comparison demonstrates that \cite{fabre2017flow}'s $\epsilon$ expansion ($\epsilon=\sqrt{Re-Re_c}/Re_c$, where $Re_c$ is the critical Reynolds number of a pitchfork bifurcation in their work) provides a better reproduction of the DNS results for both angular velocity and associated lift forces, compared to \cite{Fabre2012}'s $\omega$ expansion (where $\omega$ represents the dimensionless rotation rate normalizing the actual rotation with $U_0/D$). The $\omega$ expansion failed to predict the DNS results for $Re$ around and beyond $Re=212$, whereas the $\epsilon$-expansion accurately reproduces the DNS results up to $Re\approx225$. 
For the flow past a sphere rotating along the streamwise direction, \cite{lorite2020description} conducted a DNS study on the wake evolution of a strongly rotating sphere and observed a sequence of continuous bifurcations from periodic, quasi-periodic, and irregular states to chaos, over the parameter range $0<\alpha<3$ and $Re=250, 500, 1000$.
Later, \cite{sierra2022unveiling} employed global LSA to determine the neutral curves of three non-zero frequency global modes on the $Re$-$\alpha$ plane, and used the normal form expansion to reveal the nonlinear interactions among the global modes. Their predictions of the normal form analysis were satisfactory and close to the DNS results, which led to a more detailed phase diagram of the nonlinear patterns.
Besides, \cite{spinningbullet_2014} carried out global LSA of the wake flow past a streamwise rotating bullet-shaped body and plotted the neutral curves on the $Re$-$\alpha$ plane. Their work indicated that the streamwise rotation can also delay the Hopf bifurcation of a bluff body with a large aspect ratio ($\ar=2$) when increasing $Re$, which is different from a sphere with an aspect ratio of 1. To sum up, in addition to the sphere rotating along the streamwise direction, the aforementioned rotating bluff bodies of various shapes and aspect ratios exhibit a moderate rotation regime in the $Re-\alpha$ plane where the neutral curve of the Hopf bifurcation is significantly shifted to higher $Re$ values. Therefore, the aspect ratio plays a significant role in the wake transition of a rotating bluff body, which should be further researched.

Our literature review has identified a research gap in the understanding of the dynamics of uniform flow past a finite-length rotating cylinder along its cylinder axis. This flow configuration is common in nature and engineering applications, but its instability mechanism, bifurcation properties and transition path are still unclear.
In a recent study, we have conducted a detailed investigation of the wake flow around a non-rotating finite-length cylinder \citep{yang_feng_zhang_2022}. Building on this research, we aim to extend our investigation to the wake flow around a rotating finite-length cylinder to explore its 3-D effects, in line with other similar works focusing on the rotation effect in the wake flow such as \cite{pralits2010instability,citro2016,sierra2022unveiling,Zhao2023}, reviewed in \cite{rao2015review}.
The primary objective of this study is to clarify the effect of rotation on the unsteadiness of finite-length cylinder wakes using the (nonlinear) DNS method. Additionally, we aim to determine the instability threshold at which the unsteadiness occurs using the global stability approach. To identify the instability region responsible for the unsteadiness, we will also probe the structural sensitivity of the flow.
Our work will contribute to a deeper understanding of the wake flow around rotating finite-length cylinders and provide insights into the dynamics of this flow configuration, which has practical implications for various engineering applications.

The paper is organised as follows. Section \S~\ref{sec:problemformulation} introduces the configuration of a 3-D finite rotating cylinder flow, the boundary conditions, the governing equations (i.e. nonlinear Nav\-ier--Sto\-kes equations and their corresponding linearised direct and adjoint equations) and the numerical methodology. In section \S~\ref{sec:results}, we show the results and discuss the base states (time-averaged flow or steady flow), nonlinear wake patterns, global eigenmodes, neutral curves in parametric plane $Re$-$\alpha$, bifurcations in this flow. Finally, the results are summarised in section \S~\ref{sec:Conclusions} and conclusions are provided. In the appendices, we provide additional results of the Stuart-Landau model, the global modes at different aspect ratios $\ar$ and a verification step of the numerical codes.

\section{Problem formulation}\label{sec:problemformulation}

\subsection{Flow configuration and governing equations}\label{subsec:NSEq}
\begin{figure}
	\hfil
	\centering\includegraphics[trim=1cm 4.0cm 1cm 4.0cm,clip,width=0.4\textwidth]{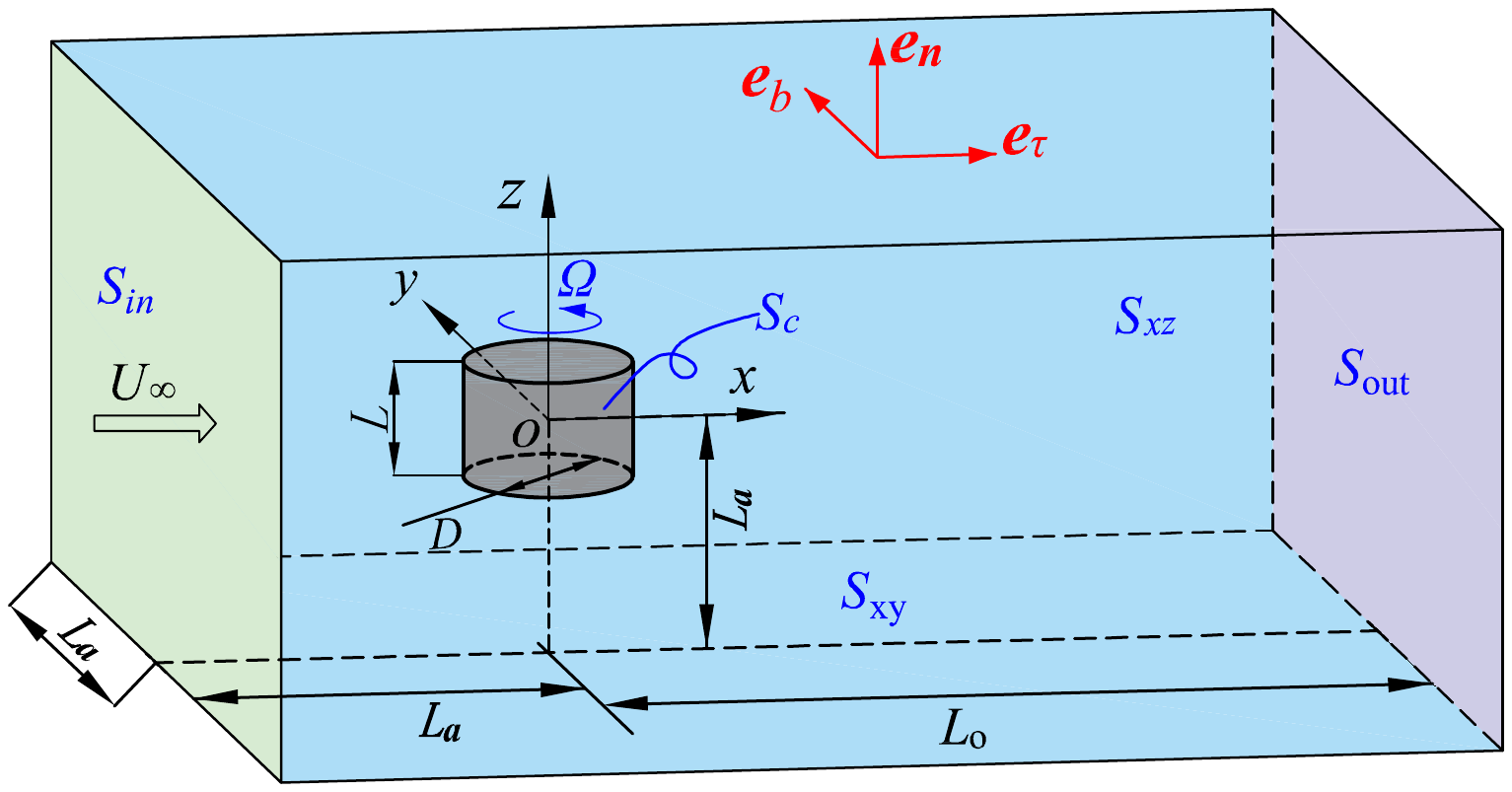} 
	\put(-162,78){($a$)} \quad
	\centering\includegraphics[trim=0cm 0.0cm 0.cm 0.0cm,clip,width=0.57\textwidth]{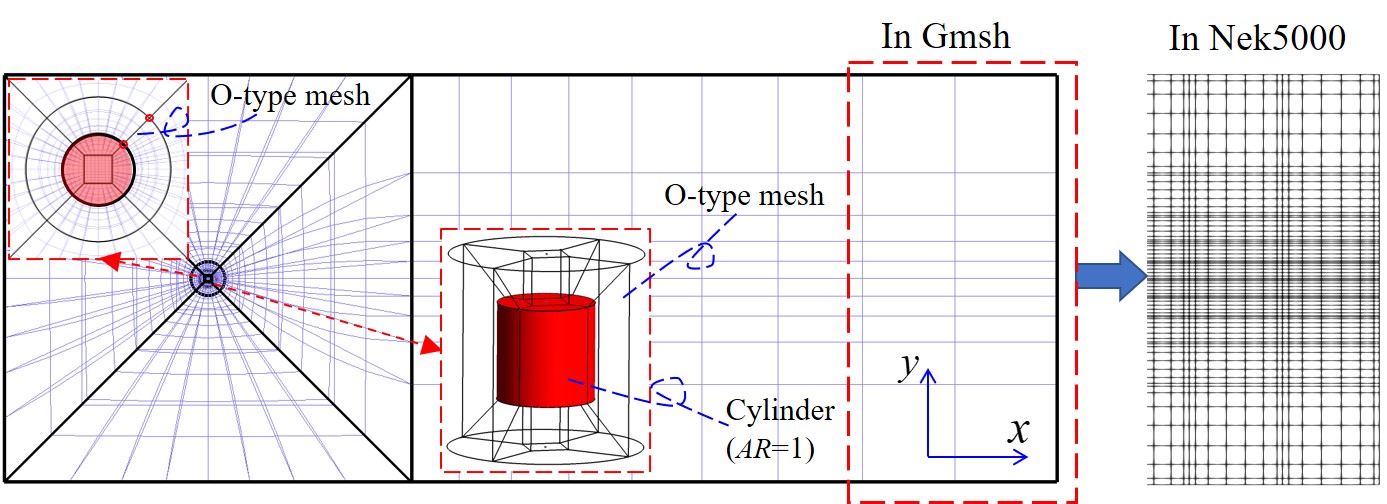}
	\put(-230,78){($b$)} 
	\caption{The computational domain and boundary conditions (not to scale) (panel $a$) and mesh design (panel $b$). The red unit vectors $(\boldsymbol{e}_{\boldsymbol{n}}, \boldsymbol{e}_{\boldsymbol{\tau}},\boldsymbol{e}_{\boldsymbol{b}})$ in panel ($a$) represent the directional vectors of the surface $S_{xy,t}$. The finite-length cylinder is rotating around its axis that is perpendicular to the incoming flow.}
	\label{fig:CD_BC}
\end{figure}

We study the 3-D stability of the flow around a finite-length rotating cylinder of length $L$, diameter $D$ and aspect ratio $\ar=L/D$, subjected to a uniform incoming flow in a Cartesian coordinate system. As shown in {\color{blue}figure~}\ref{fig:CD_BC}, the origin of the coordinate system is located at the center of the cylinder, the $x$ axis points in the flow direction, the $y$ axis represents the transverse direction and the $z$ axis extends along the center line of the cylinder. The nondimensional Nav\-ier--Sto\-kes (NS) equations for the unsteady Newtonian incompressible flow read 
\begin{align}
	\frac{\p \U }{\p t}+
	(\U \bcdot\nabla)\U =  -\nabla P+ \frac{1}{Re} \nabla^2\U, \ \ \ \ \ \  \nabla\bcdot \U=0,
	\label{eq:NS}
\end{align}
where $\U = (U_x,U_y,U_z)$ is the velocity vector and $P$ is the pressure. The Reynolds number $Re =  DU_\infty / \nu$  is defined based on cylinder diameter $D$, the velocity of the uniform incoming flow  $U_\infty$ at infinity and the kinematic viscosity coefficient $\nu$. 
Strouhal number $St=fD/{U_\infty}$ is defined based on the frequency $f$ of vortex shedding. 
The dimensionless rotation rate $\alpha=\Omega D / 2{U_\infty}$, where $\Omega$ is the rotating angular speed of the cylinder along the $z$ axis, as shown in {\color{blue}figure~}\ref{fig:CD_BC}. 
Setting $\rho U^2_\infty$ as the reference dynamic pressure, the drag and lift coefficients are defined respectively as 
\begin{align}
   C_d =C_{dp}+C_{dv}=\frac{F_{dp}+F_{dv}}{(1/2)\rho U^2_\infty A} \quad	  \rm{and} \quad \it  C_{l} =C_{lp}+C_{lv}=\frac{F_{lp}+F_{lv}}{(1/2)\rho U^2_\infty A}, 
\end{align}
where $\rho$ is the fluid density, $F_{dp}=\int_{S_c} P_x \, \rm{d}\it{S}$ and $F_{dv}=\int_{S_c} \tau_{wx} \, \rm{d}\it{S}$ are the pressure drag and friction drag on the cylinder surface $S_c$ along the streamwise direction, $F_{lp}=\int_{S_c} P_{y,z} \, \rm{d}\it{S}$ and $F_{lv}=\int_{S_c} \tau_{wy,wz} \, \rm{d}\it{S}$ are the pressure lift and  wall shear stress lift acting in either  $y$ axis or $\boldsymbol{n}_{z}$ (defined as $C_{ly}$ and $C_{lz}$, respectively) and $A$ is the reference area $A=LD$. Here $(P_{x}, P_{y}, P_{z})$ are the components of the pressure acting on the cylinder surface along the $x$, $y$ and $z$ axes, respectively. $\tau_w$ is wall (surface) shear stresses. Furthermore, we will also use the letters $\overline{C}_d$ and $\overline{C}_{l}$ to denote the time-averaged values of $C_d$ and $C_{l}$, respectively.

As shown in {\color{blue}figure~}\ref{fig:CD_BC}, $S_c$ represents the surface of the cylinder. Here $S_{in}$ and $S_{out}$ represent the inlet and outlet surfaces of the rectangular computation domain, whose normal is along the $x$ direction. $S_{xy,t}$, $S_{xy,b}$, $S_{xz,f}$ and $S_{xz,b}$ denote the surfaces of the cuboid on the top, bottom, front and back side walls, which are parallel to the $xy$, $xy$, $xz$ and $xz$ planes, respectively. 
The boundary conditions of the system \ref{eq:NS} are 
\begin{subeqnarray}
  \U=(1,0,0)\quad & \rm{on} \ \it S_{in},\ \\
  \U = \Omega \hat{\boldsymbol{e}}_z  \times \boldsymbol{r}=\Omega (-y_c,\ x_c,\ 0)\quad & \rm{on} \ \it S_c,\ \ \\  
  (-P\mathbf{I}+Re^{-1}\nabla \boldsymbol{U})\cdot\boldsymbol{e}_{\boldsymbol{n}}=0\quad & \rm{on} \ \it S_{out},\\
  \boldsymbol{U}\bcdot \boldsymbol{e}_{\boldsymbol{n}} =0,\ \  (\nabla \boldsymbol{U}\bcdot \boldsymbol{e}_{\boldsymbol{\tau}})\bcdot\boldsymbol{e}_{\boldsymbol{n}} =0,\ \ (\nabla \boldsymbol{U}\bcdot \boldsymbol{e}_{\boldsymbol{b}})\bcdot\boldsymbol{e}_{\boldsymbol{n}} =0 \quad & \rm{on} \ \it S_{xy}, S_{xz}, \label{eq:nonlinearBC}
\end{subeqnarray} 
where $\hat{\boldsymbol{e}}_z=(0,0,1)$ is a unit vector aligned with the positive $z$-axis. The vector $\boldsymbol{r}=(x_c,y_c,z_c)$ is the position vector of a point located on the cylinder surface, i.e. the vector from the origin of the coordinate system to the point. Here $\boldsymbol{e}_{\boldsymbol{n}}, \boldsymbol{e}_{\boldsymbol{\tau}},\boldsymbol{e}_{\boldsymbol{b}}$ are the unit normal, unit tangent and unit bitangent vectors, respectively. As shown in {\color{blue}figure~}\ref{fig:CD_BC}($a$), the vector $\boldsymbol{e}_{\boldsymbol{n}}$ of surfaces $S_{xy,t}, S_{xy,b}, S_{xz,f}, S_{xz,b}$ and $S_{out}$ points out the computational domain. The directions of the vector $\boldsymbol{e}_{\boldsymbol{\tau}}$ of surfaces  $S_{xy,t}, S_{xy,b}, S_{xz,f}, S_{xz,b}$ point along positive $x$-axis, negative $x$-axis, negative $x$-axis and positive $x$-axis, respectively. The vector $\boldsymbol{e}_{\boldsymbol{b}}$ points in the direction that is perpendicular to both the normal vector and the tangent vector. Here $\rm \mathbf{I}$ is the identity tensor.

\subsection{Linearisation}
The global linear stability/instability of the flows past the finite-length rotating cylinder will be studied.
Reynolds decomposition $\boldsymbol U = \boldsymbol U_b+\boldsymbol u, P=P_b+p$ will be substituted into the nonlinear governing equations \ref{eq:NS}. The base-state terms $(\boldsymbol U_b,P_b)$ satisfying the steady Navier--Stokes equations and the nonlinear terms are neglected, yielding the linearised equations for the infinitesimal perturbations $(\boldsymbol u,p)$ residing on these base states, i.e.,
\begin{align}
	\frac{\p \u }{\p t}+(\Ub \bcdot\nabla)\u + (\u \bcdot\nabla)\Ub 
	=  -\nabla p+ \frac{1}{Re} \nabla^2\u, \ \ \ \ \ \  \nabla\bcdot \u  =   0,
	\label{eq:LNS}
\end{align} 
where $\boldsymbol u$ is the 3-D perturbation velocity vector $\boldsymbol u=(u_x,u_y,u_z)$ and $p$ is the perturbation pressure. Homogeneous boundary conditions are applied for the perturbed variables as follows,

\begin{subeqnarray}
  \boldsymbol{u}=\mathbf{0} \quad  &\rm{on} \  \it\boldsymbol{S_c}\ \rm{and}\ \it{S_{in}},\\ 
  (p\mathbf{I}-Re^{-1}\nabla \boldsymbol{u})\cdot\boldsymbol{e}_{\boldsymbol{n}}=0 \quad &\rm{on} \ \it S_{out},\\
	\frac{\p u_x }{\p y}=u_y=\frac{\p u_z }{\p y}=\frac{\p p}{\p y}=0 \quad &\rm{on} \  \it\boldsymbol S_{xz,f},S_{xz,b}, \\
	\frac{\p u_x }{\p z}=\frac{\p u_y }{\p z}=u_z=\frac{\p p}{\p z}=0 \quad &\rm{on} \  \it\boldsymbol S_{xy,t},S_{xy,b}.  
	\label{eq:BCLIN}
\end{subeqnarray}

Linear equation \ref{eq:LNS} is rewritten in matrix form with $\boldsymbol q=(\boldsymbol u, p)^{\rm T}$ as
\begin{align}
	{\boldsymbol M}\frac{\p \boldsymbol q }{\p t} = {\boldsymbol L}_{\boldsymbol U_b} \boldsymbol q 
	\label{eig1}
\end{align}
where ${\boldsymbol L}_{\boldsymbol U_b}$ is the linearised Navier--Stokes operator depending on the base states ${\boldsymbol U_b}$. The elements of mass matrix $\boldsymbol M$ and the Jacobian matrix $\boldsymbol L_{\boldsymbol U_b}$ are

\begin{equation}
	\boldsymbol M =\left(\begin{array}{cc}
		\boldsymbol I & 0\\
		0           & 0
	\end{array} 
	\right), \ \ \ \
	{\boldsymbol L}_{\boldsymbol U_b}=
	\left(\begin{array}{cc}
		-\boldsymbol U_b \bcdot\nabla - \nabla\boldsymbol U_b + Re^{-1}\nabla^2 & \quad-\nabla\\ 
		\nabla\bcdot & \quad 0
	\end{array} 
	\right).
	\label{eig2}
\end{equation}
As the considered base flow states are steady, we seek the wavelike solution ${\boldsymbol{q}}(x,y,z,t)$ of the form
\begin{align}
\boldsymbol{q}(x,y,z,t)=\hat{\boldsymbol{q}}(x,y,z)e^{\lambda t}, \quad \rm{where} \  \lambda=\sigma+\rm{i}2\pi \omega.
\label{prtb}
\end{align} 
Substituting this form (Eq. \ref{prtb}) into equation \ref{eig1}, we can get the following eigenvalue problem 
\begin{align}
	{\boldsymbol L}_{\boldsymbol U_b}\hat{\boldsymbol{q}}=\lambda{\boldsymbol M} \hat{\boldsymbol{q}},
	\label{eq:eig}
\end{align}
where the stability of the base state $\boldsymbol U_b$ is dictated by the eigenvalues $\lambda$ in the linearised problem with $\sigma$ being the temporal growth/decay rate of perturbations and $\omega$ the eigenfrequency. The flow is linearly unstable if $\sigma>0$; stable otherwise. The eigenfrequency $\omega$ of the most unstable eigenvalue determines whether the base state $\boldsymbol U_b$ experiences a regular bifurcation ($\omega=0$) or a Hopf bifurcation ($\omega>0$).
Note that the flow problem considered in this work is not spatially periodic or homogeneous in either $x,y,z$ directions, and $\hat{\boldsymbol{q}}$ depends on all the three coordinates, leading to a global stability problem \citep{Theofilis2011}.

\subsection{Sensitivity analysis} \label{subsec:adjLin}
Sensitivity analyses based on the adjoint approach \citep{luchini2014adjoint} will be conducted to identify the instability mechanism responsible for the unsteadiness. Following \cite{giannetti2007}, the adjoint equations of the linearised Navier--Stokes equations read
\begin{align}
	-\frac{\p \up}{\p t} - \Ub\bcdot(\nabla\up) + (\nabla\Ub)\bcdot\up =  -\nabla \pp+ \frac{1}{Re} \nabla^2\up, \ \ \ \ \ \  \nabla\bcdot\up=0,
	\label{eq:adjLNS}
\end{align} 
where $\up$ and $\pp$ are the adjoint vector of perturbation field $\u$ and $p$, respectively. Following \cite{giannetti2007,marquet2008,citro2016}, the boundary conditions of the adjoint equations are set as
\begin{subeqnarray}
 \up= \boldsymbol{0} \quad &\rm{on} \quad \it S_c, S_{in}, S_{xz} \ \rm{and} \ \it S_{xy},\\ 
  \pp \n- Re^{-1}(\nabla \up)\bcdot\n=(\Ub\bcdot \n)\up \quad& \rm{on} \quad \it S_{out}. \label{eq:outflowadj}
\end{subeqnarray}

The identification of the core region of the instability can help to understand the instability mechanism \citep{giannetti2007,luchini2014adjoint}. According to \cite{giannetti2007}, the sensitivity wavemaker ${\zeta}$ can be identified by overlapping the direct eigenvector $\u$ and adjoint eigenvector $\up$,
\begin{align}
  {\zeta} =\frac{|\u||\up|}{\langle\u,\up\rangle} .
  \label{wavemaker}
\end{align}

\subsection{Numerical method}\label{sec:Numerical methods}
In order to obtain the accurate wake pattern and the base states of the fully 3-D flow past a short rotating cylinder at medium and low Reynolds numbers, we adopt the high-order parallelised open-source code Nek5000 \citep{nek5000} (version 19.0), which is based on the nodal spectral element method (SEM) originally proposed by \cite{patera1984SPE}. Hexahedral elements with a polynomial order $N=7$ are used, which indicates that there are eight points in each spatial dimension of the element \citep{nek5000}. 
The time step $\Delta t$ is determined by the Courant--Friedrichs--Lewy (CFL) condition with the target Courant number $\lesssim 1.0$. Following the practice in our previous work \cite{yang_feng_zhang_2022}, the boundary layer elements in the vicinity of the rotating cylinder have been refined by the O-type mesh (see {\color{blue}figure~}\ref{fig:CD_BC}$b$). The only difference lies in the enlarged computational domain and increased number of elements to accommodate the rotational effects.

We  focus on studying two types of base states (both of which are denoted as $\boldsymbol {Q}_b=(\Ub,P_b)^{\rm T}$ in the text to follow as long as there is no confusion) to analyse their global dynamics. 
The flow will become unstable in a certain region of the parametric space. The unstable base flow $\boldsymbol {Q}_b$ in this case cannot be obtained directly by time-evolving the NS equations. For these unstable base flows, the selective frequency damping (SFD) method proposed by \cite{SFD2006} was used to obtain an equilibrium solution to the NS equations \ref{eq:NS}. This type of base state will be called (SFD) base flow.
Another base state of interest is the mean flow, which is obtained by time averaging the periodic flow with vortex shedding. For the present global stability analysis, at least ten vortex shedding cycles will be used in the time-average procedure. 

For a non-parallel 3-D flow past the finite rotating cylinder, the numerical discretisation of linearised NS equation \ref{eq:LNS} will result in a large-scale Jacobian matrix $\boldsymbol{A}$ in the generalised eigenvalue problem \ref{eq:eig}. It is impractical to solve a large-scale eigenvalue problem for its whole eigenspectrum in a 3-D flow. Based on Nek5000 solver and the ARPACK package \citep{lehoucq1998arpack}, the matrix-free time-stepper method \citep{Theofilis2011,doedel2012} will be adopted in the present work, implementing the Implicitly Restarted Arnoldi Method (IRAM) \citep{radke1996MatIRAM,lehoucq1998arpack}.

The validation of the nonlinear DNS code and the linear stability code is provided in Appendix \ref{app_validation} together with a convergence study on the size of the computational domain.

\section{Result and Discussions} \label{sec:results}
\subsection{Base states} \label{subsec:BS}
\subsubsection{Base flows and pressure by SFD method} \label{subsec:BF}
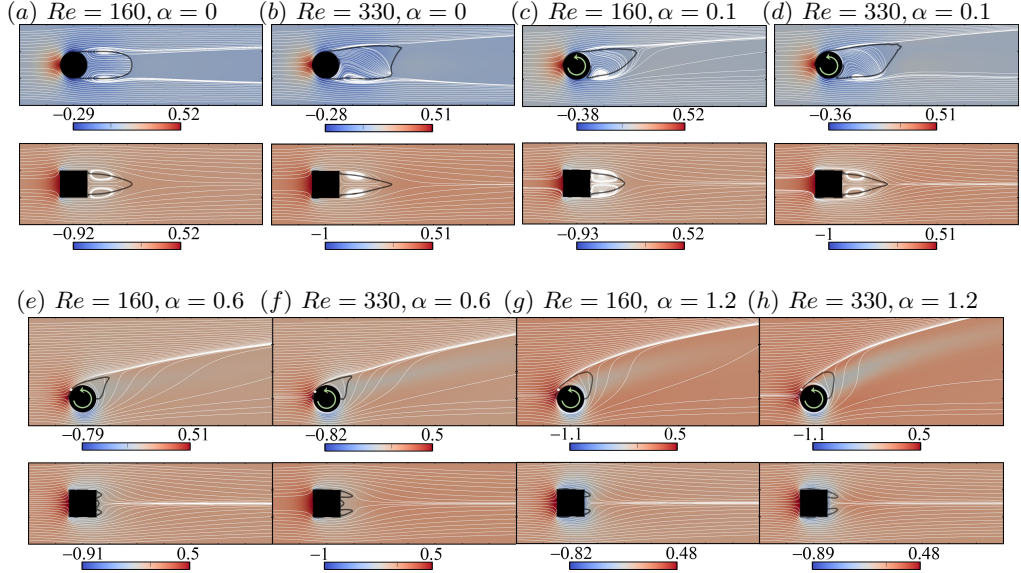
\begin{figure}
	\hfil
	\centering\includegraphics[trim=0.0cm 0cm 0.0cm 0cm,clip,width=0.24\textwidth]{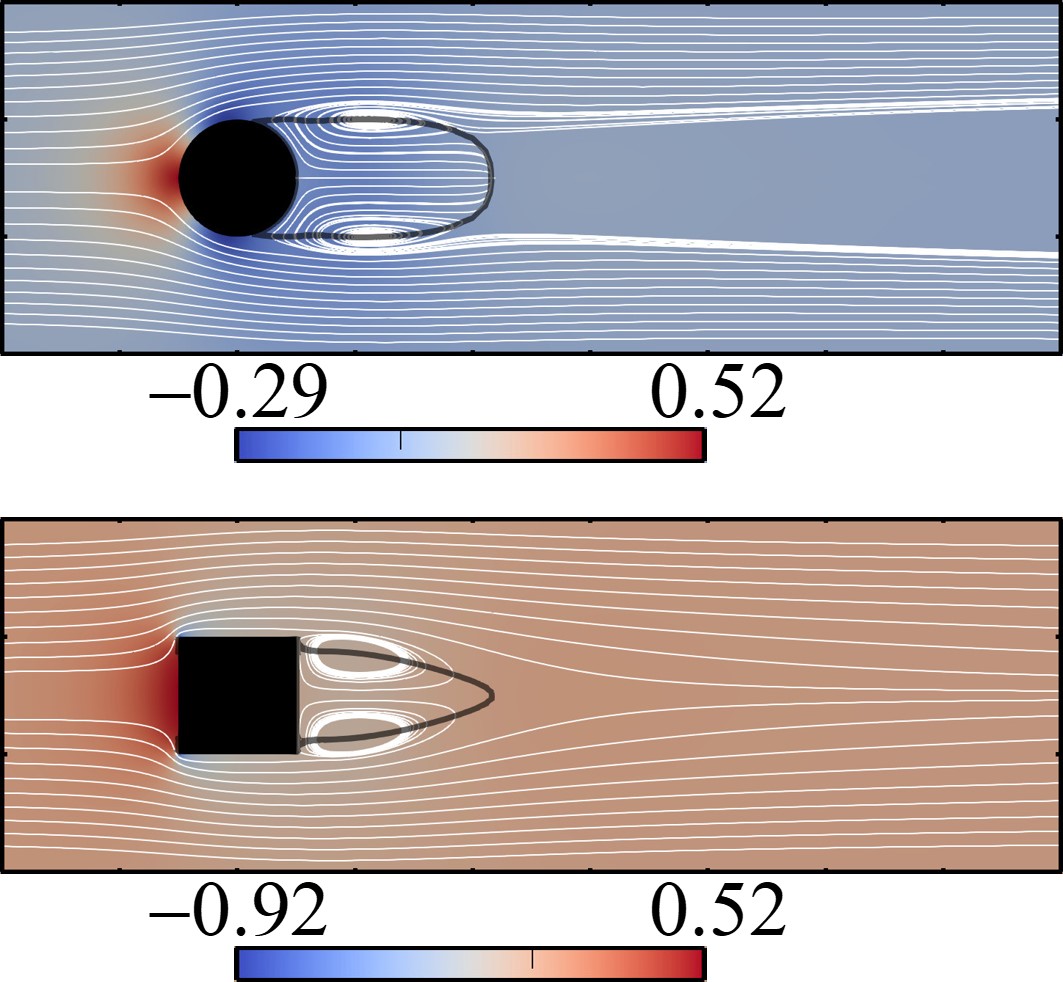}\put(-97,87.5){($a$) $Re=160, \alpha=0$}\  
	\centering\includegraphics[trim=0.0cm 0cm 0.0cm 0cm,clip,width=0.24\textwidth]{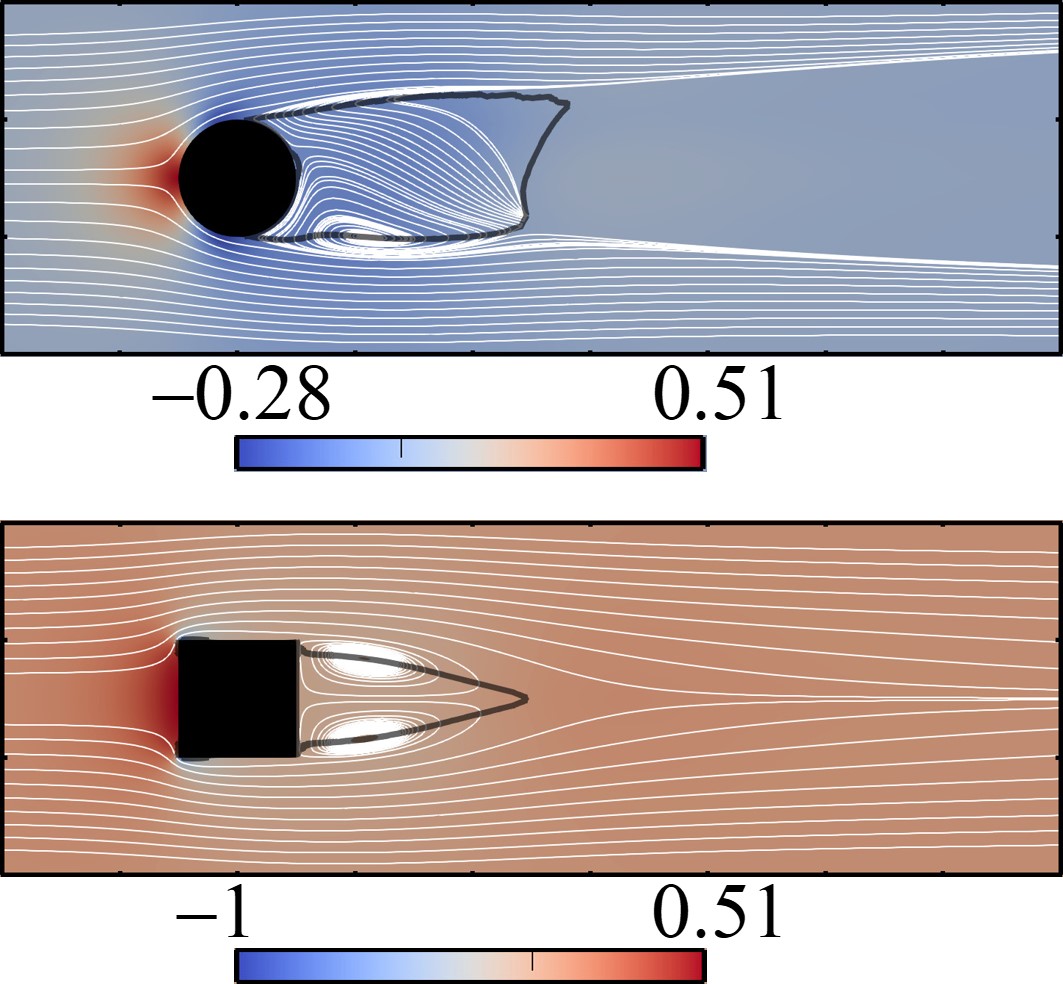}\put(-97,87.5){($b$) $Re=330, \alpha=0$}\  
	\centering\includegraphics[trim=0.0cm 0cm 0.0cm 0cm,clip,width=0.24\textwidth]{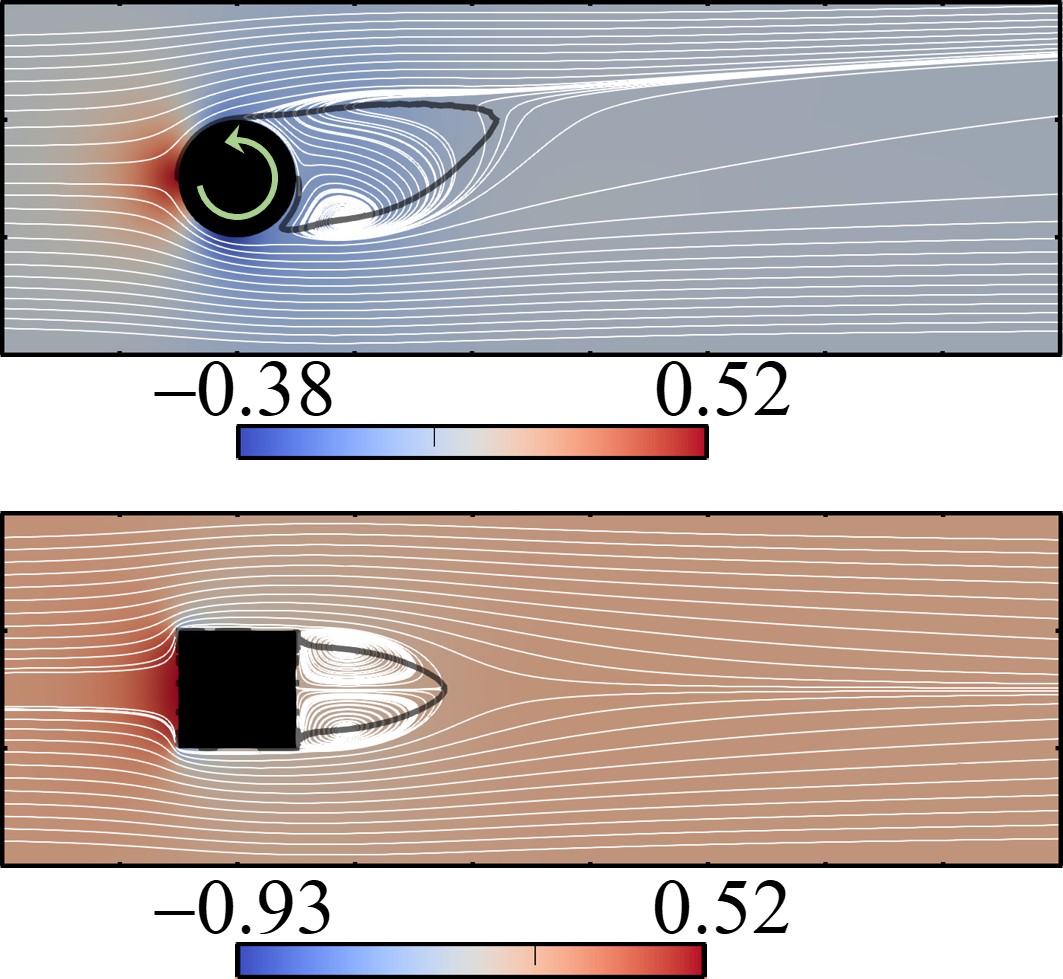}\put(-97,87.5){($c$) $Re=160, \alpha=0.1$}\ 
	\centering\includegraphics[trim=0.0cm 0cm 0.0cm 0cm,clip,width=0.24\textwidth]{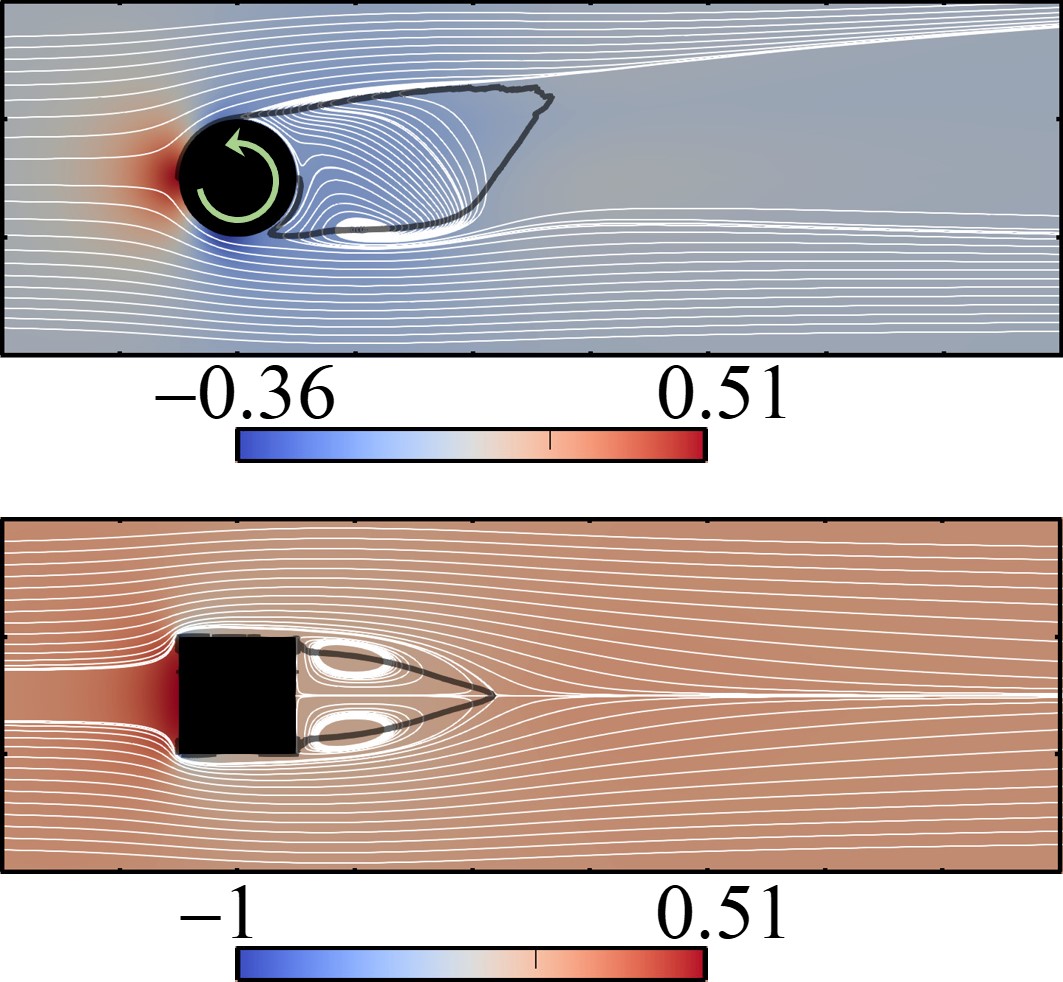}\put(-97,87.5){($d$) $Re=330, \alpha=0.1$}\\
	\vspace{0.5cm}
	
	\centering\includegraphics[trim=0.0cm 0cm 0.0cm 0.0cm,clip,width=0.24\textwidth]{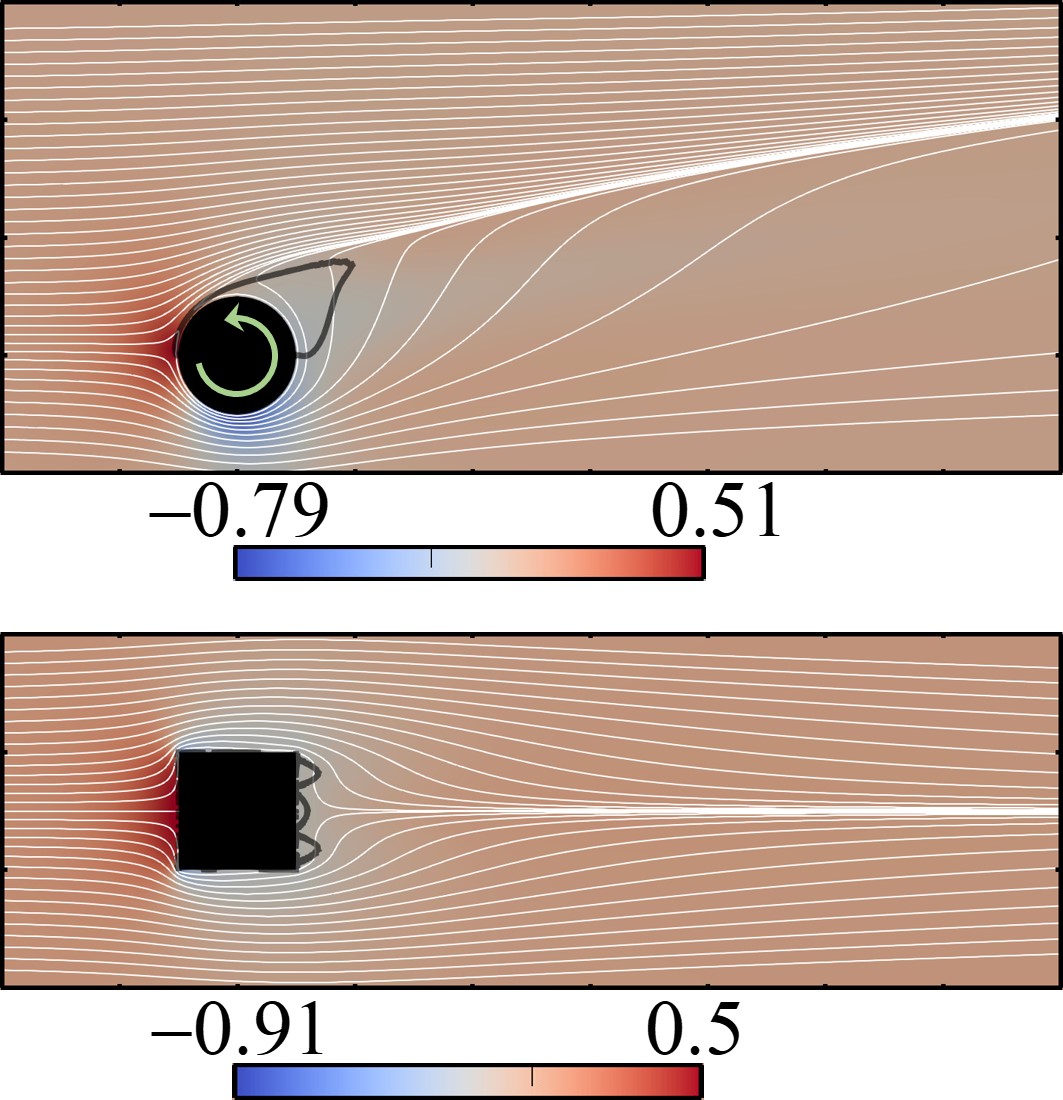}\put(-97,99){($e$) $Re=160, \alpha=0.6$} \ 
		\put(-77.4,67.3){\color{white}{.}}
	\centering\includegraphics[trim=0.0cm 0cm 0.0cm 0cm,clip,width=0.24\textwidth]{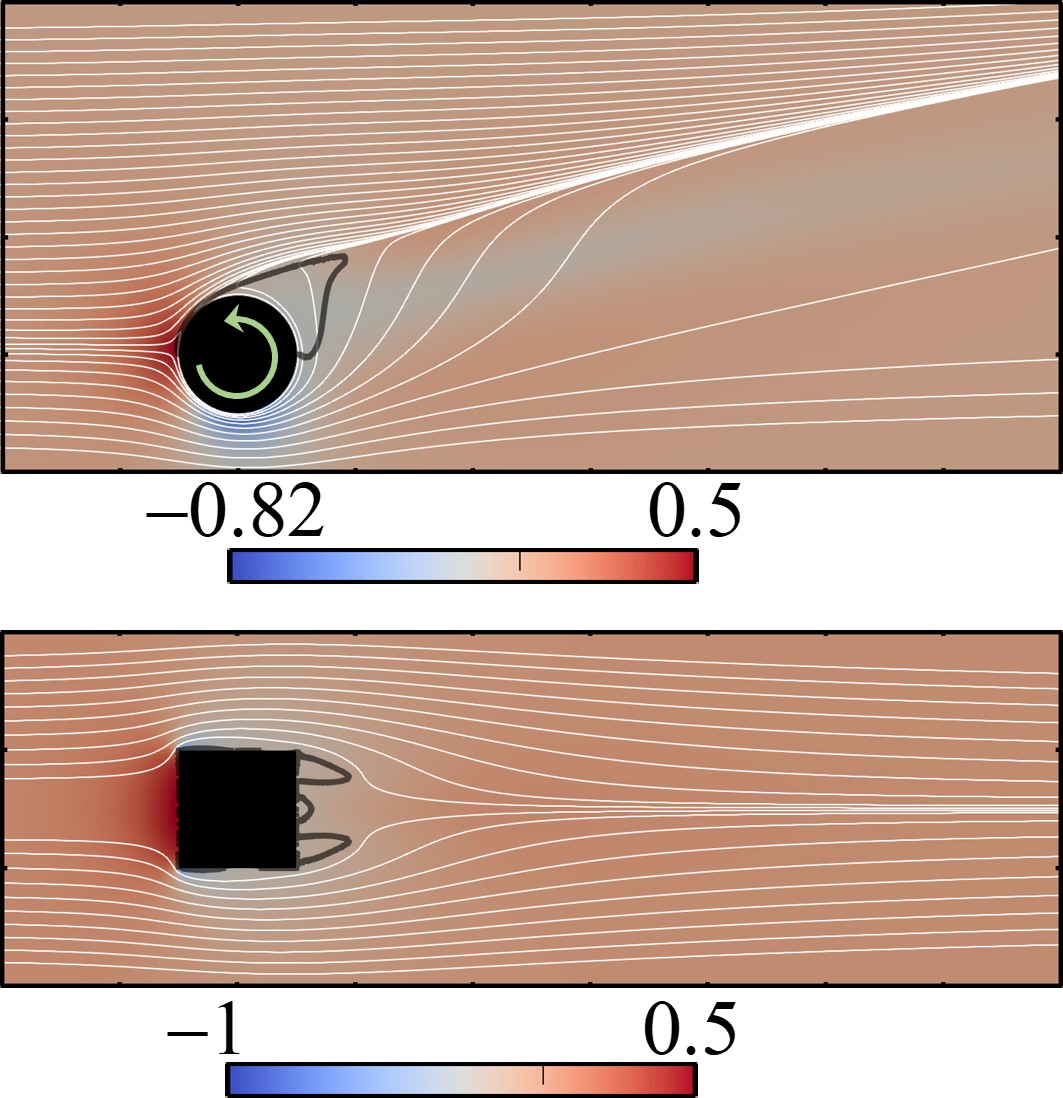}\put(-97,99){($f$) $Re=330, \alpha=0.6$} \ 
			\put(-77.3,66.9){\color{white}{.}}
	\centering\includegraphics[trim=0.0cm 0cm 0.0cm 0cm,clip,width=0.24\textwidth]{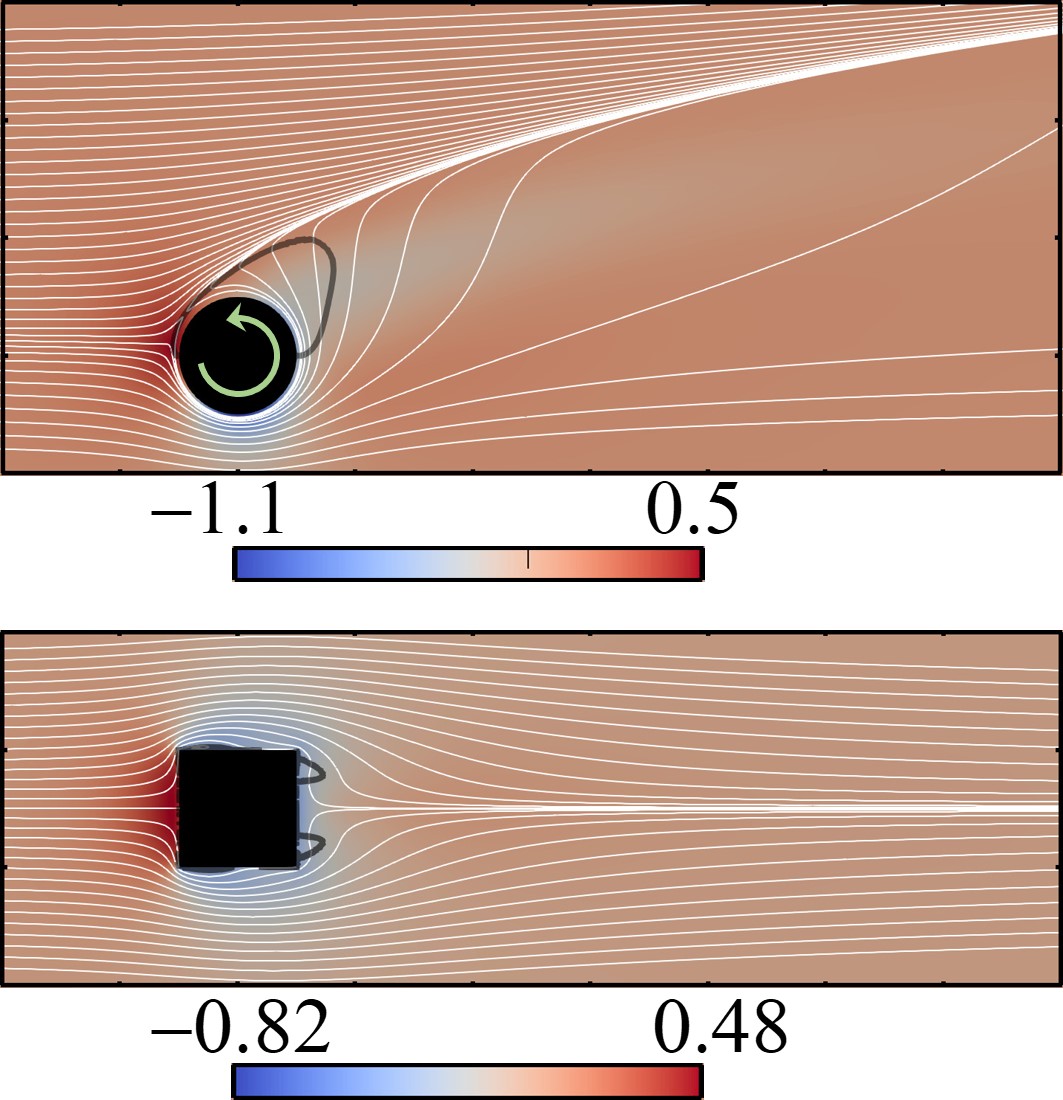}\put(-97,99){($g$) $Re=160,$ $\alpha=1.2$} \ 
			\put(-77.4,67.5){\color{white}{.}}
	\centering\includegraphics[trim=0.0cm 0cm 0.0cm 0cm,clip,width=0.24\textwidth]{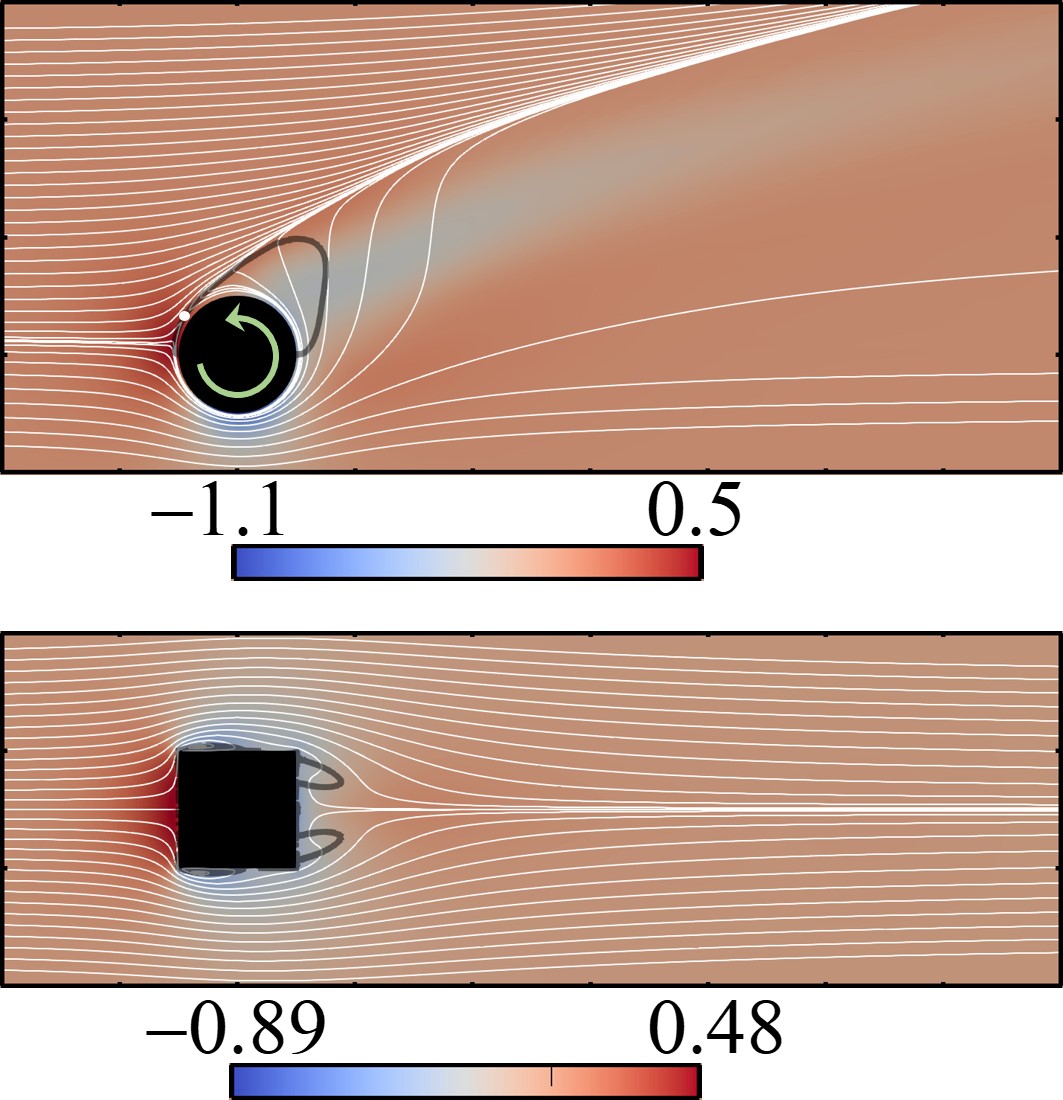}\put(-97,99){($h$) $Re=330, \alpha=1.2$}
	\caption{Steady SFD base flow past an $\ar=1$ short cylinder. The colour illustrates pressure contour and the white lines streamlines. ($a$) $Re=160, \alpha=0$;  ($b$) $Re=330, \alpha=0$; ($c$) $Re=160, \alpha=0.1$; ($d$) $Re=330, \alpha=0.1$; ($e$) $Re=160, \alpha=0.6$; ($f$) $Re=330, \alpha=0.6$; ($g$) $Re=160,$ $\alpha=1.2$; ($h$) $Re=330, \alpha=1.2$. For each subgraph, the top panel shows the flow visualisation at the plane $z=0$ and the bottom panel that at the plane $y=0$. The rotation direction of the cylinder is counterclockwise. The black translucent thick solid lines denote the recirculating region separatrix, which is identified by $U_x=0$.}
	\label{fig:BS_ar1.0}
\end{figure}

In this section, we present the base state of the flow past a short rotating cylinder. We focus on the case of $\ar=1$ and vary the values of $Re$ and $\alpha$. Both temporally mean flow and the steady base flow (solved using the SFD method if unstable) will be displayed and discussed. By extracting pressure field contours and streamlines, we show in figure \ref{fig:BS_ar1.0} the typical unstable steady base flow obtained by the SFD method.
First, we review the flow past a finite-length cylinder without rotation $(\alpha=0)$ in panels ($a,b$). For $Re<Re_c=172.2$ (figure \ref{fig:BS_ar1.0}$a$), the wake flow is steady, which has two mutually perpendicular symmetry planes $xy$ and $xz$, passing through the geometric center of the cylinder (as also shown in figure 5 of \citealt{yang_feng_zhang_2022}. The value of $Re_c=172.2$ has also been determined from this work). The wake consists of a recirculation region with closed streamlines, characterized by planar symmetrical separation points and foci.
For $Re>Re_c= 172.2$, the flow undergoes a regular bifurcation, resulting in spontaneous symmetry breaking in the $xz$ plane due to nonlinear wake dynamics, as shown in figure \ref{fig:BS_ar1.0}$b$ for the case $Re=330, \alpha=0$. 
When viewed from the remaining symmetric $xy$ plane with $z=0$ (top panel), one vortex structure in the wake becomes stronger than the other. 
Besides, the positions of the two separation points slightly change, and the pressure contours indicate that the pressure near the upper part of the cylinder arc surface is larger than that of the lower part, which results in a negative lift coefficient $C_{ly}$. 
The sign of the $C_{ly}$ in the case of no rotation is unimportant because of the symmetric setting.

Figure \ref{fig:BS_ar1.0}$(c,d)$ show respectively the base flow structures of cases $Re=160$ (stable) and $Re=330$ (unstable) at a low rotation rate $\alpha =0.1$. Even for a low $Re=160$, it can be seen that the rotation breaks the symmetry of the wake, leaving the flow only symmetric with respect to the $oxy$ plane. Compared with the non-rotation case (figure \ref{fig:BS_ar1.0}$a$), the positions of the upper and lower separation points shift significantly along the rotation direction. Overall, the asymmetric recirculation region generated by the weak rotation is similar to the asymmetric wake generated by regular bifurcation in the non-rotated case. The counterclockwise rotation makes the flow velocity near the cylinder arc surface on the lower side increase, and the velocity near the upper wall surface decrease. It can be deduced that the pressure near the lower side wall will decrease, and the pressure near the upper side wall will increase, so a negative lift $C_{ly}$ will be obtained, i.e. the Magnus effect. It can be seen from figures \ref{fig:BS_ar1.0} $(c-h)$ that the numerical results conform to the classical Magnus effect. In the parameter space studied in this paper (especially relatively low $Re$), there is no boundary layer transition from a laminar flow to turbulence, that is, there is no inverse Magnus effect \citep{kim2014inverse}. 
As mentioned above, the symmetry breaking effect caused by the regular bifurcation resembles that due to the rotation (see also the rotating sphere by \citealt{citro2016}). Therefore, the rotation 'strengthens' the symmetry breaking caused by the inherent wake mechanism observed in the regular bifurcation without rotation, resulting in the lift force in the rotating cases in figures \ref{fig:BS_ar1.0} $(c-h)$ being larger, to be shown in figure \ref{fig:Re_cdcl_alpha}($b$). On the other hand, the drag coefficient $C_{d}$ is not very sensitive to the rotation when at low rotation rate $\alpha<0.3$, as shown in figure \ref{fig:Re_cdcl_alpha} (a).

Figures \ref{fig:BS_ar1.0}($e$) $(Re=160)$ and \ref{fig:BS_ar1.0}($f$) $(Re=330)$ show the base states at moderate rotation $\alpha=0.6$, and both their nonlinear saturation states are steady. So the base flow and mean flow are the same for these cases. Compared with the previous low rotation case, the flow topology further changes, i.e. the recirculation zone in the wake almost disappears, in either the $xy$ or $xz$ plane. But a stagnation point (white point in the panel) similar to that observed in the 2-D rotating cylinder (see figure 3$b$ in \citealt{sierra2020bifurcation}) appears, which is located in the second quadrant of the cylinder. 

Continuing to increase the rotation rate, the figures \ref{fig:BS_ar1.0}($g$) ($Re=160$) and \ref{fig:BS_ar1.0}($h$) ($Re=330$) show the cases of a high rotation rate ($\alpha=1.2$), and their wake structure and pressure distribution are similar to those of figures \ref{fig:BS_ar1.0}($e$) and \ref{fig:BS_ar1.0}($f$), but the wakes are swung further upwards. The hyperbolic stagnation point (see the white dot) on the upper side moves forward along the direction of tangential velocity, which is similar to the results of \cite{citro2016} on a rotating sphere at a high rotation rate.

\begin{figure}
%	\hfil
	\centering\includegraphics[trim=5.3cm 1cm 6.7cm 1.0cm,clip,width=0.46\textwidth]{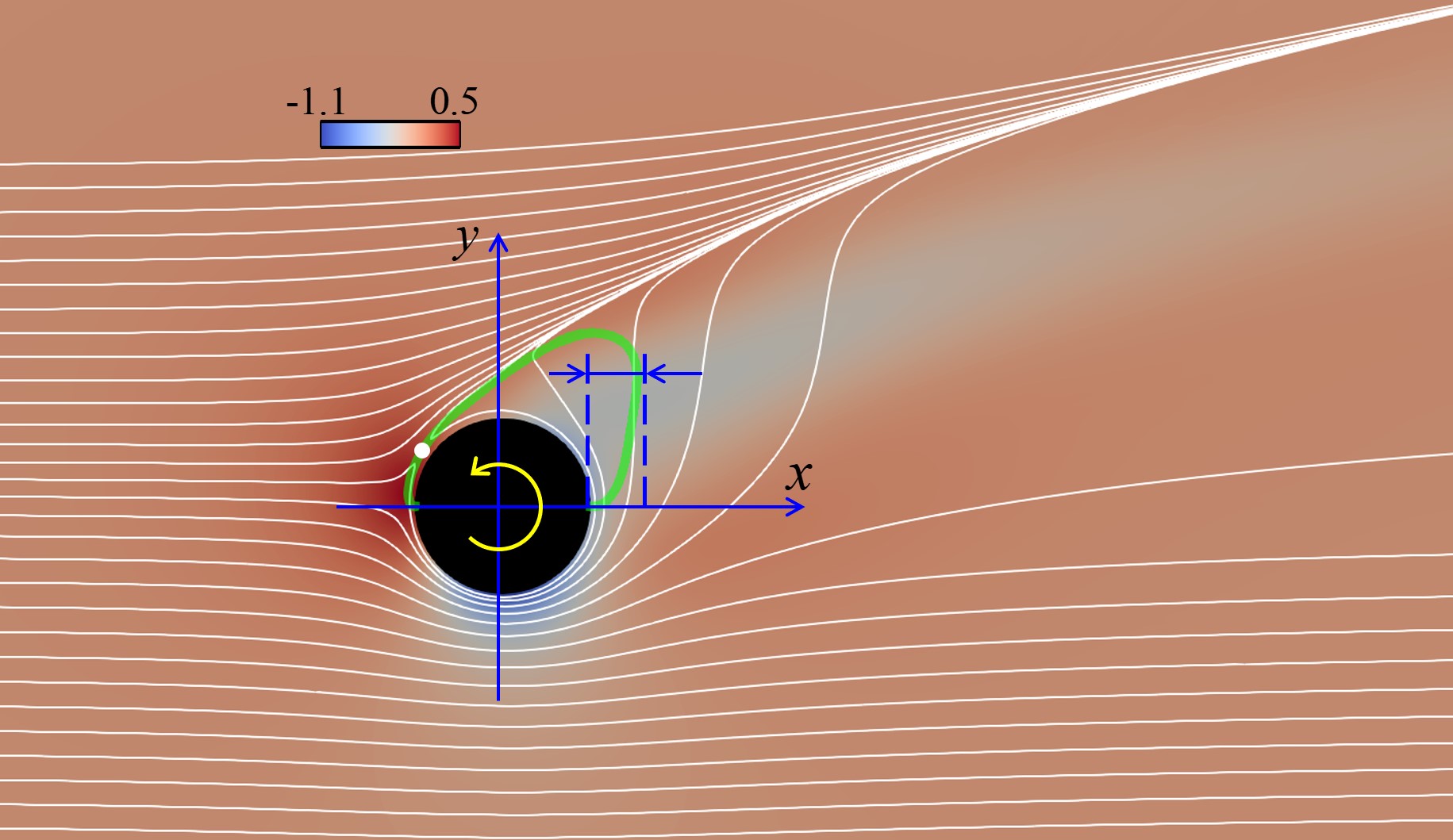} \put(-165,80){$U_x=0$}
	\put(-155,48){$\theta_s$}  \put(-96,86){$x_s$}  \put(-105,48){$\theta_s=\pi$} \put(-188,152){($a$)}\quad
	\centering\includegraphics[trim=0.0cm 0.0cm 0.0cm 0cm,clip,width=0.475\textwidth]{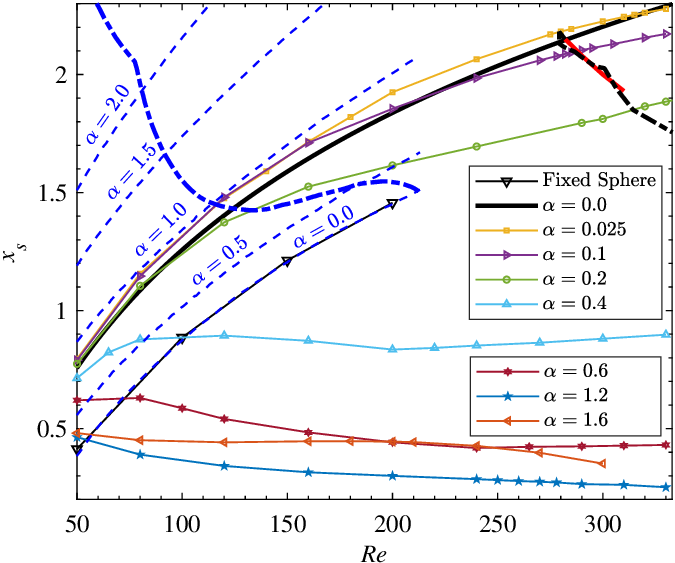}	\put(-188,151){($b$)}  \\
	\caption{$(a)$ A separation bubble under the effect of rotation in plane $oxy$, whose area is surrounded by the green curve ($U_x(x,y,z=0)=0$) and the wall of the rotating cylinder. The white dot shows the position of the hyperbolic stagnation point. The colour represents the pressure field. $(b)$ Separation bubble length $x_s$ of the SFD base flow as a function of $Re$, compared with a fixed sphere \citep{johnson1999flow} and a sphere \citep{sierra2022unveiling} rotating along the streamwise (blue dashed lines). The blue and black dash-dotted lines represent the $x_s$ obtained from the neutrally stable SFD base flow of rotating sphere \citep{sierra2022unveiling} and the present cylinder, respectively. The black and red thick lines represent the $x_s$ obtained from the SFD base flow and the mean flow of a fixed finite cylinder ($\alpha=0$) \citep{yang_feng_zhang_2022}, respectively.}
	\label{fig:Re_xs_alpha}
\end{figure}
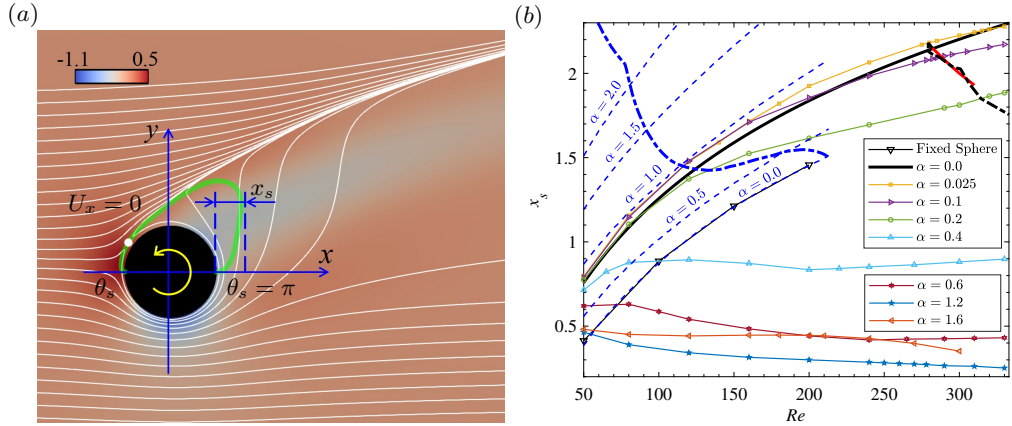

\subsubsection{Separation bubble}
Next, we discuss figure \ref{fig:Re_xs_alpha} where the streamlines are further analysed.
Here, the separation point is defined as the intersection between the envelope of the separation bubble (indicated by the green curve with $U_x(x,y,z=0)=0$ in figure \ref{fig:Re_xs_alpha}$a$) and the surface of the cylinder, and thereby, the separation angle $\theta_s$ is defined as the angle from the negative $x$-direction to the separation point. In the present rotating cylinder, the values of parameters Reynolds number, rotation rate, and aspect ratio have no effect on the separation angle. The separation bubble is always confined between angles $\theta_s=0$ and $\theta_s=\pi$, as shown in figure \ref{fig:Re_xs_alpha}$(a)$. Note that in the case of non-rotation cylinder \citep{yang_feng_zhang_2022}, the separation angles may be varying depending on $Re$ and aspect ratio.
We define the separation bubble length as $x_s=\max\{x-D/2 \  \vert\  U_x(x,y,z=0) \leq 0 \}$, see panel \ref{fig:Re_xs_alpha}($a$). Panel \ref{fig:Re_xs_alpha}$(b)$ shows $x_s$ as a function of $Re$ for the $\ar=1$ cylinder, compared with the separation length of a fixed sphere \citep{johnson1999flow} and a non-rotating finite-length cylinder \citep{yang_feng_zhang_2022}.
The effect of Reynolds number on $x_s$ is different for the high- and low-rotation-rate cases. When the rotation rate is low, the increase of Reynolds number makes $x_s$ increase monotonically, which is similar to the (mean flow) cases of fixed finite-length cylinder \citep{yang_feng_zhang_2022} at $0.5\leq \ar \leq2$ prior to Hopf  bifurcation, sphere \citep{johnson1999flow}, and rotating sphere \citep{sierra2022unveiling}. The increase of $\alpha$ decreases the value of $x_s$ at $0.025\leq \alpha \leq1.2$, which means that the rotation shortens the recirculation length. It has been proposed that shortening the recirculation length stabilises the flow; for example, \cite{sierra2022unveiling} discussed such a stabilising control strategy.

When the rate $\alpha$ is high, $x_s$ slightly decreases with the increase of $Re$, for example, see $\alpha=0.6,1.2$ in figure \ref{fig:Re_xs_alpha}$(b)$. According to the above discussion on the relation between rotation, recirculation length and flow stability, we cannot make the present rotating cylinder flow more stable at high $\alpha \geq 0.6$-$1.2$ by shortening the base flow recirculation region along the streamwise direction because the value of $x_s$ bounces back when $\alpha \geq 0.6$-$1.2$. We provide a further mechanistic explanation of this result in the structural sensitivity analysis section \ref{subsec:SSA}. Besides, the higher the rotation rate, the less sensitive the value of $x_s$ is to the variation of the Reynolds number.

For fixed $Re$, the effect of rotation rate $\alpha$ on $x_s$ is not monotonous, unlike the sphere rotating along the streamwise axis \citep{sierra2022unveiling}.
\cite{sierra2022unveiling} also suggested that $x_s$ does not have to increase monotonically as $\alpha$ increases after the bifurcation, which can also be seen in the time-averaged recirculation length of the sphere rotating along the streamwise \citep{kim_choi_2002,lorite2020description}. 
 
\begin{figure}
	\hfil
	\centering\includegraphics[trim=0.0cm 0cm 0.0cm 0cm,clip,width=0.48\textwidth]{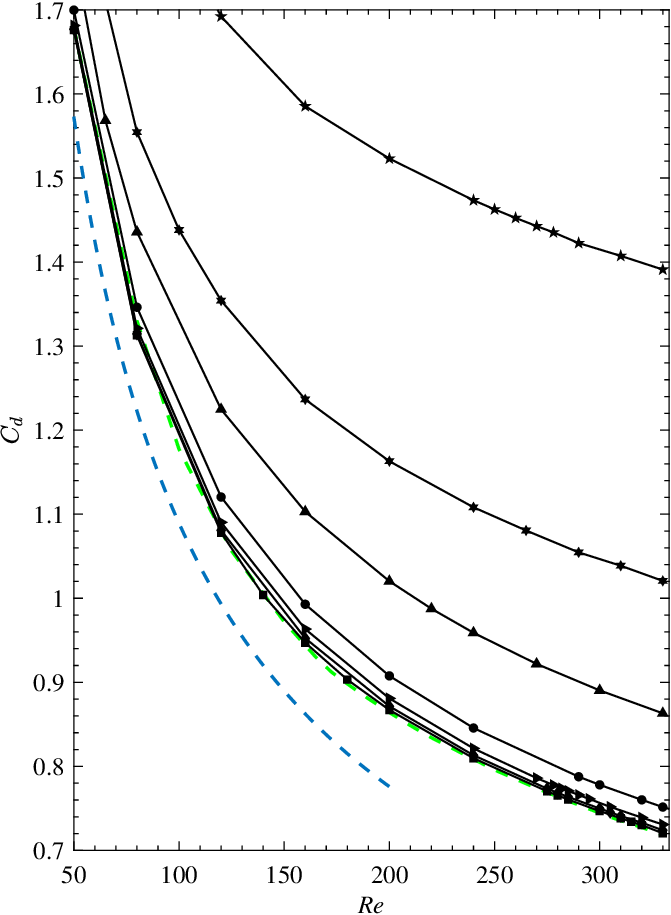} \put(-187,255){($a$)}\quad
	\centering\includegraphics[trim=0.0cm 0cm 0.0cm 0cm,clip,width=0.48\textwidth]{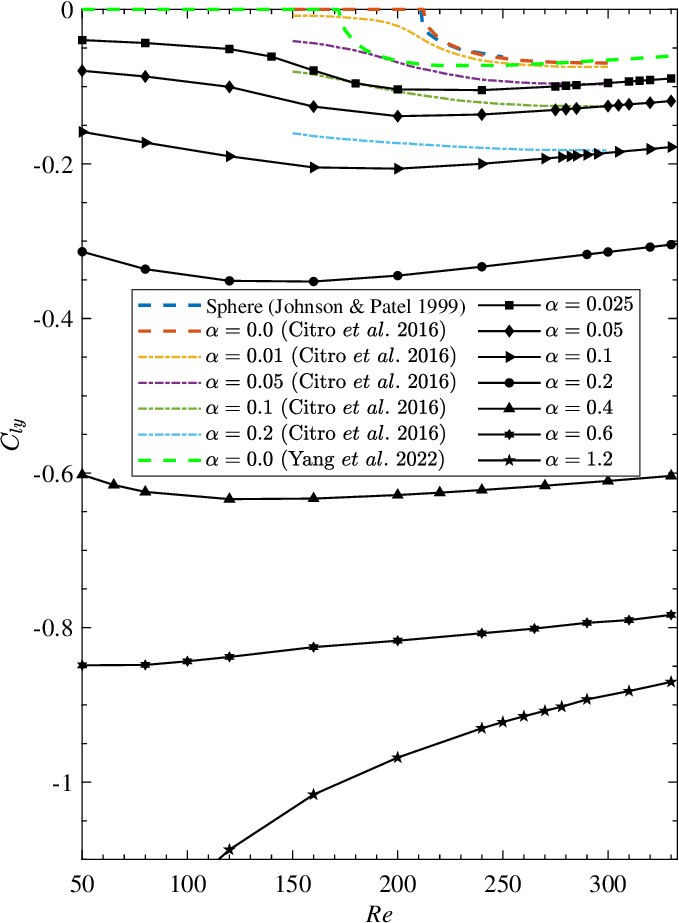}  \put(-188,255){($b$)} 
	\caption{Drag (panel $a$) and lift (panel $b$) coefficients of the steady SFD base flow as a function of $Re$ at $\ar=1$.  Comparison with a fixed sphere \citep{johnson1999flow}, a fixed finite cylinder \citep{yang_feng_zhang_2022}, and a sphere rotating about the transverse direction \citep{citro2016}.}
	\label{fig:Re_cdcl_alpha}
\end{figure} 
 
\subsubsection{Drag and lift coefficients}
Now, we turn to figure \ref{fig:Re_cdcl_alpha} to discuss the drag and lift coefficients in the short rotating cylinder flow. 
{Starting from the fixed finite-length cylinder case \citep{yang_feng_zhang_2022}, panel ($a$) shows that the drag coefficient increases as the rotation rate increases, which is caused by the fact that the rotation makes the negative pressure in the leeward area smaller, but has little effect on the maximum positive pressure at the stagnation point on the windward side. 
These can be seen from the pressure contour in panels $a,b$ and $c,d$ of figures \ref{fig:BS_ar1.0}, and more quantitative results on the lift and drag coefficients varying with $\alpha$ are shown in figures \ref{fig:alpha_Coefficient}($a$) and ($b$) in Appendix \ref{app_validation}. A comparative analysis reveals that the augmentation of $\alpha$ increases the lift coefficient of both the 2-D infinite cylinder and present 3-D finite-length cylinder. 
However, in the 3-D case, the lift coefficient ceases to increase further when $\alpha$ approaches 2, whereas that in the 2-D case increases monotonically with approaching 2.5 \citep{stojkovic2003new,mittal2003flow}.
The effect of $\alpha$ on drag shows contrasting behaviors between the 2-D infinite cylinder and present 3-D finite-length cylinder cases within $\alpha<2.5$. Note that our analysis is based on the drag and lift in the SFD base flow, whereas the drag and lift for the 2-D infinite cylinder are time-averaged \citep{stojkovic2003new,mittal2003flow}.

\begin{figure}
\includegraphics[trim=0.0cm 0cm 0.6cm 0cm,clip,width=0.495\textwidth]{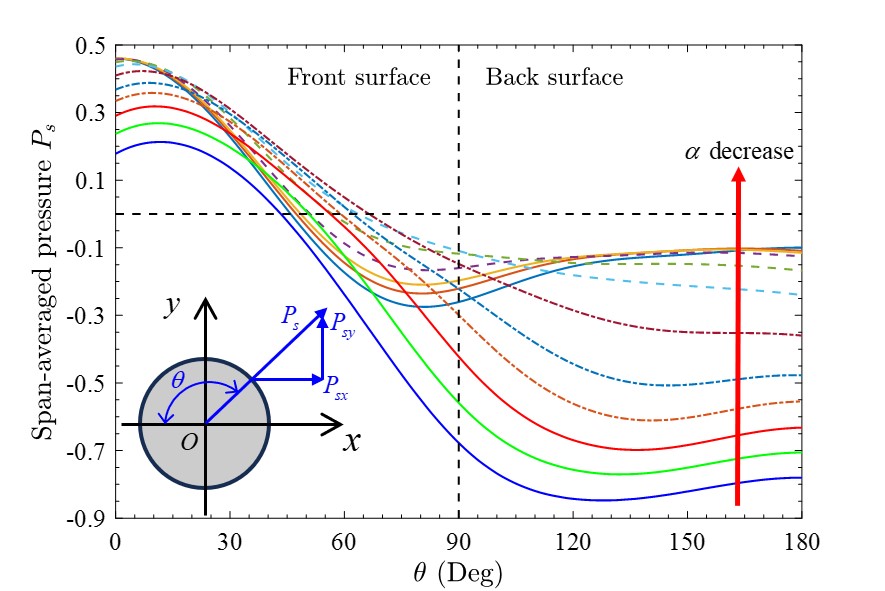}	\put(-190,125){($a$)} \
\includegraphics[trim=0.0cm 0cm 0.6cm 0cm,clip,width=0.495\textwidth]{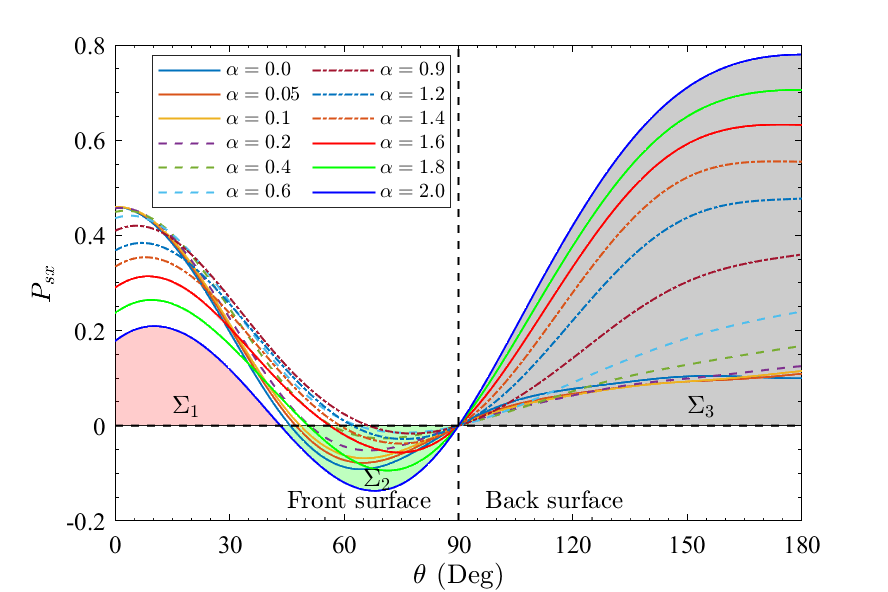}	\put(-190,125){($b$)}
\caption{The distribution of span-averaged pressure $P_s$ (panel $a$) acting on the upper surface, as well as its component $P_{sx}=P_s \rm{cos}\it\theta$ (panel $b$) along the $x$-axis, is characterized with respect to the angle $\theta$. The conditions considered involve cases of ($Re=160, AR=1, \alpha<2)$. As an example, the areas corresponding to the shaded regions in panel ($b$) are denoted as  $\Sigma_1$ (colored by red),$\Sigma_2$  (colored by green) and $\Sigma_3$ (colored by grey) for the case $\alpha=2$.}
\label{fig:pr_theta}
\end{figure}

Despite the presence of rotation, it is evident from figure \ref{fig:alpha_Coefficient} ($a$) that the predominant contributor to the drag is the pressure differential.
As the rotational speed increases, the proportion of pressure drag $\overline{\it C}_{dp}$ to total drag $\overline{\it C}_{d}$ becomes greater. Similarly, the amplitude of the time-averaged lift coefficient $|\overline{\it C}_{lp}|$ generated by the pressure differential remains a predominant component of the total lift coefficient $|\overline{\it C}_{l}|$.
Furthermore, emphasis is placed on the primary determinant of drag, namely, the pressure differential. As shown in figure \ref{fig:pr_theta}, the distribution of spanwise-averaged pressure $P_s$ (panel $a$) acting on the upper surface, as well as its component $P_{sx}=P_s \rm{cos}\it\theta$ (panel $b$) along the $x$-axis, is characterized with respect to the angle $\theta$.  As an example, the areas corresponding to the shaded regions in panel ($b$) are denoted as  $\Sigma_1,\Sigma_2$ and $\Sigma_3$ for the case $\alpha=2$.
So, the drag resulting from the pressure differential can be regarded as the net area enclosed between the blue solid line and the $P_{sx}=0$ axis, specifically $\Sigma_1-\Sigma_2+\Sigma_3$. Consequently, it can be observed that the increase in drag is primarily attributed to a significant reduction (see panel $a$) in pressure $P_{s}$ on the back surface of the cylinder due to rotation, resulting in a substantial enlargement (see panel $b$) of the area  $\Sigma_3$. }

As with a fixed infinite cylinder \citep{schlichting2016boundary}, a fixed sphere \citep{johnson1999flow} and a fixed finite-length cylinder \citep{yang_feng_zhang_2022}, within the parameter range shown in figure \ref{fig:Re_cdcl_alpha}$(a)$, an increase in $Re$ generally leads to a decrease in drag coefficient.
On the other hand, unlike the drag coefficient $C_d$, the values of the lift coefficient $C_{ly}$ in figure \ref{fig:Re_cdcl_alpha}($b$) present more variation when $Re$ changes. At low rate $\alpha\leq0.3$, $C_{ly}$ first decreases with the increase of $Re$ and then increases; at high rate $\alpha\geq0.6$, $C_{ly}$ increases monotonically as $Re$ increases from 50 to 330. Besides, in the interval $0.2<\alpha<0.6$ of figure \ref{fig:Re_cdcl_alpha}($b$), $C_{ly}$ is relatively insensitive to $Re$, and the wake is steady (see the neutral stability curves in the next section).
It is worth noting that the lift coefficient $C_{ly} \vert_{\alpha=c}$ of the rotating cylinder and the $C_{ly}\vert_{\alpha=2c}$ of the rotating sphere gradually coincide for $Re\gtrsim270$ (with $c\leq0.1$ in present works), see the top-right corner in panel ($b$). For example, the values of $C_{ly}\vert_{\alpha=0.025}$ of a rotating cylinder and $C_{ly}\vert_{\alpha=0.05}$ of a rotating sphere, or $C_{ly}\vert_{\alpha=0.05}$ of a rotating cylinder and $C_{ly}\vert_{\alpha=0.1}$ of a rotating sphere, are approximately the same for $Re\gtrsim270$.

\subsection{Global stability analysis} \label{subsec:GSA}
In this section, we will discuss the Hopf bifurcation diagram for the $\ar=1$ case; the effect of aspect ratio will be discussed in section \ref{subsec:SSA}. As the low- and high-rotation-rate flows present different behaviours, we will discuss them separately.

\subsubsection{At low rotation speed $0\le\alpha\leq0.3$} \label{subsec:GSAatlowrotation}

\begin{figure}
	\hfil
	\centering\includegraphics[trim=0.0cm 0cm 0.0cm 0cm,clip,width=0.485\textwidth]{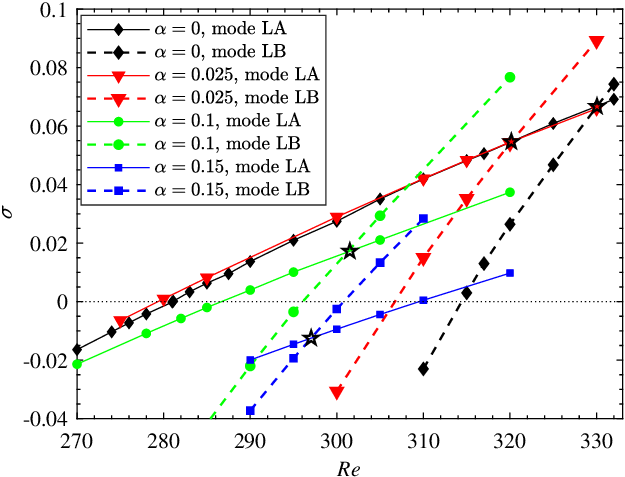}	\put(-190,140){($a$)} \ 
	\centering\includegraphics[trim=0.0cm 0cm 0.0cm 0cm,clip,width=0.485\textwidth]{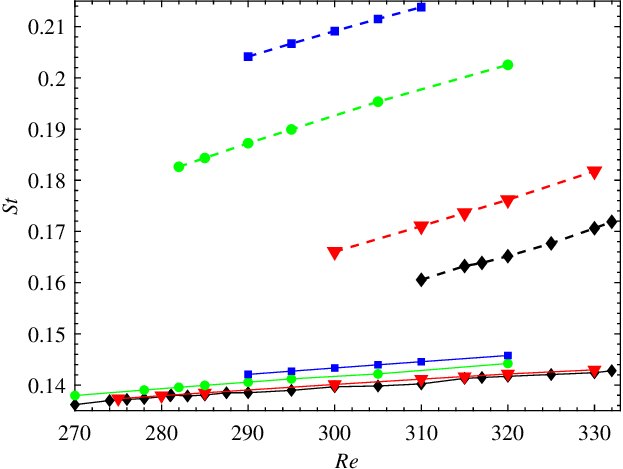}	\put(-190,140){($b$)}\\
	\centering\includegraphics[trim=0.0cm 0cm 0.0cm 0cm,clip,width=0.485\textwidth]{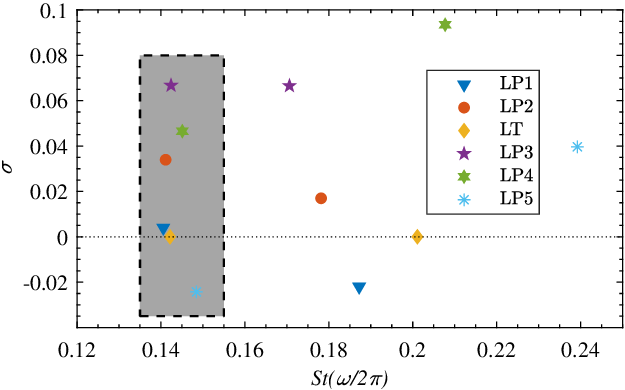}	\put(-190,120){($c$)} \ 
	\caption{The growth rate $\sigma$ (panel $a$) and frequency (panel $b$) of the leading global modes (of the SFD base flow) as a function of $Re$ at $\ar=1$. The position of the black star symbol '\ding{73}' in panel $(a)$ indicates that $\sigma_{LA}=\sigma_{LB}$. $(c)$ The eigenspectra of global modes LA (gray shaded area) and LB at points LP1, LP2, LT, LP3, LP4 and LP5 in figure \ref{fig:alpha_Re_AR1} $(a)$.}
	\label{fig:Re_GR_St}
\end{figure}

\begin{figure}	
	\centering\includegraphics[trim=0.0cm 0cm 4.3cm 0cm,clip,width=0.48\textwidth]{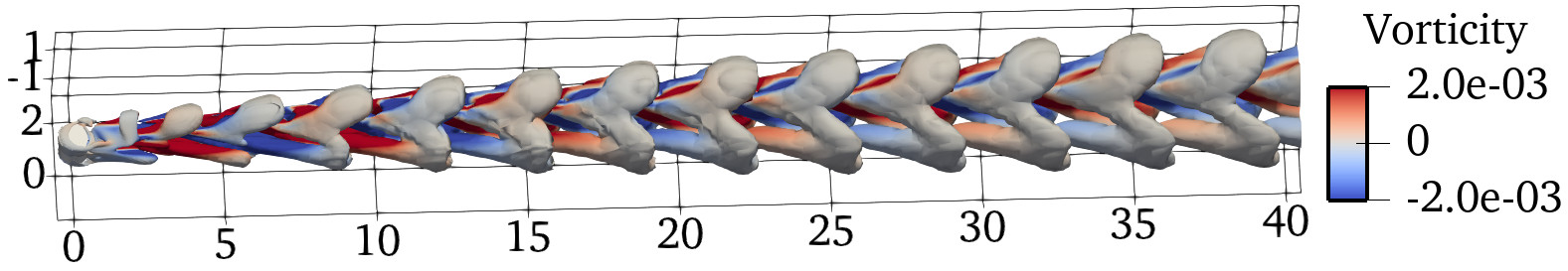}
	\put(-192,38){($a$)}  \quad
	\put(-172,38){LA mode $\lambda_{\rm LA}=4.120\times10^{-3}+\rm{i}0.8833$}  \quad
	\centering\includegraphics[trim=0.0cm 0cm 4.3cm 0cm,clip,width=0.485\textwidth]{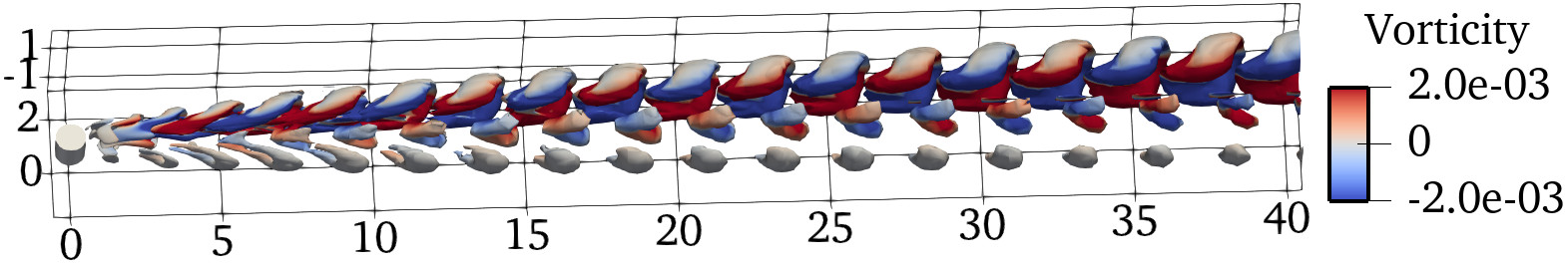}
	\put(-192,38){($b$)}
	\put(-172,38){LB mode $\lambda_{\rm LB} = -2.202\times10^{-2} + \rm{i}1.176$}  
	\vspace{0.4cm}
	\centering\includegraphics[trim=0.0cm 0cm 4.3cm 0cm,clip,width=0.48\textwidth]{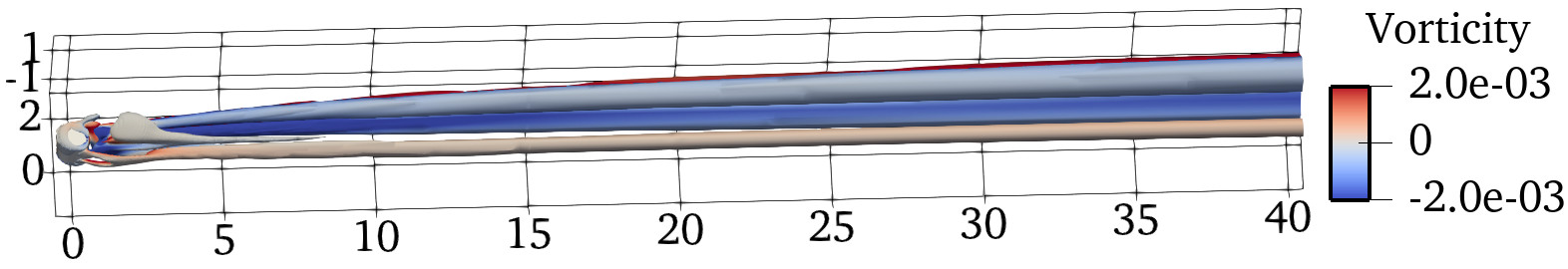}
	\put(-192,38){($c$)} \quad
	\put(-172,38){LC mode $\lambda_{\rm LC} = -6.295\times10^{-2} + \rm{i}0.0$}  \quad	
	\centering\includegraphics[trim=0.0cm 0cm 4.3cm 0cm,clip,width=0.48\textwidth]{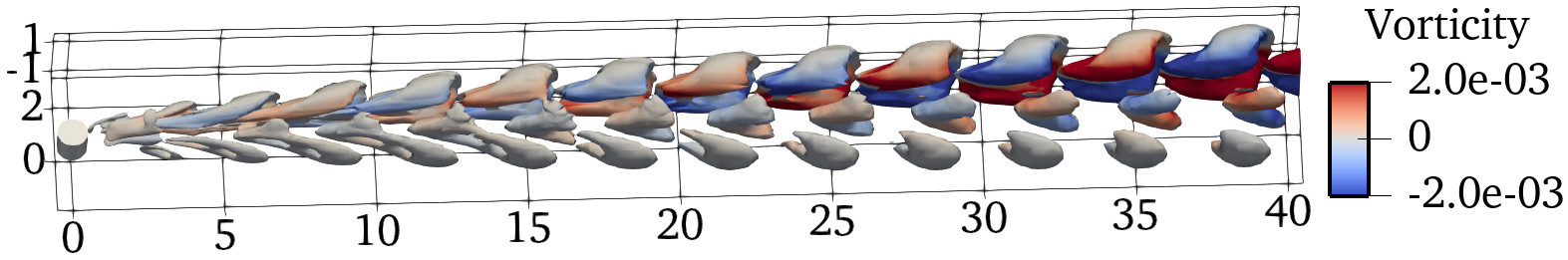}
	\put(-192,38){($d$)}
	\put(-172,38){LD mode $\lambda_{\rm LD} = -8.683\times10^{-2} + \rm{i}0.8687$}  
	\caption{Four representative eigenmodes with weak asymmetry in the flow past a short rotating cylinder at $\alpha=0.1$ and $Re=290$. Mode LA (in panel $a$) and mode LD (in panel $d$) are similar to the mode A and mode B reported by \cite{yang_feng_zhang_2022}, respectively. The new mode LB (in panel $b$) with high frequency is more asymmetric caused by rotation. The mode LC (in panel $c$) has zero frequency. The $Q$-criterion isosurfaces $Q=0$ are colored by the $x$-component of the vorticity ranging from $-2\times 10^{-3}$ to $2\times 10^{-3}$.}
	\label{fig:eigenmodeABCD}
\end{figure}
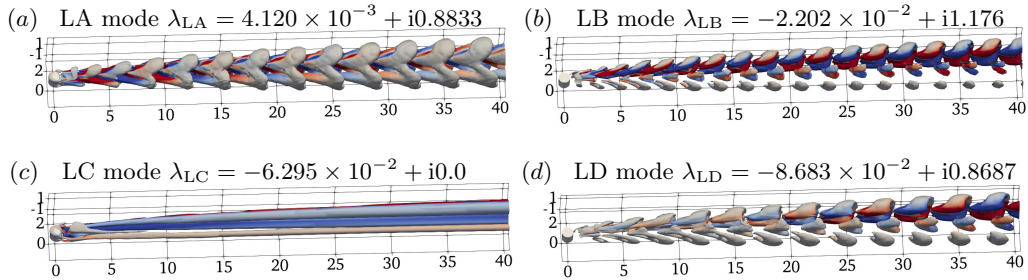

\begin{figure}
	\centering\includegraphics[trim=0.0cm 0cm 0.0cm 0cm,clip,width=0.485\textwidth]{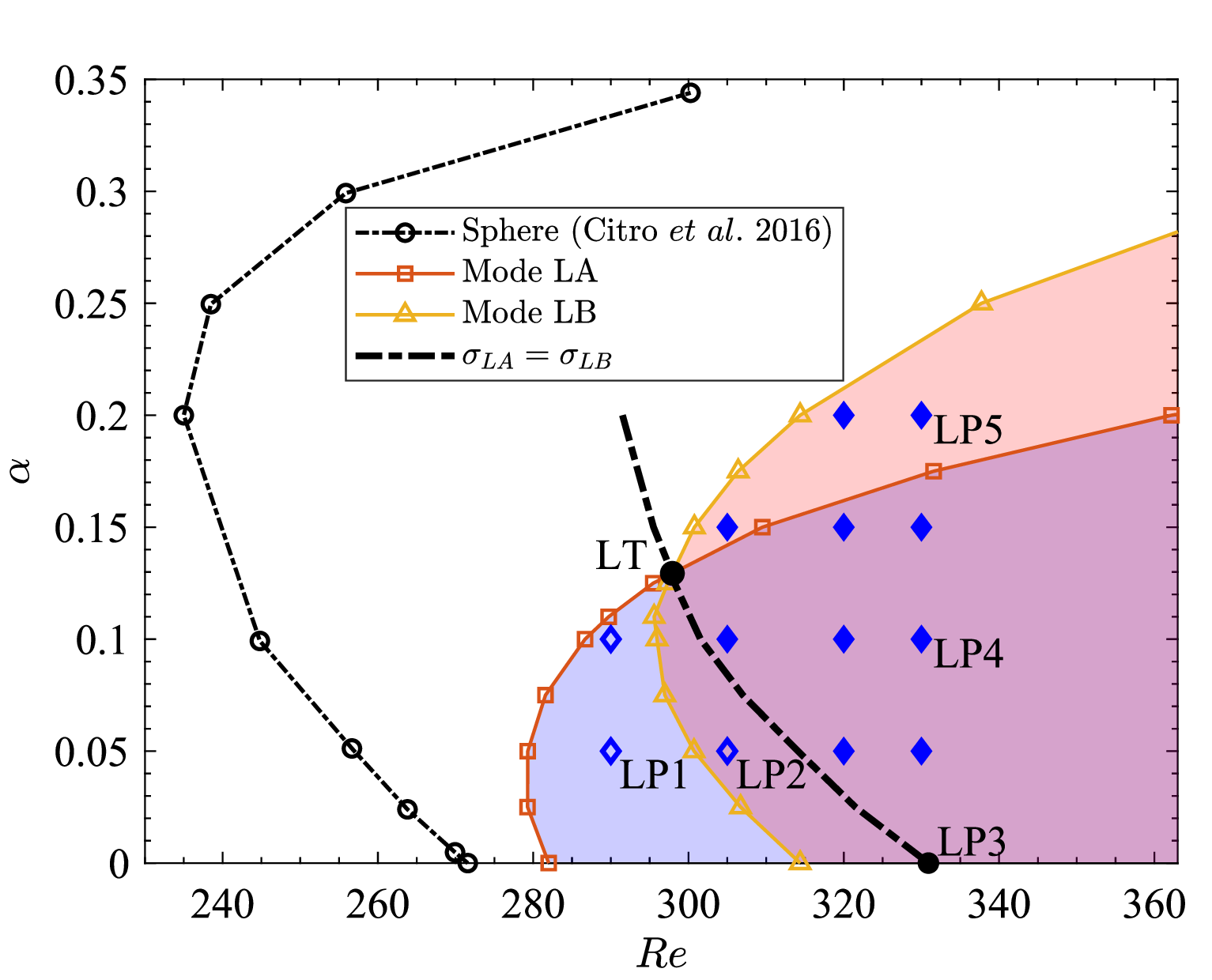}  \put(-192,140){($a$)} \quad 
	\centering\includegraphics[trim=0.0cm 0cm 0.0cm 0cm,clip,width=0.485\textwidth]{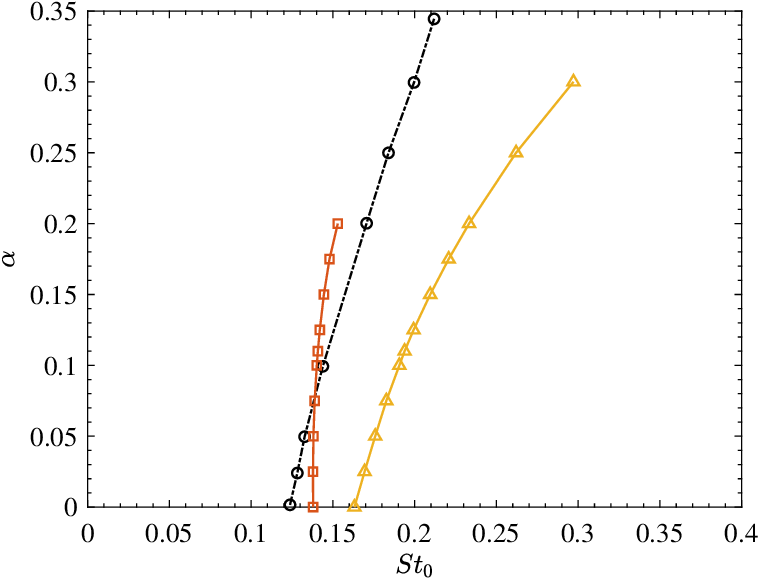}  \put(-192,140){($b$)} 
	\caption{($a$) Neutral stability curves undergoing the Hopf bifurcation for the flow past a finite rotating cylinder at $\ar=1$ and the rotating sphere (the mode I by \citealt{citro2016}). The specifications  of the points LP1-5 and the codimension-two point LT are shown in table \ref{tab:pointseigevalues}. The hollow symbols represent the corresponding nonlinearly saturated wake looks similarly to the global mode LA, whereas the solid symbols denote the nonlinearly saturated wake resembles more the global mode LB.
	($b$) Plot of critical Strouhal numbers $St_0$ on the neutral stability curves against rotating ratio $\alpha$.}
	\label{fig:alpha_Re_AR1}
\end{figure}

We discuss figure \ref{fig:Re_GR_St} and figure \ref{fig:eigenmodeABCD} collectively, showing the eigenspectra and the eigenfunctions respectively.
Figure \ref{fig:Re_GR_St} illustrates the influence of parameters ($Re,\alpha$) on the leading eigenmodes at low rotation rate $\alpha\leq0.3$. The global LSA of the flow past the short rotating cylinder shows that there exist two linear unstable modes (LA $\&$ LB) for Reynolds numbers $Re<360$. The linear growth rates $\sigma$ of modes LA and LB increase linearly with $Re$ in the vicinity of the instability, as shown in Figure \ref{fig:Re_GR_St}$(a)$. When increasing $Re$ for the cases ($\alpha=0,\ 0.025$ and 0.1), both mode LA and mode LB become unstable through a Hopf bifurcation because the frequency at the neutrally stable condition is non-zero, as shown in figure \ref{fig:Re_GR_St}($b$). Mode LA undergoes the Hopf bifurcation prior to mode LB at a lower $Re$. At $\alpha=0.15$, the critical $Re$ for mode LB is smaller. From panel $b$, one can also see that the frequency of global mode LA is not sensitive to $\alpha$ or $Re$ with an approximate value around 0.14, while the frequency of global mode LB increases rapidly with the increase of rotation ratio $\alpha$. Combining the growth rate $\sigma$ and the frequency $St$, we plot in panel \ref{fig:Re_GR_St}($c$) the modes LA and LB for the representative points LP1-LP5 and the LT point (which can be found in figure \ref{fig:alpha_Re_AR1} for their $Re,\alpha$ values). The implication of panel $c$ is similar and, thus, we will not discuss it further.
 
Figure \ref{fig:eigenmodeABCD} plots four eigenmodes LA, LB, LC, LD at $\alpha=0.1$ and $Re=290$. Only mode LA is unstable; see also the solid green lines in figure \ref{fig:Re_GR_St}$(a)$ and one of the unnamed hollow diamonds on the $Re$-$\alpha$ plane in figure \ref{fig:alpha_Re_AR1}$(a)$.
The structure of the low-frequency global mode LA is shown in figure \ref{fig:eigenmodeABCD}$(a)$, corresponding to the periodic wake LA (to be discussed in figure \ref{fig:wakes-NL}$a$ obtained by DNS) and is mainly caused by the vortices shedding from the flat ends of cylinder \citep{yang_feng_zhang_2022}. We loosely use the same name for the global mode and the corresponding wake pattern observed in DNS, if their frequencies are close.
 Global mode LB, as shown in figure \ref{fig:eigenmodeABCD}$(b)$, with a higher frequency is caused by the vortices shedding from the curved surface and the associated nonlinear flow pattern wake LB is shown in figure \ref{fig:wakes-NL}($b$). In figure \ref{fig:eigenmodeABCD}, we have additionally shown two additional modes (named LC and LD) at $\alpha=0.1$ and $Re=290$. 
Mode LC in panel $c$ has zero frequency and presents a long smooth streamwise structure. The leading eigenmode in some other cases, such as LT, may take this form. LT, as shown in figure \ref{fig:alpha_Re_AR1}($a$), represents a low-rotation-rate transition state where the LA and LB modes are both neutral, to be discussed further below. 
Mode LD in panel $d$ appears similar to the mode B identified in \cite{yang_feng_zhang_2022} for the non-rotating short cylinder.
Each global mode in figure \ref{fig:eigenmodeABCD} maintains the same symmetry as the base flow, namely, symmetry with respect to the $x$$y$ plane, consistent with that of the non-rotating short cylinder \citep{yang_feng_zhang_2022}. 

Examining closely the solid and dashed lines in figure \ref{fig:Re_GR_St}($a$), when $\alpha$ is relatively small, mode LA becomes unstable before mode LB when increasing $Re$. In the range of larger $\alpha$, mode LB starts to become unstable before mode LA. The intersection point of the $\sigma$-$Re$ lines of mode LA and mode LB denotes the condition where $\sigma_{LA}=\sigma_{LB}$, as represented by the black stars in figure \ref{fig:Re_GR_St}$(a)$. This competitive phenomenon also exists in spheres rotating along the streamwise direction \citep{sierra2022unveiling}, but differs from the single-mode instability observed in spheres rotating along the transverse direction \citep{citro2016}. Note that our cylinder is rotating along its axis which is in the transverse direction.

The neutral curves in the $Re-\alpha$ plane associated to the two most unstable modes LA and LB are displayed in figure \ref{fig:alpha_Re_AR1}. The present asymmetric steady state is linearly stable in the blank region and linearly unstable in the shaded region, as shown in panel ($a$).
With the rotation speed increasing, the threshold Reynolds number for the instability first decreases slightly and then increases. The critical Reynolds number approximately reads $Re_c=279$ at $\alpha=0.0375$. As the rotational speed continues to increase, the threshold Reynolds number becomes less sensitive to the rotation speed. Thus, the result indicates that the rotation of a short cylinder can influence and control the stability properties of the flow.

 The present work designates the overlapping area between the unstable regions of modes LA and LB in figure \ref{fig:alpha_Re_AR1}$(a)$ as the bi-unstable region, where points LP2, LP3 and LP4 are located.
 In this study, the intersection of the neutral curves of mode LA and mode LB is denoted as the codimension-two point LT, where two different bifurcations (due to mode LA and mode LB) occur simultaneously. Similar to the rotating sphere \citep{sierra2022unveiling}, the present point LT, as the organizing center of the linear system, represents a turning point of competition between different modes and results in the generation of three distinct wake patterns around it in the nonlinear system (to be discussed in figure \ref{fig:wakes-NL}).
The dotted line in figure \ref{fig:alpha_Re_AR1}$(a)$ passes the codimension-two point LT in the bi-unstable region, on which the growth rate $\sigma_{\rm LA}=\sigma_{\rm LB}$. Above this dotted line, the growth rate of mode LB is greater than that of mode LA; $vice\ versa$.

\begin{table}\centering	
\begin{tabular}{p{0.7cm} p{1.1cm} p{1.0cm} p{3.9cm} p{3.9cm} p{3.5cm}}
No.&$Re$&$\alpha$ &\qquad $(\sigma+\rm{i}\omega/2\pi)_{LA}$     &\qquad $(\sigma+\rm{i}\omega/2\pi)_{LB}$ & $St_{\rm DNS}$\\ %\ (In)stability \\ 
LP1 &290   & 0.05  &$\ \ 1.4642\times10^{-2}+$i$\mathbf{0.1394}$&$-4.7063\times10^{-2}+$i$0.1703$   & 0.1406\\
LP2 &305   & 0.05  &$\ \ 3.3957\times10^{-2}+$i$\mathbf{0.1411}$&$\ \ 1.6972\times10^{-2}+$i$0.1782$& 0.1443\\
LT  &297.9 & 0.1294&$\ \ 0.0+$i$0.1422$                         &$\ \ 0.0+$i$0.2011$                & $/$ \\
LP3 &330.33& 0.0   &$\ \ 6.6731\times10^{-2}+$i$0.1424$         &$\ \ 6.7834\times10^{-2}+$i$0.1706$& $/$ \\
LP4 &330   & 0.1   &$\ \ 4.6550\times10^{-2}+$i$0.1451$         &$\ \ 9.3466\times10^{-2}+$i$\mathbf{0.2077}$& 0.1977\\
LP5 &330   & 0.2   &$-2.4274\times10^{-2}+$i$0.1485$            &$\ \ 3.9667\times10^{-2}+$i$\mathbf{0.2392}$& 0.2358\\
\end{tabular}
\caption{The specifications of the five typical points in the parameter space $(Re,\alpha)$ (see also figure \ref{fig:alpha_Re_AR1} $a$) for the cylinder $\ar=1$. Subscripts $_{LA}$ and $_{LB}$ represent global modes LA and LB (figure \ref{fig:eigenmodeABCD} $a$ and $b$), respectively. The eigenfrequencies marked in bold are close to the frequency $St_{\rm DNS}$ in DNS  (e.g., see also figure \ref{fig:PSD}) of the saturated nonlinear wake.}
\label{tab:pointseigevalues}
\end{table}

The frequencies $St_0$ corresponding to the neutral conditions in panel $(a)$ are reported in figure \ref{fig:alpha_Re_AR1}($b$) as a function of $\alpha$. As the frequencies are non-zero, the unstable flows will undergo Hopf bifurcation, that is, the unstable mode will oscillate at a certain frequency. 
Similar to the flow past a sphere rotating along the transverse \citep{citro2016} and  streamwise directions \citep{sierra2022unveiling}, the $St_0$ of present rotating cylinders increase rapidly as $\alpha$ increases in the regime of low rotation rates $\alpha<0.3$.

\begin{figure}
	\centering\includegraphics[trim=0.0cm 0cm 0.0cm 0cm,clip,width=0.5\textwidth]{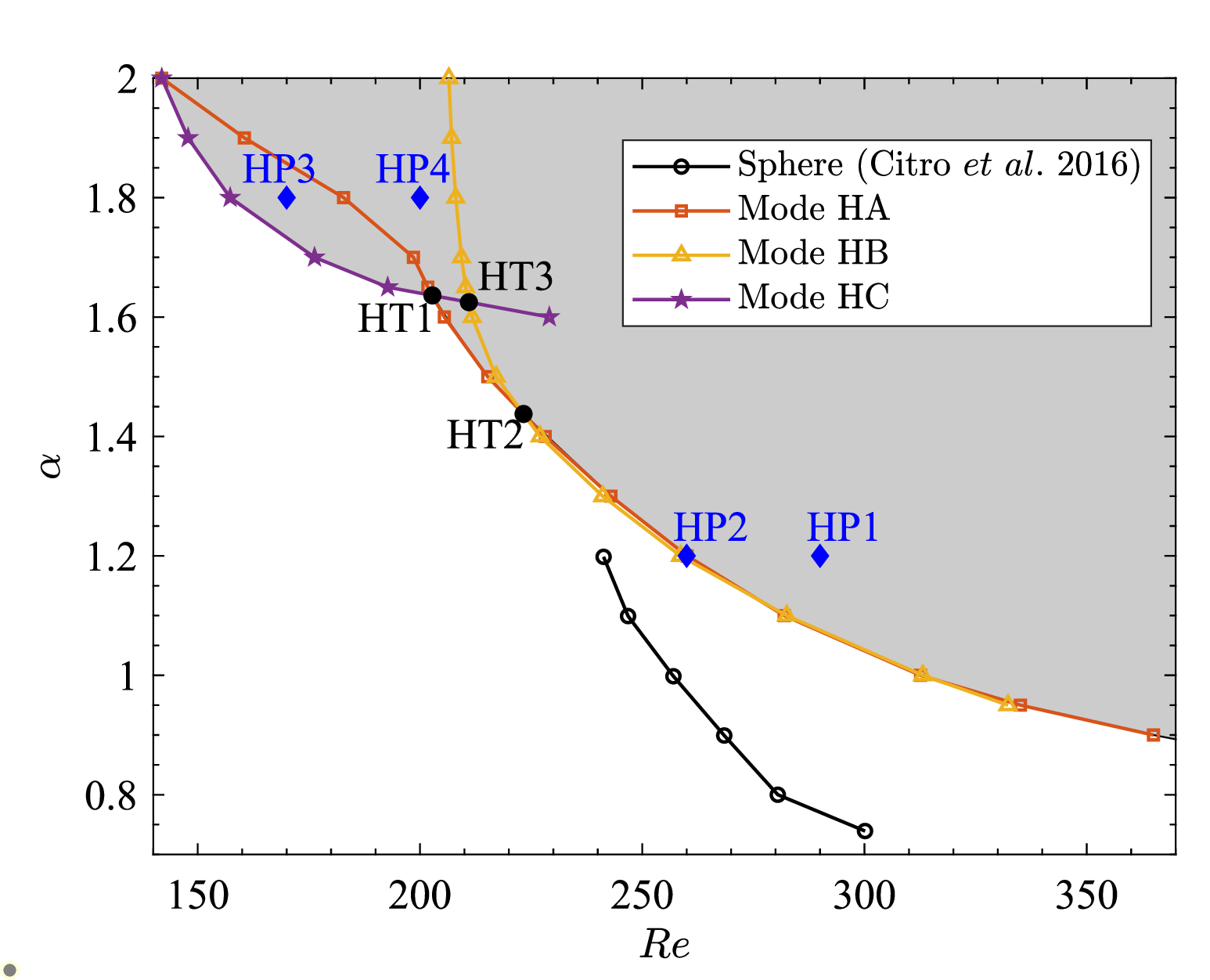}   \put(-192,145){($a$)} \ 
	\centering\includegraphics[trim=0.0cm 0cm 0.0cm 0cm,clip,width=0.485\textwidth]{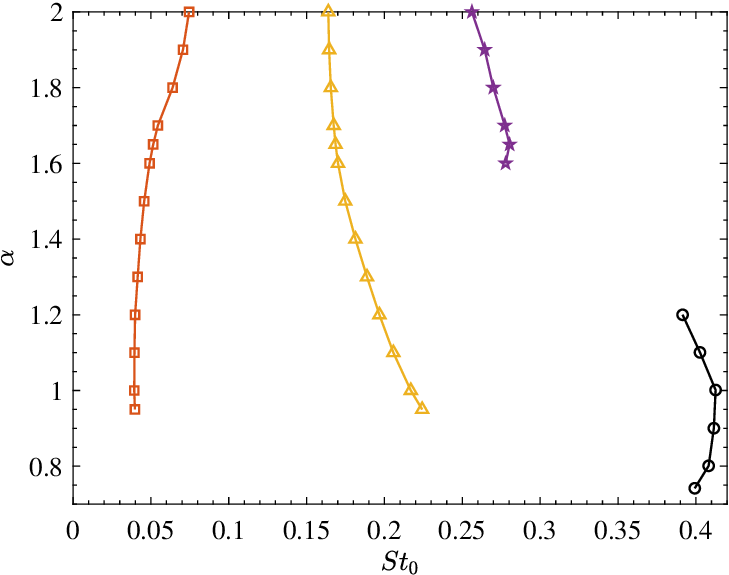} \put(-192,145){($b$)} 
	\caption{($a$) Neutral stability curves  undergoing the Hopf bifurcation for the flow past a finite rotating cylinder at $\ar=1$ and the rotating sphere (the mode I by \citealt{citro2016}). The shaded area indicates a region of linear instability. The eigenvalues of points HP1-HP4 are shown in table \ref{tab:pointseigevaluesHighM}.
	($b$) Plot of critical Strouhal numbers $St_0$ against rotating ratio $\alpha$.}
	\label{fig:highalpha_Re_AR1}
\end{figure}

\begin{table}\centering	
	\begin{tabular}{p{0.6cm} p{0.7cm} p{0.5cm} p{3.7cm} p{3.5cm} p{3.5cm}}
	No.&$Re$  &$\alpha$&\qquad $(\sigma+\rm{i}\omega/2\pi)_{HA}$   &\qquad $(\sigma+\rm{i}\omega/2\pi)_{HB}$ &\qquad $(\sigma+\rm{i}\omega/2\pi)_{HC}$ \\ 
	HP1 &290   & 1.2  & $\ \ 2.6560\times10^{-2}+$i\textbf{0.04063} & $\ \ 3.4548\times10^{-2}+$i$0.1971$ &  \qquad / \\
    HP2 &260   & 1.2  & $-7.4985\times10^{-5}+$i\textbf{0.04025}    & $\ \ 1.9047\times10^{-3}+$i\textbf{0.1968} &  \qquad / \\
	HP3 &170   & 1.8  & $-2.2614\times10^{-2}+$i\textbf{0.06755} & $-7.3908\times10^{-2}+$i$0.1611$    & $\ \ 3.0499\times10^{-2}+$i\textbf{0.2646} \\
	HP4 &200   & 1.8  & $\ \ 4.0641\times10^{-2}+$i\textbf{0.05973} & $-1.3588\times10^{-2}+$i$0.1651$    & $\ \ 8.8217\times10^{-2}+$i$0.2542$ \\
	HT1 &202.8 & 1.64 & $\ \ 0.0+$i$0.05462$                 & \qquad /                           & $\ \ 0.0+$i$0.2966$ \\
	HT2 &223.3 & 1.44 & $\ \ 0.0+$i$0.04673$                 & $\ \ 0.0+$i$0.1852$                 & \qquad / \\
	\end{tabular}
	\caption{The location and eigenvalues of typical points in the parameter space $(Re,\alpha)$ (figure \ref{fig:highalpha_Re_AR1} $a$). Subscripts $_{HA}$, $_{HB}$ and $_{HC}$ represent global modes HA HB, and HC (figure \ref{fig:eigenmodeEF}), respectively. The eigenfrequencies marked in bold are identified in the frequency spectrum (figure \ref{fig:PSD} $b$-$f$) of the saturated nonlinear wake.}\label{tab:pointseigevaluesHighM}
\end{table}

\subsubsection{At high rotation speed $0.9<\alpha<2$ } \label{subsec:GSAathighrotation}
We consider relatively low $Re$ in our work. Consequently, in the medium range of the rotation rate $\alpha$ from 0.3 to 0.9 and $Re<360$, the wake is steady without vortex shedding. Thus, we will not discuss this range of parameters.

We continue to study high rotation speeds in the range of $0.9<\alpha<2$ and $Re<360$, where the flow may become linearly unstable as shown in figure \ref{fig:highalpha_Re_AR1}$(a)$. From this figure, one can observe that the global LSA based on the asymmetric SFD steady state (see figure \ref{fig:BS_ar1.0}) indicates three unstable modes in this space of parameters, namely HA, HB and HC, whose critical conditions are depicted as three lines in the figure. Four typical pairs of $Re$ and $\alpha$ in this figure, denoted as HP1 to HP4, are highlighted and their HA, HB and HC eigenvalues are quantified in table \ref{tab:pointseigevaluesHighM}. The four probed points will be further analyzed in figure \ref{fig:wakes-NL}. From panel $a$, one can see that higher rotation speeds in the range of $0.9<\alpha<2$ can significantly lower the critical Reynolds number for instability, which is different from the low-rotation-rate cases $0\le\alpha\le0.3$. 

As shown in panel \ref{fig:highalpha_Re_AR1}$b$, the three eigenmodes HA, HB, HC are characterised by different frequencies $St_0$, namely, low frequency for HA, immediate frequency for HB and high frequency for HC. The non-zero frequencies of these modes again indicate that the unstable wake flow experiences Hopf bifurcation. By comparison to figure \ref{fig:alpha_Re_AR1}($b$), the influence of the rotation on the eigenfrequencies of HA, HB and HC is less significant compared with the mode LB at low rotation rate. This indicates that in the high-rotation-speed regime, changing the value of $\alpha$ affects less the frequency in the flow. For the following discussions, when a nonlinear wake flow possesses multiple characteristic frequencies of these global modes at the same time, we will name the wake flow by combining the modes; for example, if both the eigenfrequencies of HA and HB modes are observed in a nonlinear wake, we will call it HAB (see figure \ref{fig:wakes-NL} to be discussed). 

\begin{figure}
\centering\includegraphics[trim=0.0cm 0cm 0.0cm 0cm,clip,width=0.48\textwidth]{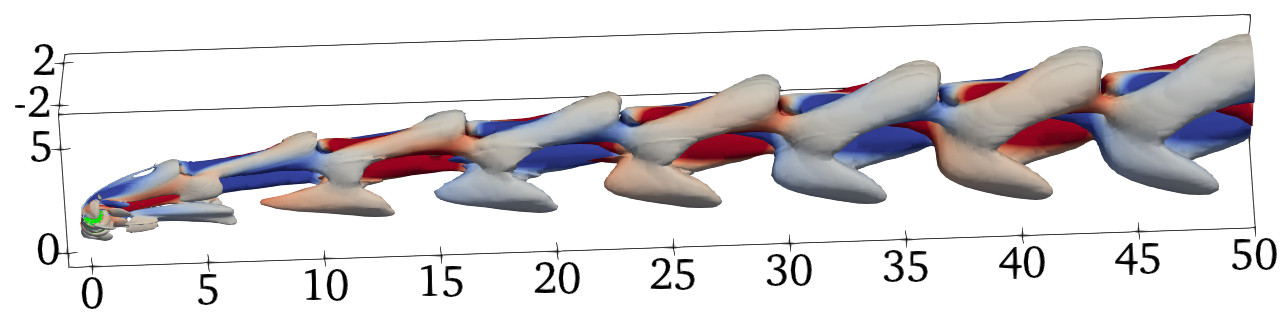}
\put(-190,50){($a$)} \ 
	\put(-162,50){HA mode}  
\centering\includegraphics[trim=0.0cm 0cm 0.0cm 0cm,clip,width=0.50\textwidth]{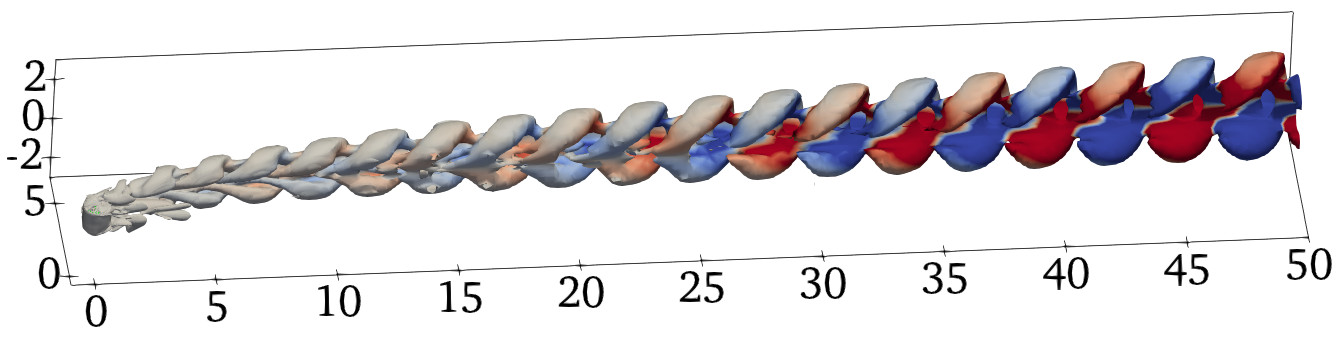}
\put(-195,50){($b$)} \ 
	\put(-167,50){HB mode}  \\
\vspace{0.3cm}
\centering\includegraphics[trim=0.0cm 0cm 0.0cm 0cm,clip,width=0.68\textwidth]{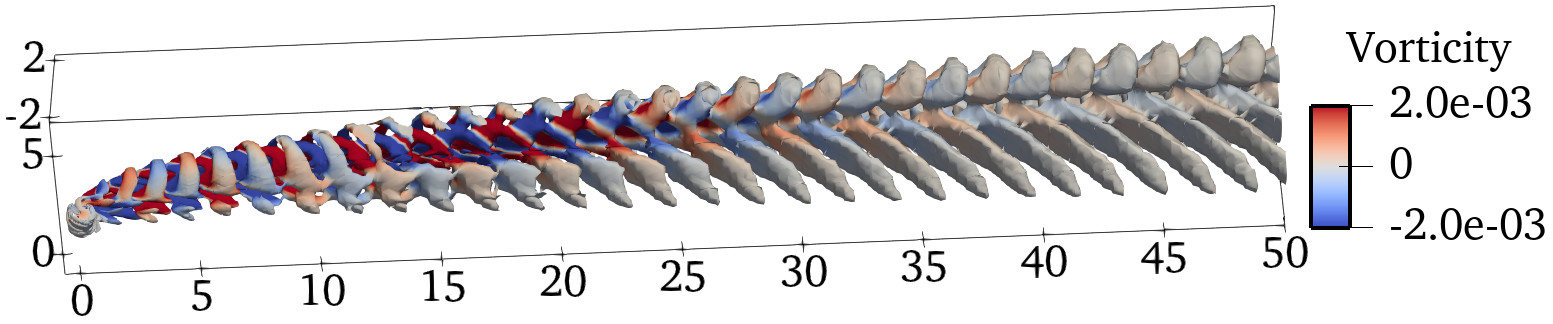}
\put(-265,52){($c$)} \quad
	\put(-232,52){HC mode}  
	\caption{The global modes with strong asymmetry for the flow past a rotating finite cylinder at high rotation rate at $Re=170, \alpha=1.8$ (point HP3). Where panels ($a$), ($b$), and ($c$) represent the modes HA, HB, and HC at point HP3.
	The $Q$-criterion isosurfaces $Q=0$ are colored by the $x$-component of the vorticity ranging from $-2\times 10^{-3}$ to $2\times 10^{-3}$. The cylinder centroid is located at $(0, 0, 0)$.}
	\label{fig:eigenmodeEF}
\end{figure}
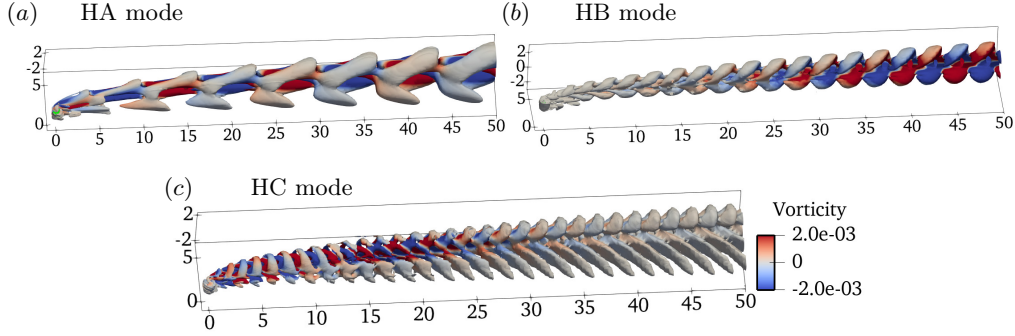

Besides, the three unstable modes can also interact with each other, resulting in the three turning points HT1, HT2, HT3 as shown in figure \ref{fig:highalpha_Re_AR1}$(a)$. In the parameter ranges of $0.9<\alpha<1.64$ and $203<Re<335$, the neutral curves of global modes HA and HB almost coincide, that is, the low-frequency mode HA and mid-frequency mode HB simultaneously become linear unstable. For $\alpha>1.64$ (see e.g. the point HT1 in figure \ref{fig:highalpha_Re_AR1}$a$), the Hopf bifurcation in the flow begins to be dominated by the high-frequency mode HC. In general, the differences of the Hopf bifurcation in our case and that in the rotating sphere are summarised as follows. (1) When $\alpha>0.7$, \cite{citro2016} reports only one unstable mode in the flow past a rotating sphere. However, there are three unstable modes in our case, indicating that the Hopf bifurcation process in the present finite cylinder may be more complex. (2) The eigenfrequency of a sphere is in general higher than that of the rotating cylinder (see the black line with circles in panel \ref{fig:highalpha_Re_AR1}$b$).

In figure \ref{fig:eigenmodeEF}, we show the flow structure of the global eigenfunctions for the three eigenmodes HA, HB, HC at the point HP3 in figure \ref{fig:highalpha_Re_AR1}. It can be seen that global mode HA (figure \ref{fig:eigenmodeEF}$a$) and mode LA (figure \ref{fig:eigenmodeABCD}$a$) have a similar wake structure, and the difference is that the greater rotation rate makes the transverse offset of mode HA larger. The mediate-frequency mode HB and the high-frequency mode HC have smaller flow structures than those in mode HA.

\begin{figure}
\hfil
\centering\includegraphics[trim=0.0cm 0cm 0.0cm	0cm,clip,width=0.22\textwidth]{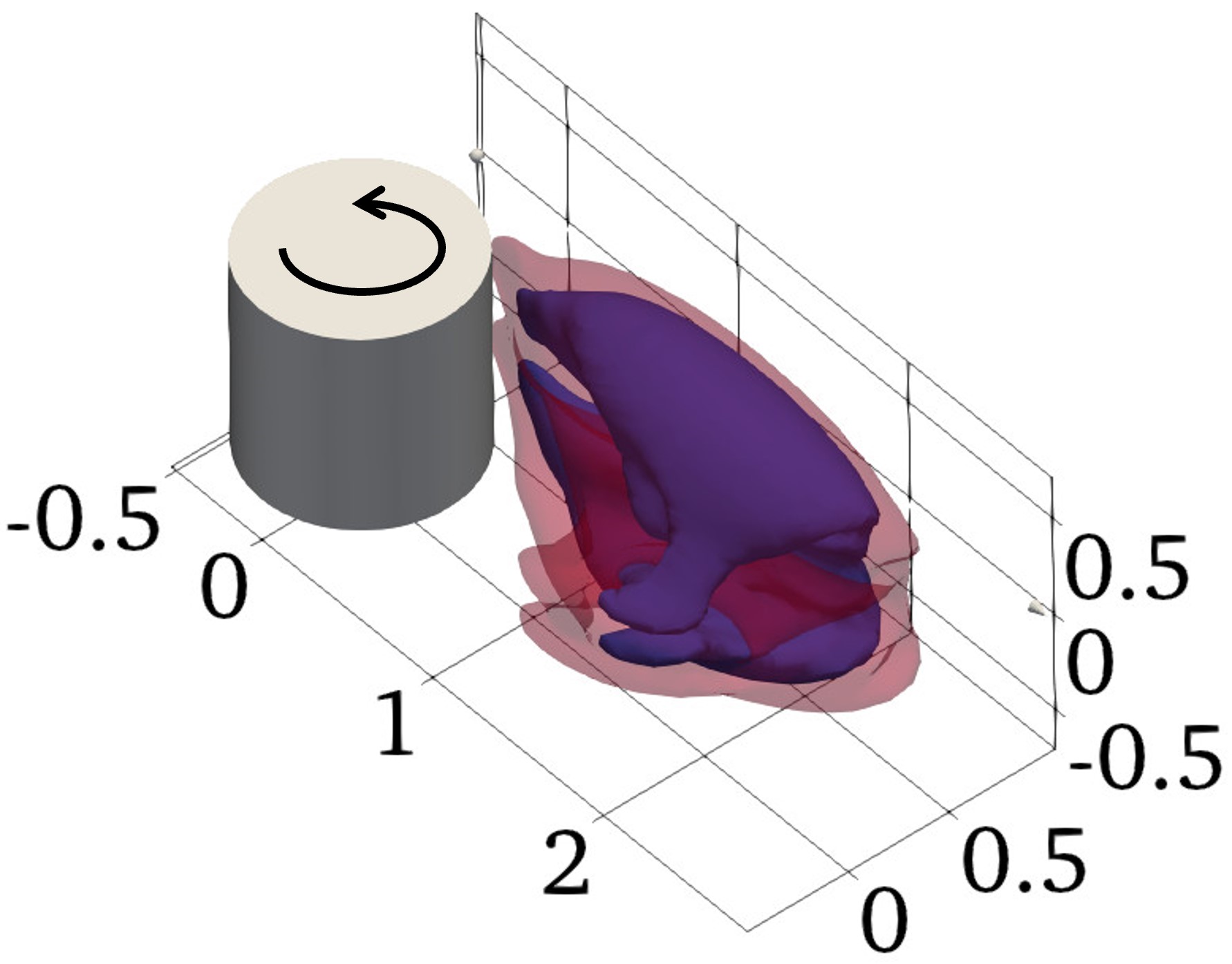} \put(-90,58){($a$)} \quad
\centering\includegraphics[trim=0.0cm 0.0cm 0.0cm 0cm,clip,width=0.22\textwidth]{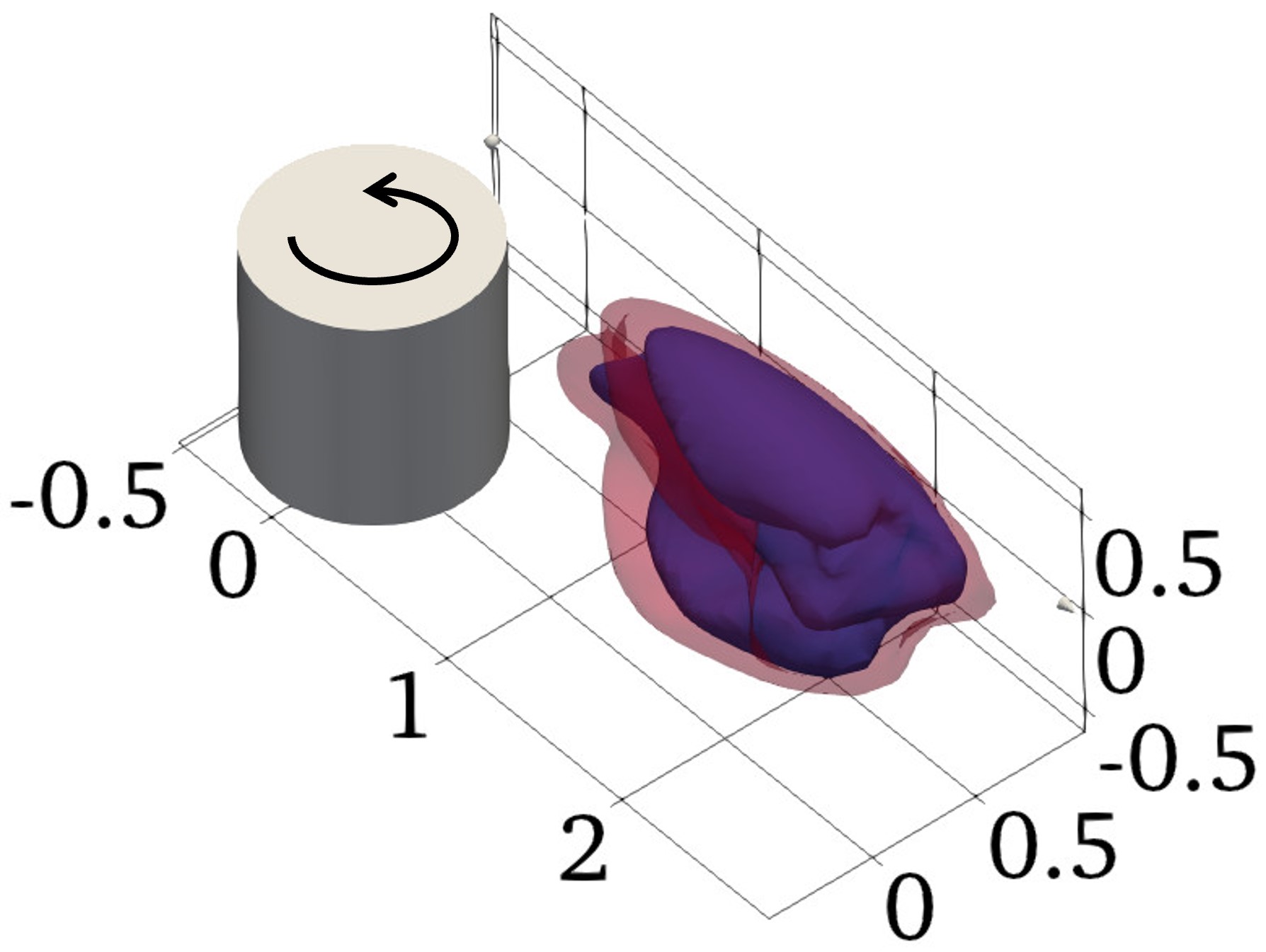} \put(-90,58){($b$)} \quad
\centering\includegraphics[trim=0.0cm 0cm 0.0cm 0cm,clip,width=0.22\textwidth]{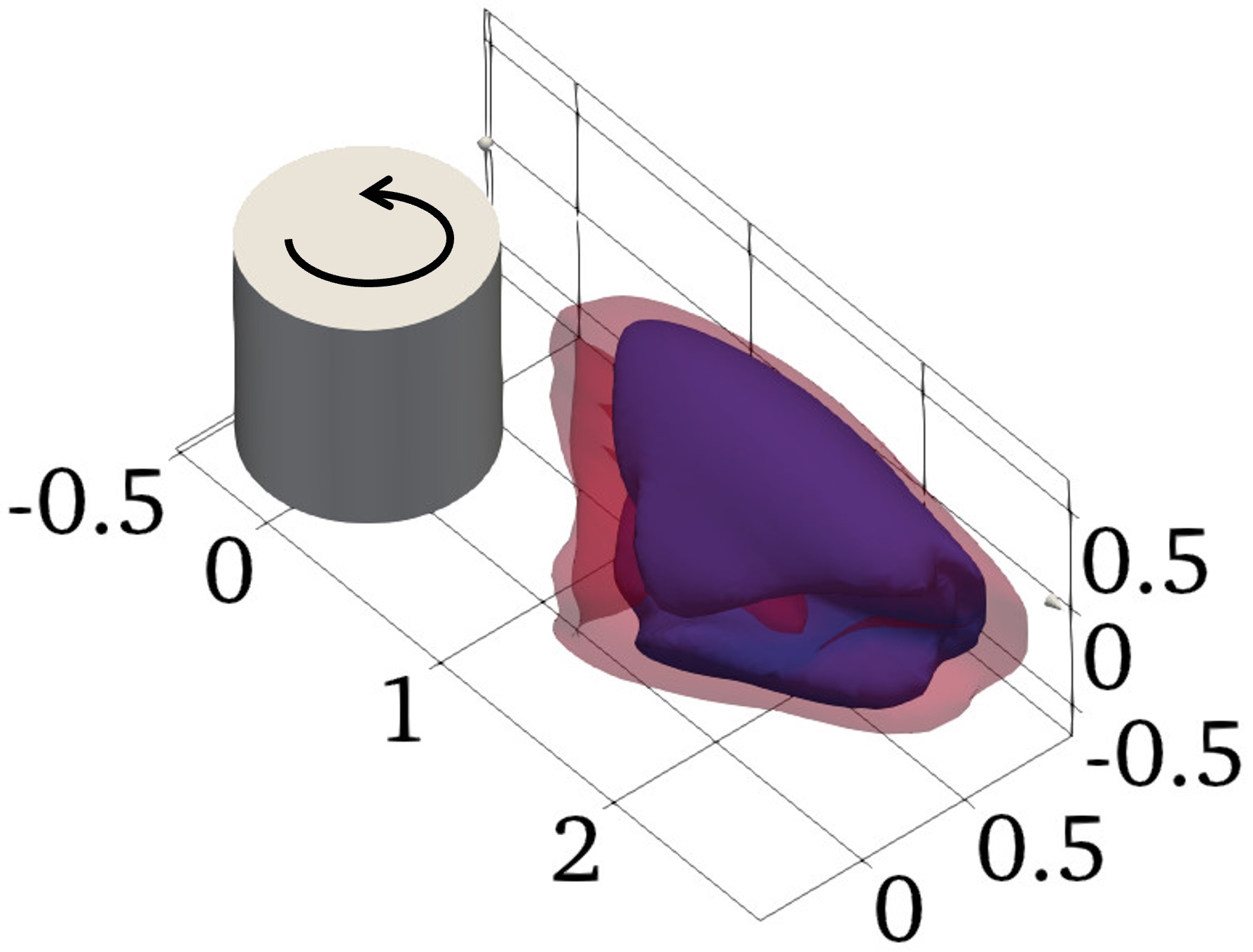} \put(-90,58){($c$)} \quad 
\centering\includegraphics[trim=0.0cm 0.0cm 0.0cm 0cm,clip,width=0.22\textwidth]{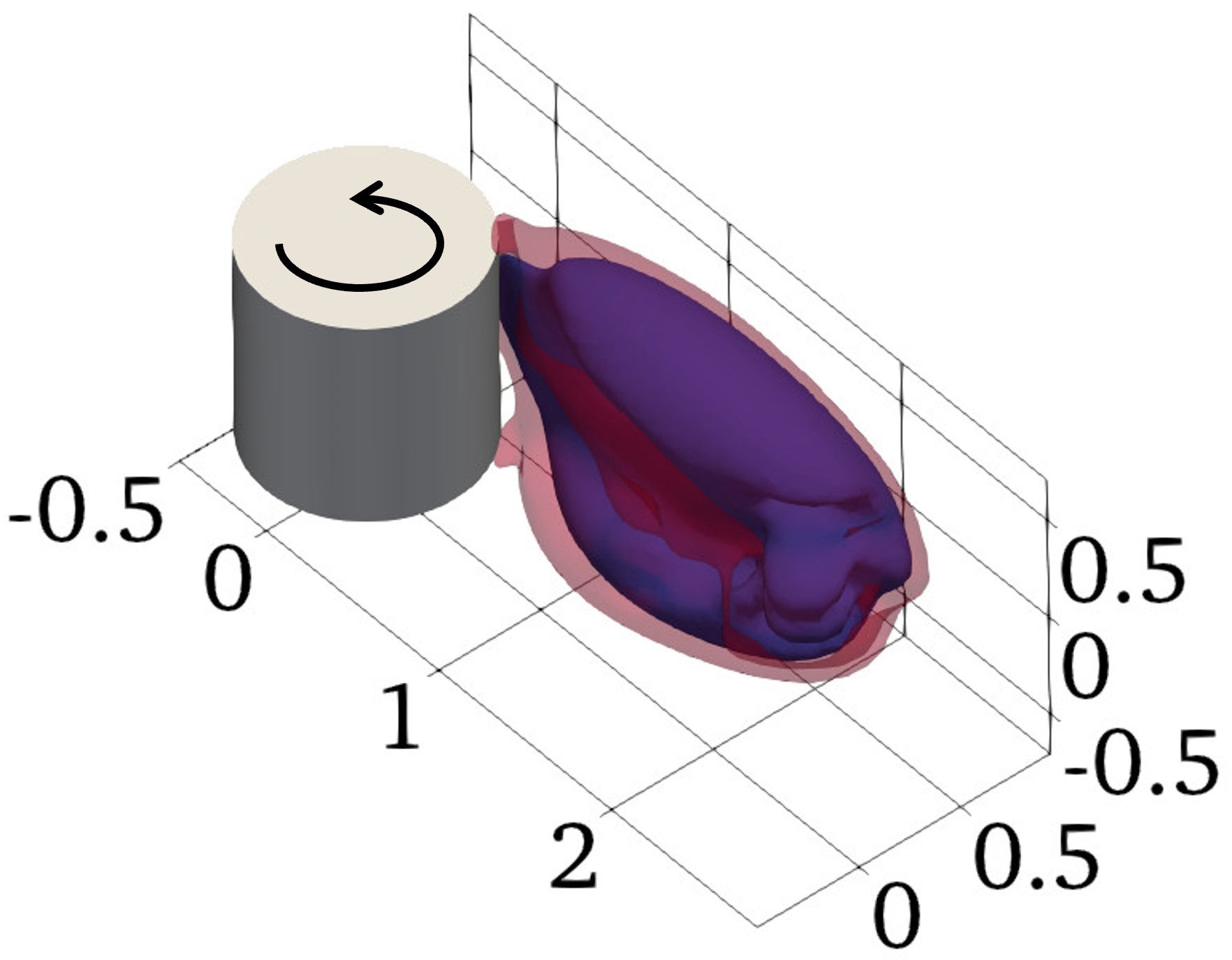}	\put(-95,58){($d$)}
\caption{Flow sensitivity in a low-rotation-rate case. The wavemaker isosurfaces are plotted for the first two unstable modes LA ($a, c$) and LB ($b, d$) at $Re=290,\alpha=0.1,\ar=1$. Transparent red is for $\zeta=0.2$ and opaque blue is for $\zeta=0.4$. Panels ($a,b$) show the results of the global LSA based on the SFD base flow and panels ($c,d$) on the time-mean flow.}
\label{fig:wavemakerRe290}
\end{figure}

\begin{figure}
\hfil
\centering\includegraphics[trim=0cm 0cm 0cm 0cm,clip,width=0.315\textwidth]{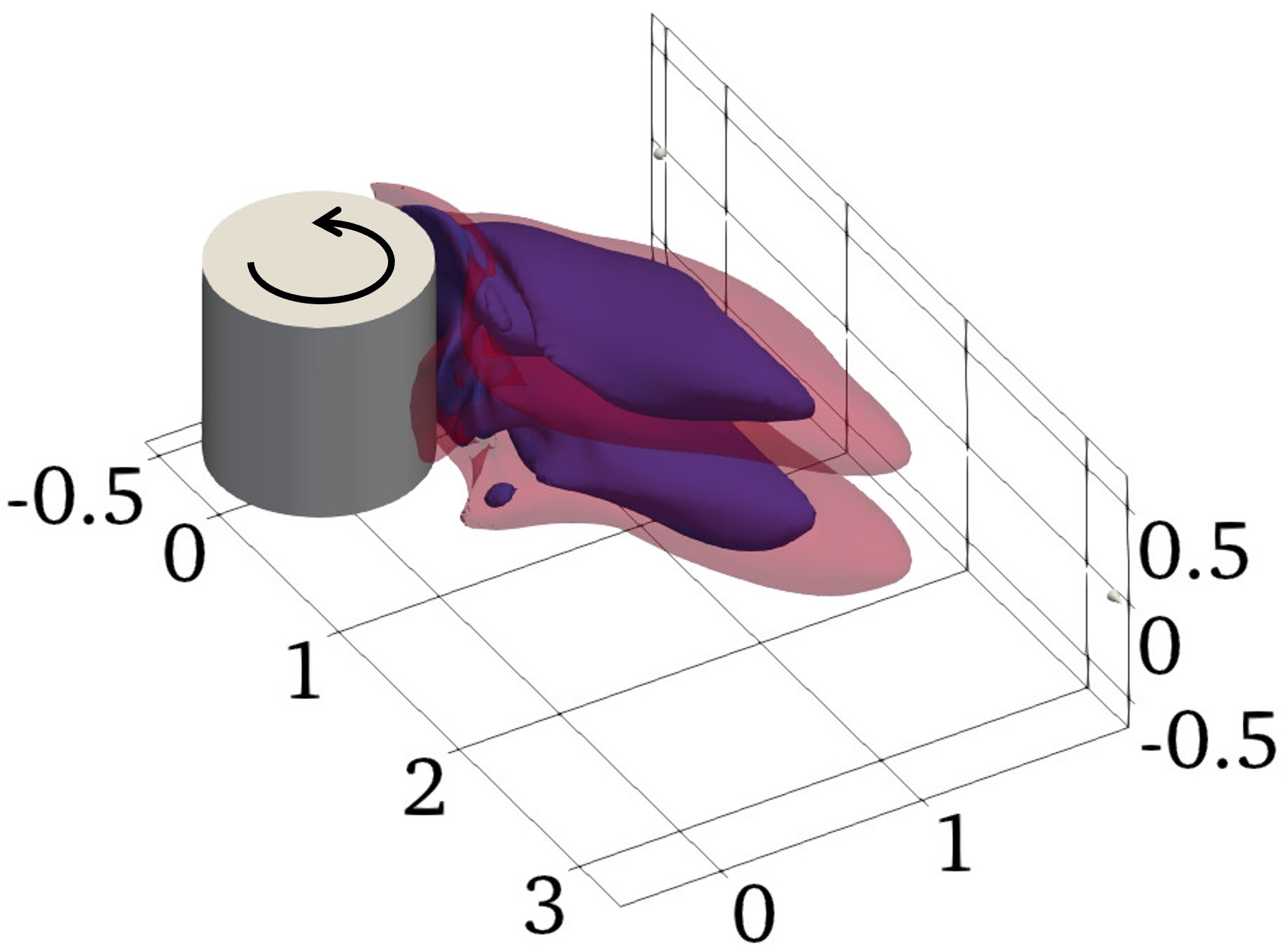} \put(-125,82){($a$)} \quad
\centering\includegraphics[trim=0cm 0cm 0cm 0cm,clip,width=0.315\textwidth]{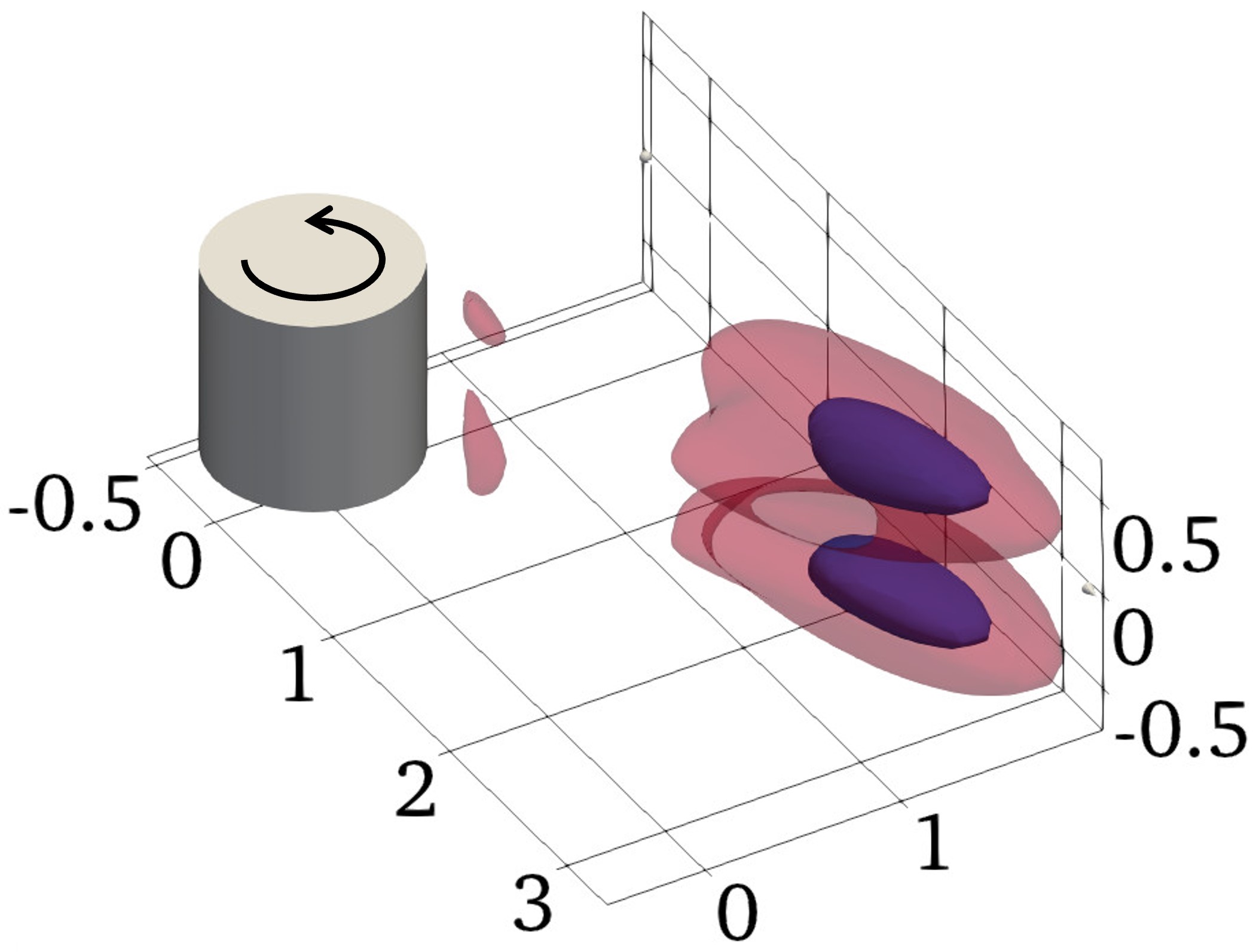} \put(-125,82){($b$)} \quad
\centering\includegraphics[trim=0cm 0cm 0cm 0cm,clip,width=0.315\textwidth]{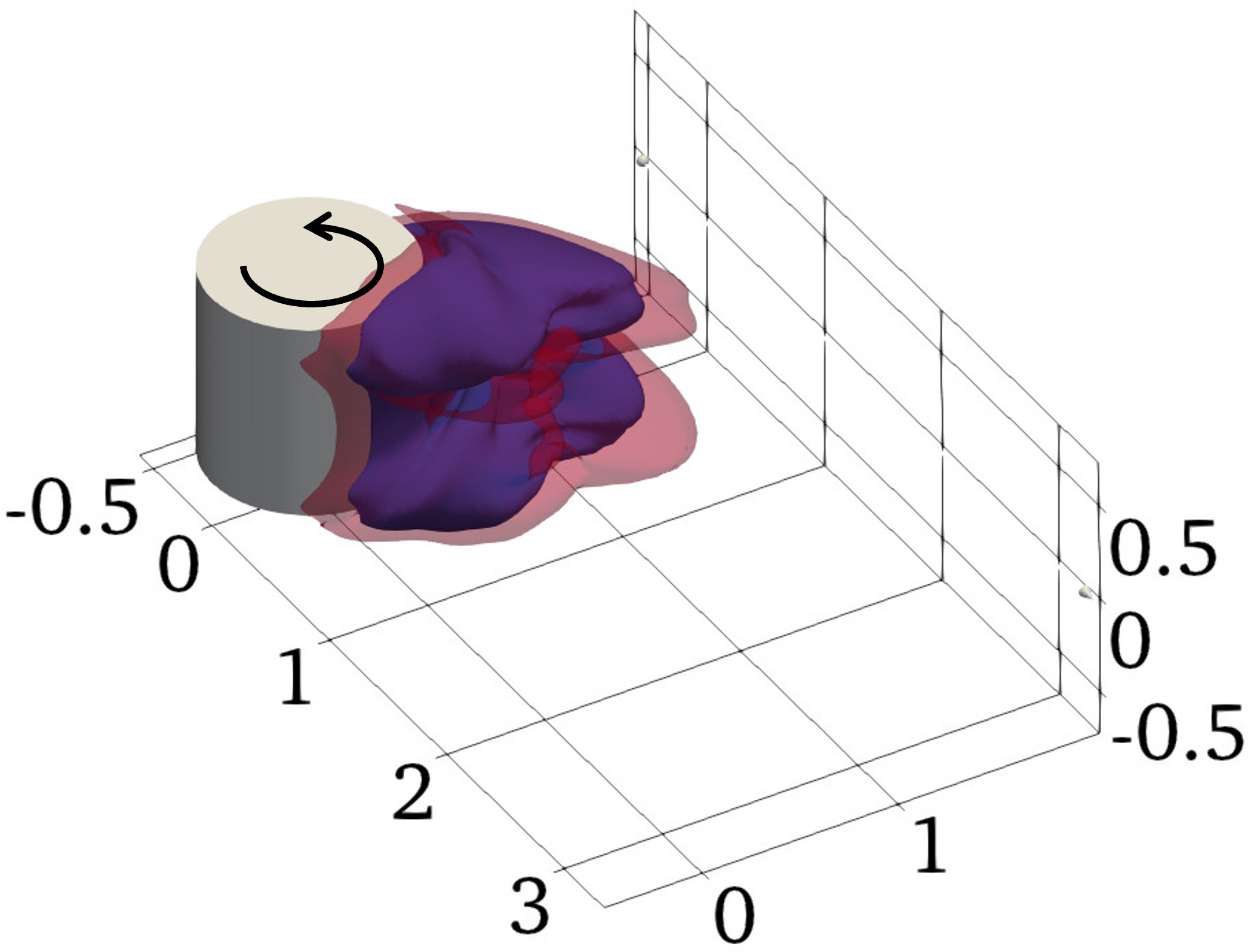} \put(-125,82){($c$)}
\caption{Flow sensitivity in a high-rotation-rate case. The most unstable wavemaker isosurfaces are plotted based on the SFD base flow of rotating cylinder at point HP3. Transparent red is for $\zeta=0.2$ and opaque blue is for $\zeta=0.4$. Panels ($a-c$) correspond to global modes HA, HB and HC, respectively.}
\label{fig:wavemakers}
\end{figure}

\subsubsection{Structure sensitivity} \label{subsec:SSA}
The global modes detailed above characterise the perturbation growth. The sensitivity of the flow to the perturbation can be studied via the structural sensitivity analysis. 
To further study the destabilization mechanism, such an analysis for the selected low-rotation and high-rotation flows is conducted. Figure \ref{fig:wavemakerRe290} depicts the case of $Re=290, \alpha=0.1, \ar=1$ for a low-rotation case. The transparent red and opaque blue isosurfaces, computed according to $\zeta$ in equation \ref{wavemaker}, delimit the wavemaker region, which is the superposition of the leading global mode and its adjoint mode. For both the unstable LA and LB modes, we compute their adjoint modes in the global LSA based on the SFD base flow and time-mean base flow, respectively. In general, the wavemaker region is located in the near wake region of the cylinder (in the top-right region of the $xy$ plane). Its structure remains the same symmetry as the base state, being symmetric with respect to the $\mathcal{O}xy$ plane. The spatial distributions of wakemakers based on mean flow and base flow are similar. Since the wavemaker region indicates the most sensitive region in the flow, one can infer from these observations that (i) the region responsible for the instability is located in the recirculation region behind the cylinder, and (ii) the instability mainly amplifies the perturbations near the cylinder surface \citep{citro2016}.

Figure \ref{fig:wavemakers} presents the flow sensitivity for the high-rotation case of HP3, showing the superposition of the global modes HA, HB, HC based on the SFD base flow with their respective adjoint modes. One can see that the three wavemaker regions are differ sigificantly, in contrast to the low-rotation case. The wavemaker region for the HB mode is located further downstream of the rotating cylinder compared to the other two modes, whose sensitivity regions are close to the cylinder. This implies that the control of the unstable modes in the high-rotation case can be treated separately; especially, control of the HB mode may be achieved more easily as it is not mingled with other modes in the spatial distribution.

\subsection{Comparison with nonlinear results} \label{subsec:Non-Lin}
In the previous sections, we have identified the global modes in the linearised wake flow past a short rotating cylinder. The relevance of these global modes in the nonlinear simulations of the flow should be established and confirmed. Thus, in this section, we will analyze the results of 3-D nonlinear simulations of the flow past a short rotating cylinder and find the trace of the identified global modes therein. Both high and low rotating speeds are considered.
\subsubsection{Wakes behind the short rotating cylinder} \label{subsubsec:wakes}
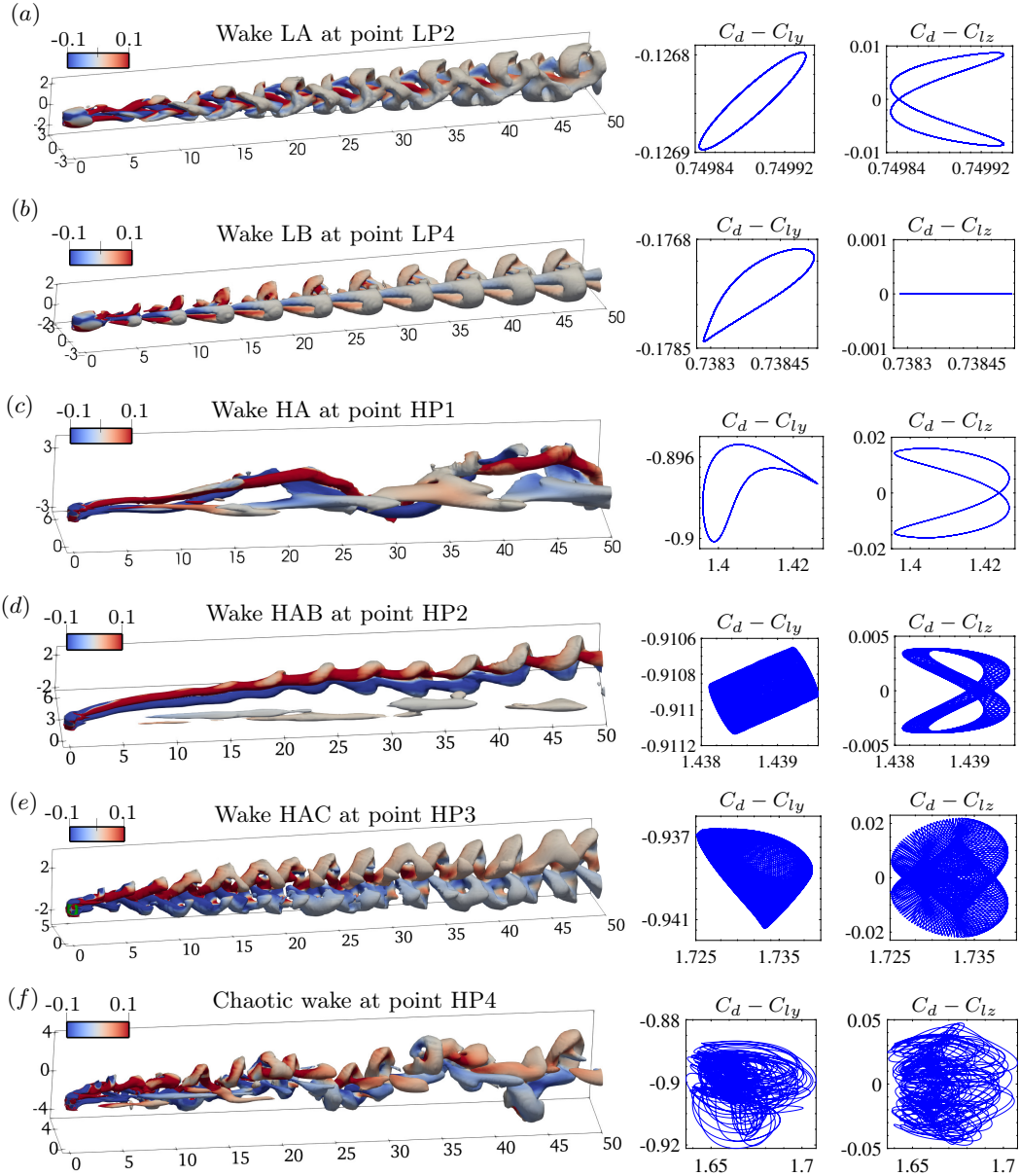
\begin{figure}
	\hfil
	\centering\includegraphics[trim=0.2cm 0cm 0.0cm 
	0cm,clip,width=0.6\textwidth]{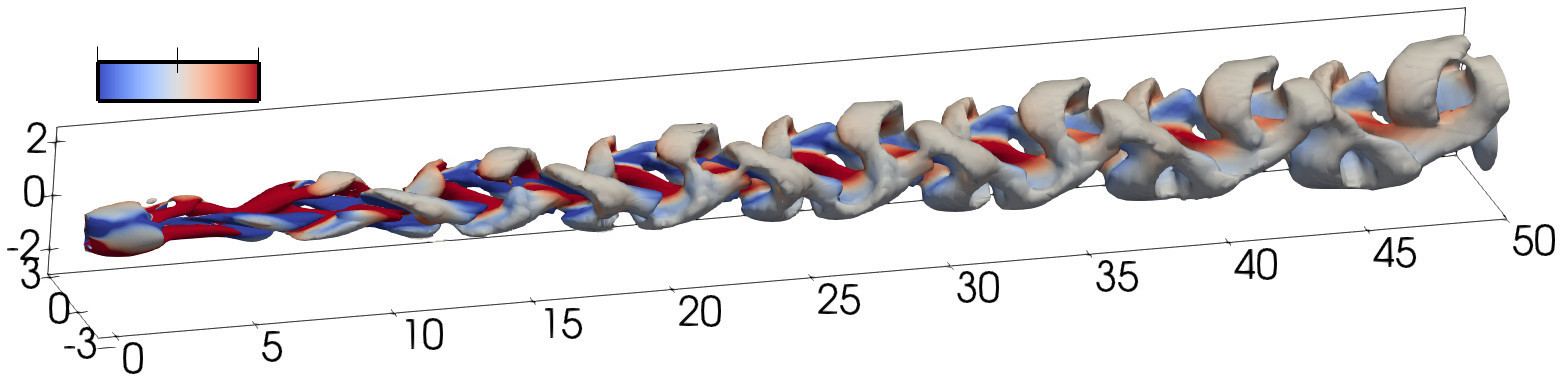}	\put(-240,60){($a$)}
	\put(-224,50){\fontsize{8pt}{\baselineskip}\selectfont {-0.1 \ \ \ 0.1}}  \put(-160,52){Wake LA at point LP2}
	\ \centering\includegraphics[trim=0.0cm 0.0cm 0.0cm 0cm,clip,width=0.188\textwidth]{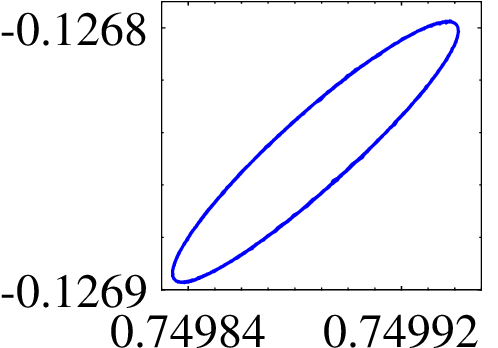} \put(-51,53) {\fontsize{8pt}{\baselineskip}\selectfont \quad \ $C_{d}-C_{ly}$}
	\quad\centering\includegraphics[trim=0.0cm 0.0cm 0.0cm 0cm,clip,width=0.175\textwidth]{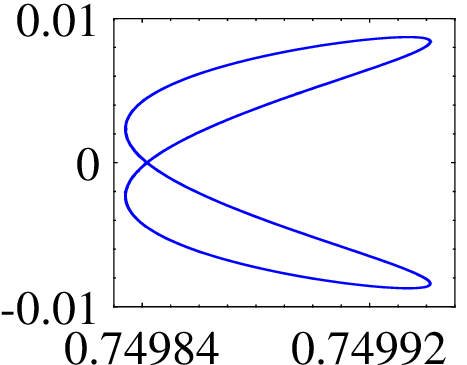} \put(-53,53) {\fontsize{8pt}{\baselineskip}\selectfont \quad \ $C_{d}-C_{lz}$} \\
	\vspace{0.3cm}
	\centering\includegraphics[trim=0.0cm 0cm 0.0cm 0cm,clip,width=0.6\textwidth]{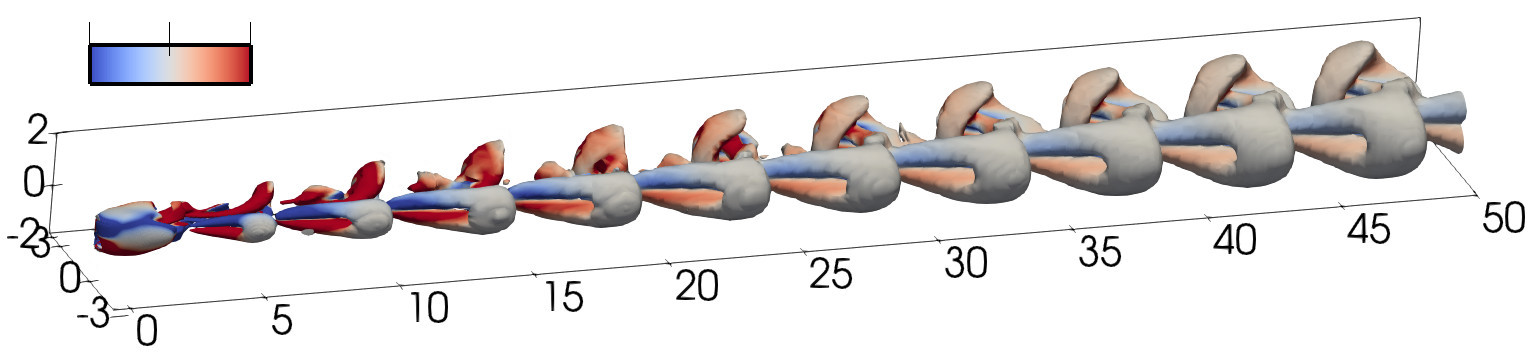}	\put(-240,60){($b$)}
	\put(-224,51){\fontsize{8pt}{\baselineskip}\selectfont {-0.1 \ \ \ 0.1}}  \put(-160,50){Wake LB at point LP4}
	\ \centering\includegraphics[trim=0.0cm 0.0cm 0.0cm 0cm,clip,width=0.185\textwidth]{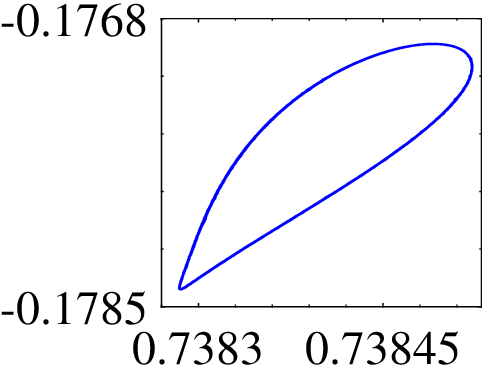} \put(-51,54) {\fontsize{8pt}{\baselineskip}\selectfont \quad \ $C_{d}-C_{ly}$}
	\quad\centering\includegraphics[trim=0.0cm 0.0cm 0.0cm 0cm,clip,width=0.175\textwidth]{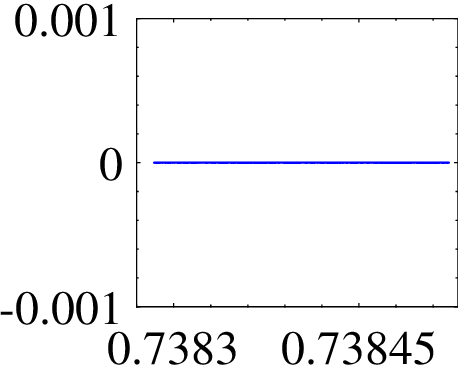} \put(-53,54) {\fontsize{8pt}{\baselineskip}\selectfont \quad \ $C_{d}-C_{lz}$}
	\vspace{0.3cm}
	\centering\includegraphics[trim=0.0cm 0cm 0.0cm 0cm,clip,width=0.595\textwidth]{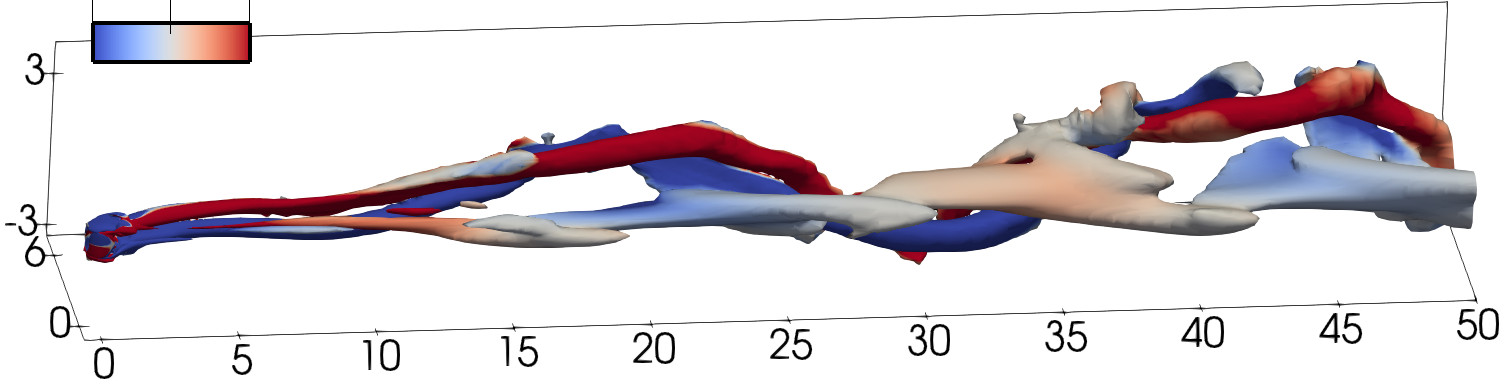}	\put(-240,62){($c$)}\put(-220,60){\fontsize{8pt}{\baselineskip}\selectfont {-0.1 \ \ \ 0.1}}  
	\put(-160,60){Wake HA at point HP1}
	\quad \centering\includegraphics[trim=0.0cm 0.0cm 0.0cm 0cm,clip,width=0.18\textwidth]{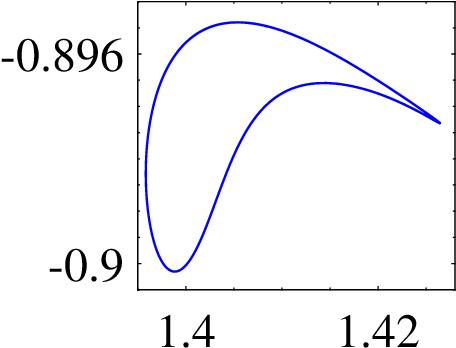} \put(-51,56) {\fontsize{8pt}{\baselineskip}\selectfont \quad \ $C_{d}-C_{ly}$}
	\quad\centering\includegraphics[trim=0.0cm 0.0cm 0.0cm 0cm,clip,width=0.17\textwidth]{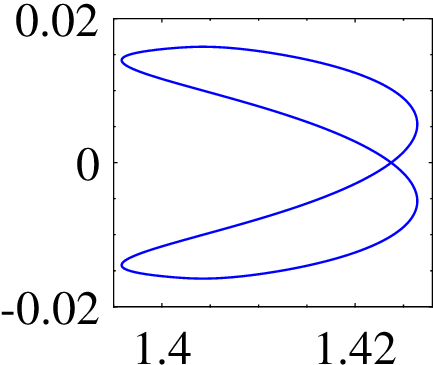} \put(-53,56) {\fontsize{8pt}{\baselineskip}\selectfont \quad \ $C_{d}-C_{lz}$}
	\vspace{0.3cm}
	\centering\includegraphics[trim=0.0cm 0cm 0.0cm 0cm,clip,width=0.59\textwidth]{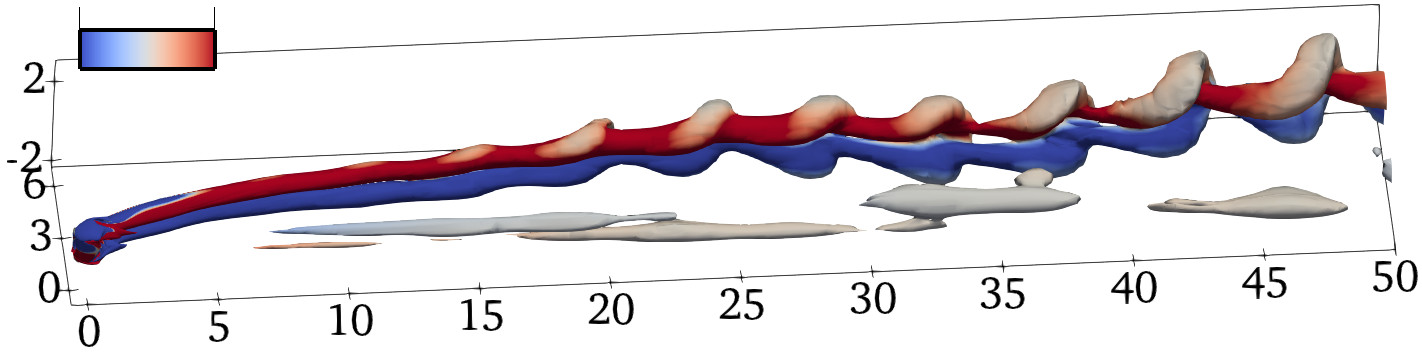} \put(-240,60){($d$)}
	\put(-223,56){\fontsize{8pt}{\baselineskip}\selectfont {-0.1 \ \ \ 0.1}}  \put(-160,57){Wake HAB at point HP2} 
	\quad \centering\includegraphics[trim=0.0cm 0.0cm 0.0cm 0cm,clip,width=0.18\textwidth]{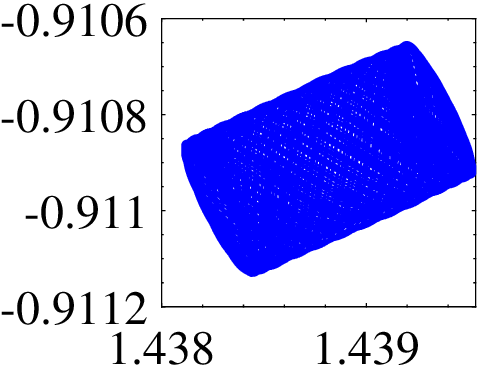} \put(-51,54) {\fontsize{8pt}{\baselineskip}\selectfont \quad \ $C_{d}-C_{ly}$}
	\quad\centering\includegraphics[trim=0.0cm 0.0cm 0.0cm 0cm,clip,width=0.175\textwidth]{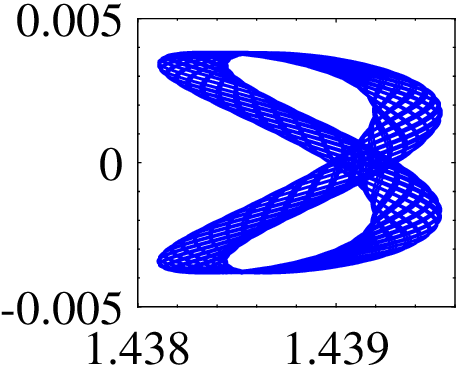} \put(-53,54) {\fontsize{8pt}{\baselineskip}\selectfont \quad \ $C_{d}-C_{lz}$}   
	\vspace{0.3cm}
	\centering\includegraphics[trim=0.0cm 0cm 0.0cm 0cm,clip,width=0.6\textwidth]{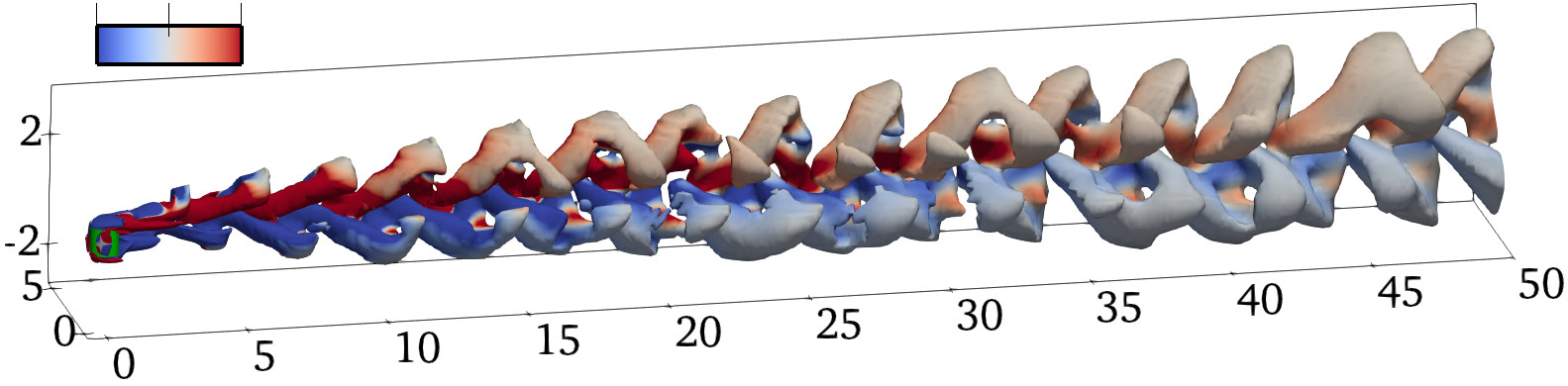}\put(-240,60){($e$)}
	\put(-226,56){\fontsize{8pt}{\baselineskip}\selectfont {-0.1 \ \ \ 0.1}} \put(-160,55){Wake HAC at point HP3}
	\ \ \centering\includegraphics[trim=0.0cm 0.0cm 0.0cm 0cm,clip,width=0.182\textwidth]{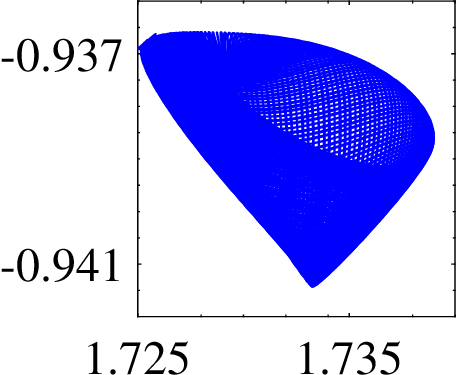} \put(-51,61) {\fontsize{8pt}{\baselineskip}\selectfont \quad \ $C_{d}-C_{ly}$}
	\quad\centering\includegraphics[trim=0.0cm 0.0cm 0.0cm 0cm,clip,width=0.174\textwidth]{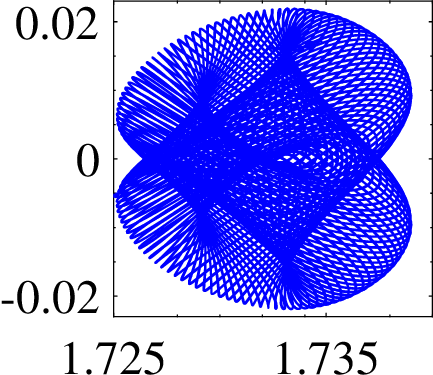} \put(-53,61) {\fontsize{8pt}{\baselineskip}\selectfont \quad \ $C_{d}-C_{lz}$} \\	
	\vspace{0.3cm}
	\centering\includegraphics[trim=0.0cm 0cm 0.0cm 0cm,clip,width=0.6\textwidth]{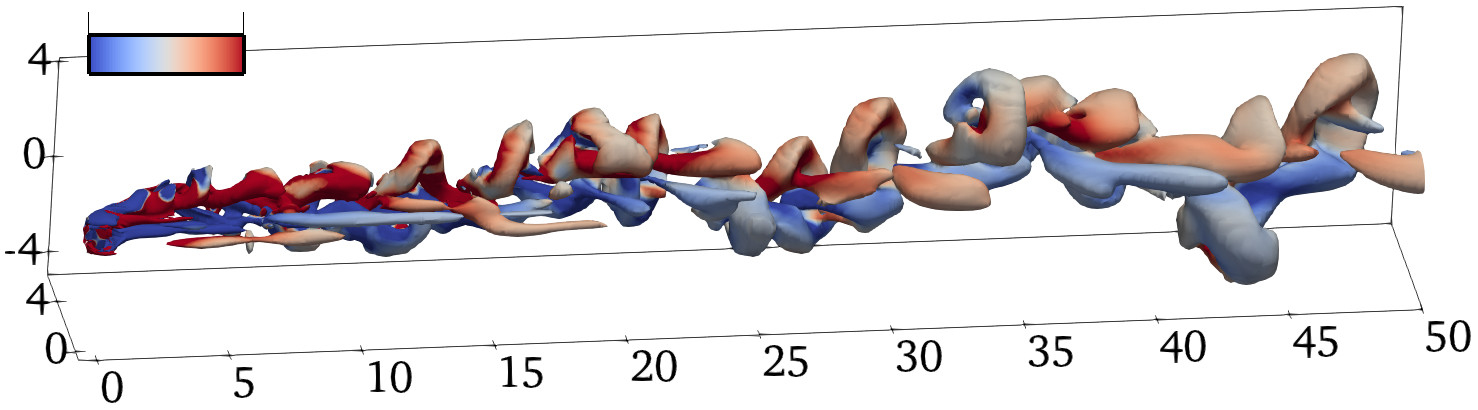}\put(-240,65){($f$)}
	\put(-225,63){\fontsize{8pt}{\baselineskip}\selectfont {-0.1 \ \ \ 0.1}}  \put(-160,64){Chaotic wake at point HP4}
	\ \ \centering\includegraphics[trim=0.0cm 0.0cm 0.0cm 0cm,clip,width=0.18\textwidth]{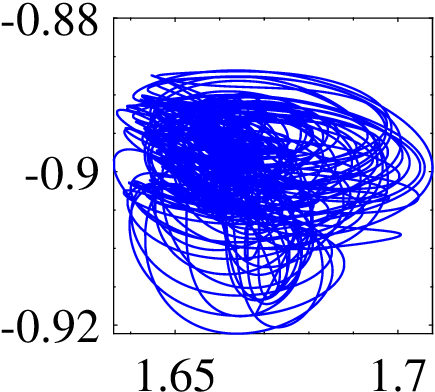} \put(-51,62.5) {\fontsize{8pt}{\baselineskip}\selectfont \quad \ $C_{d}-C_{ly}$}
	\quad\centering\includegraphics[trim=0.0cm 0.0cm 0.0cm 0cm,clip,width=0.18\textwidth]{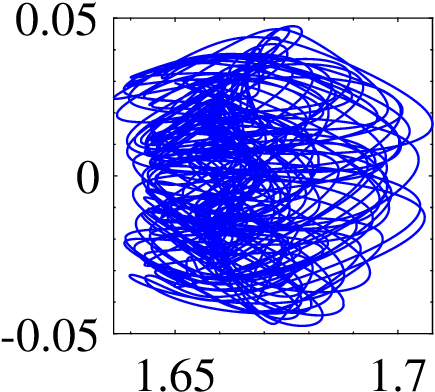} \put(-53,62.5) {\fontsize{8pt}{\baselineskip}\selectfont \quad \ $C_{d}-C_{lz}$} \\
\caption{Left panel: the nonlinear wakes spatial structure obtained by DNS. The $Q=0$ isosurfaces are colored by the streamwise vorticity ranging from -0.1 to 0.1. Right panels: the corresponding phase diagrams of $C_d-C_l$ for rotating cylinder. ($a$) wake LA at point LP2, ($b$) wake LB at point LP4, ($c$) wake HA at point HP1, ($d$) wake HAB at point HP2, ($e$) wake HC at point HP3 and ($f$) wake HAC at point HP4. The corresponding $Re$, $\alpha$ and eigenvalues for each case are shown in tables \ref{tab:pointseigevalues} and \ref{tab:pointseigevaluesHighM}.
	}\label{fig:wakes-NL}
\end{figure}

\begin{figure}
	\hfil
	\centering\includegraphics[trim=0.0cm 0.0cm 0.0cm 
	0.0cm,clip,width=0.30\textwidth]{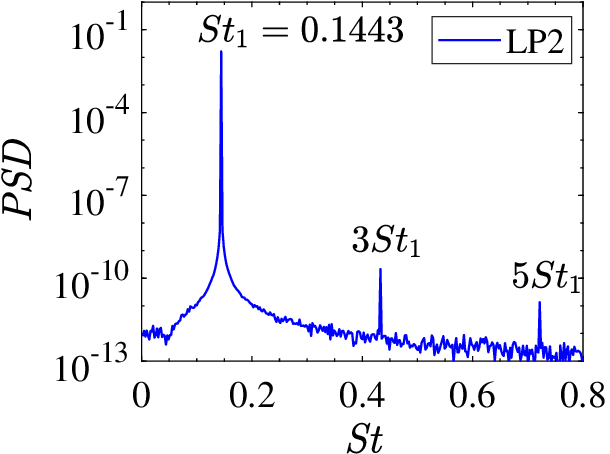}	\put(-120,85){($a$)}
	\quad\ \centering\includegraphics[trim=0.0cm 0.0cm 0.0cm 
	0.0cm,clip,width=0.30\textwidth]{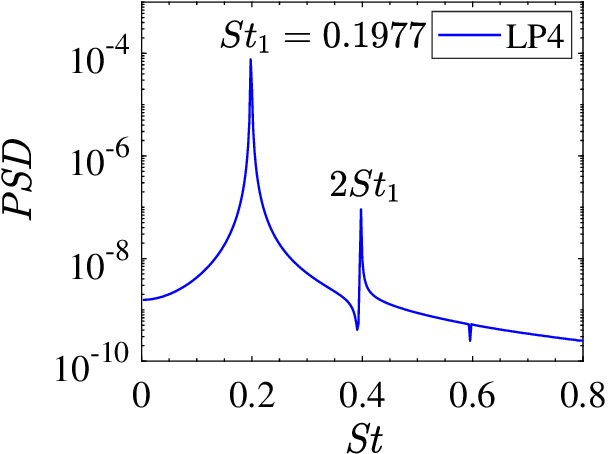}	\put(-120,85){($b$)}
	\quad\ \centering\includegraphics[trim=0.0cm 0cm 0.0cm 
	0cm,clip,width=0.30\textwidth]{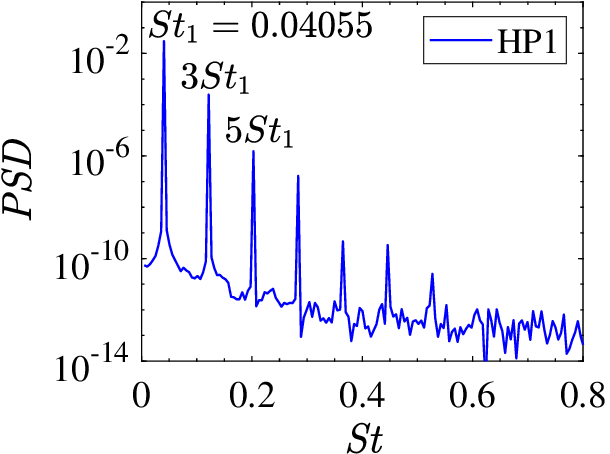}	\put(-120,85){($c$)}
	\vspace{0.3cm}
	\centering\includegraphics[trim=0.0cm 0cm 0.0cm 
	0cm,clip,width=0.30\textwidth]{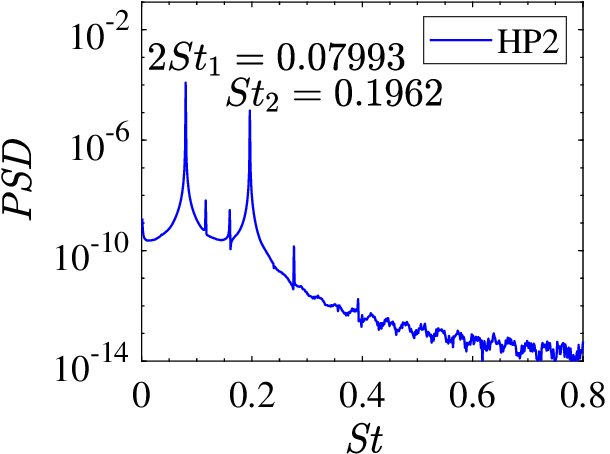}	\put(-120,85){($d$)}
	\quad\ \centering\includegraphics[trim=0.0cm 0cm 0.0cm 
	0cm,clip,width=0.30\textwidth]{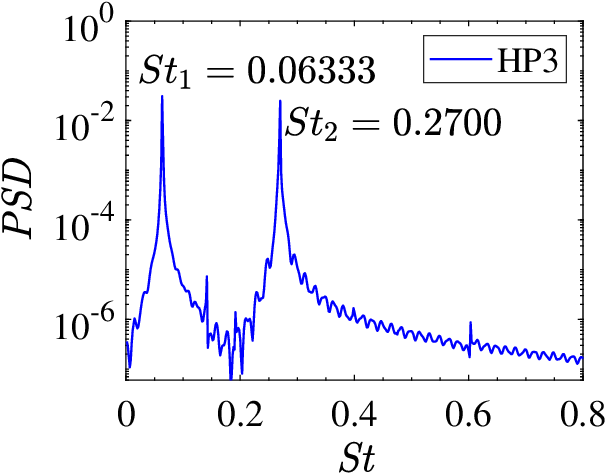}	\put(-120,85){($e$)}
	\quad\  \centering\includegraphics[trim=0.0cm 0cm 0.0cm 
	0cm,clip,width=0.30\textwidth]{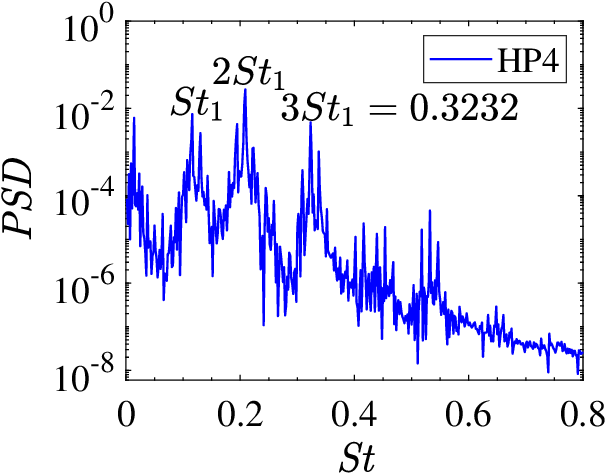}	\put(-120,85){($f$)} 
	\caption{The power spectral density (PSD) for the cases ($a$) LP2, ($b$) LP4, ($c$) HP1, ($d$) HP2, ($e$) HP3 and ($f$) HP4, The PSD is calculated based on the oscillatory part of the time series of the drag coefficient.  }
	\label{fig:PSD}
\end{figure}

Some representative spatial structures and the phase diagrams of the lift-drag coefficients are depicted in figure \ref{fig:wakes-NL} for the short rotating cylindrical wake flow, obtained by the nonlinear DNS. To analyze the wake structure and compare with the results of the global stability analysis, the frequency in the nonlinear saturated system is also computed, i.e., the power spectral density (PSD) in figure \ref{fig:PSD}. 
The spectra are obtained by calculating the oscillatory part in the time series of the drag coefficient, e.g. $C_d'=C_d-\overline{C}_d$. We calculate the PSD from the drag coefficient, instead of the lift coefficient, because the vortex shedding can take place from the end plates and also the curved surface of the short cylinder, as depicted in figure \ref{fig:CD_BC}. When the vortices shedding from the end plates (causing spanwise oscillations) are much weaker than those shedded from the curved surface, we noticed that the corresponding oscillation frequencies cannot be observed clearly in the fast Fourier transform (FFT) spectra of the lift coefficient $C_{ly}$. Compared to the lift coefficient, the drag coefficient seems to be a more robust option for analysing the time-history data in our case. When the vortices shed alternately, the frequency of the drag coefficient is twice that of the lift coefficient.

At low rotation, panels \ref{fig:wakes-NL}($a,b$) show that the saturated states at points LP2 and LP4 are a limit cycle. The wakes LA and LB represent the wake dominated by vortices shedding from the cylinder' flat ends and the circular arc surface, respectively. 
Both types of wake structures LA and LB are also observed in the fixed cylinder flow (see wake patterns P3-2 and P3-1 in figure 10 of \citealt{yang_feng_zhang_2022}, respectively). The difference is that, due to the rotation effect, wake LB undergoes a Hopf bifurcation and becomes a saturated wake state. In the non-rotating cylinder flow, wake LB is only an intermediate transitional state.

The results of high rotation speeds are shown in panels ($c$-$f$). Three frequencies of oscillations, including low (HA), medium (HB), and high (HC), are identified in the present nonlinear wakes, see previous discussions on figure \ref{fig:highalpha_Re_AR1}. 
The wake HA pertaining to the point HP1 (figure \ref{fig:wakes-NL}$c$) is characterised by a low-frequency oscillation in spanwise direction and its higher-order harmonics (see figure \ref{fig:PSD}$c$). Its phase diagram of the lift-drag coefficient is also a limit cycle.
The wake HAB (figure \ref{fig:wakes-NL}$d$) is identified with both a low and a medium frequency oscillation at point HP2. The phase diagram indicates a limit torus with two incommensurate frequencies, see also the PSD result in figure \ref{fig:PSD}($d$). From the perspective of the flow structure, the medium frequency oscillation (HB) is caused by the vortices shedding from the cylindrical arc surface, while the low frequency (HA) is associated to the oscillation of the vortices in the $z$-direction. 
Currently, we are not able to identify a monochromatic wake HB with only a medium frequency. From the vortex street structure of the wake HAB at point HP2, it can be seen that the medium-frequency content exists (referring to the gray-coloured structure in the upper part). In the next subsection \ref{subsubsec:DMDmodes}, we will further use the DMD method to decompose the main components in HAB wake to understand the wake HB.
The wake HAC (figure \ref{fig:wakes-NL}$e$) is identified with a high frequency and a low frequency at point HP3, and its phase diagram of the lift-drag coefficient is also a limit torus (see figure \ref{fig:PSD}$e$ for the PSD result).
Finally, in figure \ref{fig:wakes-NL}$(f)$, a main low-frequency oscillation and its higher harmonics are identified at point HP4, along with a broadband of high-frequency oscillations, leading to a very chaotic signal. HP4 is located at the region where the modes HA and HB are both unstable from the linear analysis. The corresponding phase diagram is also more chaotic compared to the previous cases.

\subsubsection{Comparison with DMD modes} \label{subsubsec:DMDmodes}
According to the global LSA results of a non-rotating 3-D finite-length cylinder flow \citep{yang_feng_zhang_2022}, at the Hopf bifurcation point, we can accurately predict the vortex shedding frequency through the linear stability analysis of the time-mean flow. Moreover, the eigenfrequency of the steady base flow (solved using the SFD method if unstable) does not differ too much from the nonlinear vortex shedding frequency. 
In the present work, we want to establish a qualitative/quantitative relationship between nonlinear and linear systems by comparing the frequency and shape of the dynamic mode decomposition (DMD) modes with the linear global modes. 

The DMD \citep{ROWLEY2009,SCHMID2010,schmid2022dynamic} and its extensions have been applied extensively in flow analyses, and have been tested and proven useful to identify the spatiotemporal patterns of nonlinear flow associated with periodic \citep{bagheri2013koopman,bagheri2014effects} and quasiperiodic oscillations \citep{sierra2022unveiling}, transitional regimes \citep{LeHODMD2017} and turbulent channel flows \citep{le2020coherent}.
The DMD analysis enables a better understanding of the influence of the linear instability on the onset of vortex shedding of the short rotating cylinder flow. For the data to be processed in the DMD analyses, we typically utilised 150 snapshots over five periods of vortex shedding. We have made sure that the simulations reached a steady-state vortex shedding state before capturing the data. In cases where the wake exhibited multiple cycles, we considered the longest cycle when determining the period of interest.

\begin{figure}
	\centering\includegraphics[trim=0.0cm 0cm 0.0cm 0cm,clip,width=0.24\textwidth]{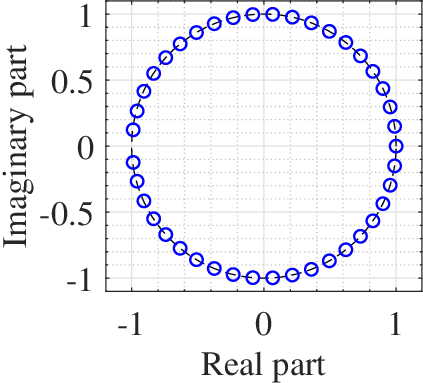}  \put(-95,90){($a$)} \put(-55,86){\fontsize{8pt}{\baselineskip}\selectfont{Point LP2}} \ 
	\centering\includegraphics[trim=0.0cm 0cm 0.0cm 0cm,clip,width=0.241\textwidth]{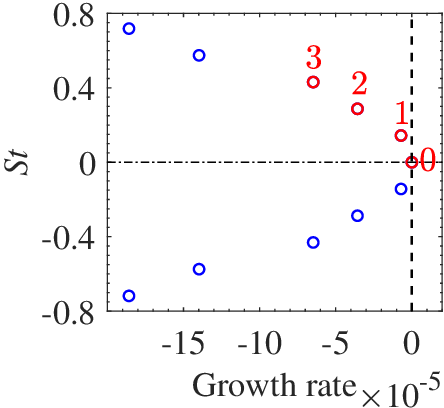} \put(-95,90){($b$)} \put(-55,86){\fontsize{8pt}{\baselineskip}\selectfont{Point LP2}}
	\quad \centering\includegraphics[trim=0.0cm 0cm 0.0cm 0cm,clip,width=0.24\textwidth]{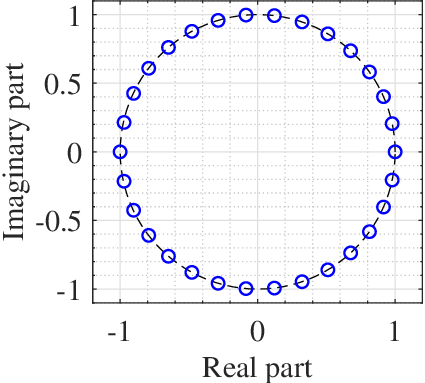} \put(-95,90){($c$)} \put(-55,86){\fontsize{8pt}{\baselineskip}\selectfont{Point LP4}} \ 
	\centering\includegraphics[trim=0.0cm 0cm 0.0cm 0cm,clip,width=0.24\textwidth]{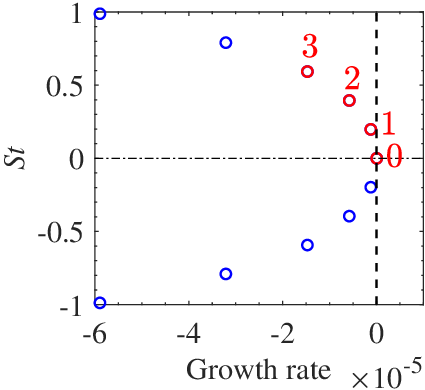} \put(-95,90){($d$)}	\put(-55,86){\fontsize{8pt}{\baselineskip}\selectfont{Point LP4}}
	\caption{The DMD spectra of point LP2 in panels ($a, c$) and of point LP4 in panels ($c, d$). See figure \ref{fig:alpha_Re_AR1}$(a)$ and table \ref{tab:pointseigevalues} for the definitions of points LP2 and LP4. Panels ($a, c$) are on the unit circle and panels ($b, d$) are on the growth-rate-$St$ plane, where the DMD modes 0, 1, 2 and 3 are marked in red, to be discussed in figure \ref{fig:dmdmode_p2}.}
	\label{fig:dmdeigenvalue_p2p4}
\end{figure}
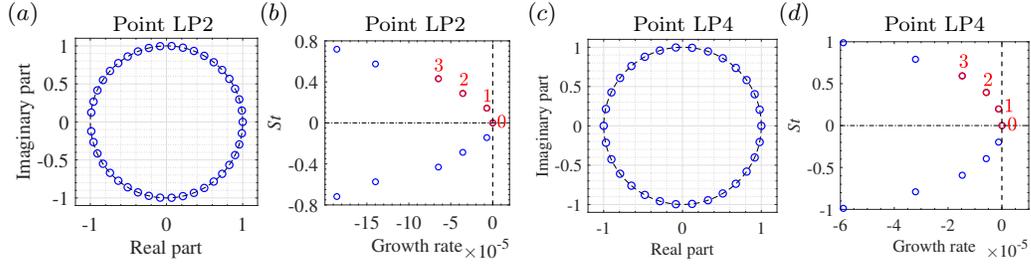

We first recall the results in figure \ref{fig:alpha_Re_AR1}(a) that below the codimension-two point LT for $\alpha<0.1294$, the steady-state flow transitions supercritically to a wave LA as $Re$ exceeds $Re_c$ (represented by the LA curve in this range of $\alpha$). To demonstrate the supercriticality, we have calculated the Landau coefficient $c_1$ and $c_3$ in Appendix \ref{appA}, see figure \ref{fig:landauDNS}. Above the codimension-two point, i.e. $\alpha>0.1294$, the steady-state flow transitions to a wave LB with high frequencies. That is, the codimension-two point corresponds to a double Hopf bifurcation, which characterizes the interaction between mode LA and mode LB.
In the overlap shaded area shown on the $Re-\alpha$ plane in figure \ref{fig:alpha_Re_AR1}$(a)$, where the modes LA and LB both are linearly unstable in the rotating cylinder flow, we cannot obtain single-periodic state corresponding to LA or LB separately for low $\alpha<0.3$ by using different initial conditions. That is, regardless of the initial conditions, our nonlinear flow in the low-rotation-rate regime always converges to the most unstable mode. 
This is in contrast to the results of \cite{sierra2022unveiling} on a rotating sphere, where they have identified a bi-stable region where a single-mode state can be obtained separately and different initial conditions may lead to different flow modes.
In figure \ref{fig:dmdeigenvalue_p2p4}, the DMD spectra of the flows with low rotation rates corresponding to the points LP2 and LP4 in figure \ref{fig:alpha_Re_AR1} are displayed. The results feature fully saturated modes located on the unit circle.

\begin{figure}
	\centering\includegraphics[trim=0.0cm 0cm 0.0cm 0cm,clip,width=0.485\textwidth]{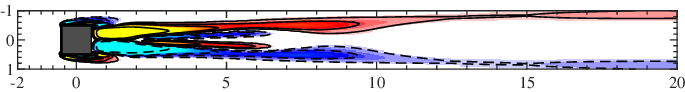}\put(-195,27){($a$)}  \put(-180,25){\fontsize{8pt}{\baselineskip}\selectfont{Mode 0, $\sigma+$i$\omega=4.5919\times10^{-8}+$i$0.0$}} \quad 
	\centering\includegraphics[trim=0.0cm 0cm 0.0cm 0cm,clip,width=0.485\textwidth]{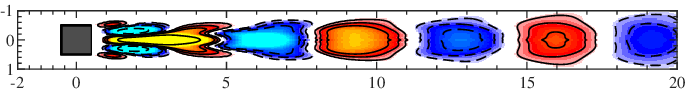}\put(-195,27){($b$)}  \put(-180,25){\fontsize{8pt}{\baselineskip}\selectfont{Mode 1, $\sigma+$i$\omega=-1.436\times10^{-6}+$i$0.1436$}}  \vspace{0.2cm}
	\centering\includegraphics[trim=0.0cm 0cm 0.0cm 0cm,clip,width=0.485\textwidth]{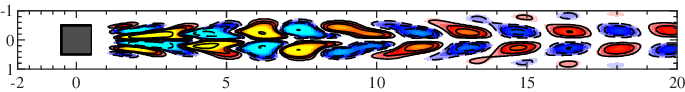}\put(-195,27){($c$)} \put(-180,25){\fontsize{8pt}{\baselineskip}\selectfont{Mode 2, $\sigma+$i$\omega=-3.5685\times10^{-5}+$i$0.2872$}}  \quad 
	\centering\includegraphics[trim=0.0cm 0cm 0.0cm 0cm,clip,width=0.485\textwidth]{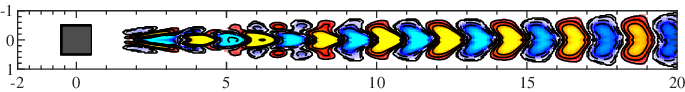}\put(-195,27){($d$)} \put(-180,25){\fontsize{8pt}{\baselineskip}\selectfont{Mode 3, $\sigma+$i$\omega=-6.4764\times10^{-5}+$i$0.4309$}} \vspace{0.2cm} 
	\qquad	\centering\includegraphics[trim=0.0cm 0.0cm 0.0cm 0cm,clip,width=0.47\textwidth]{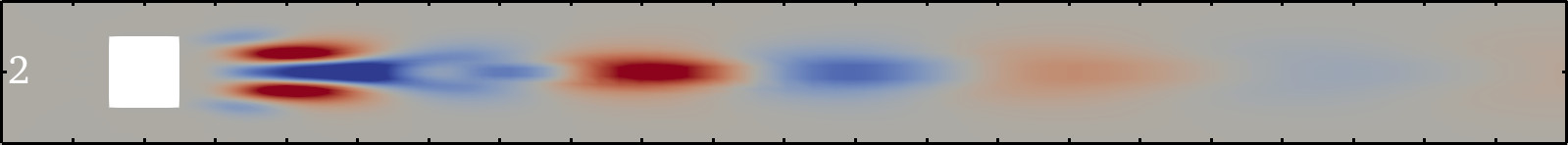}\put(-190,22){($e$)} \put(-175,20){\fontsize{8pt}{\baselineskip}\selectfont {$\rm Mode^{LA}$ $(\sigma+$i$\omega)_{\rm BF}=3.3957\times10^{-2}+$i$0.1411$} }\quad \ \ 
	\centering\includegraphics[trim=0cm 0cm 0cm 0cm,clip,width=0.47\textwidth]{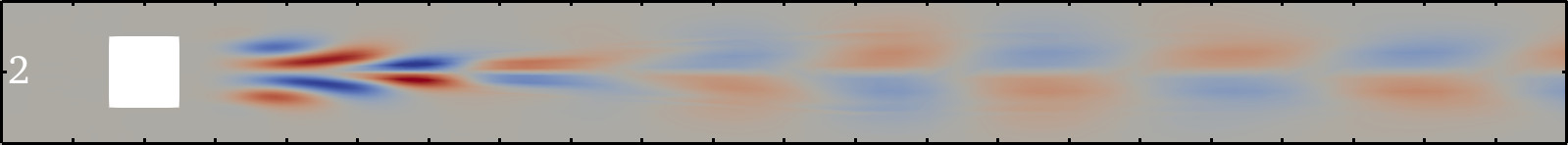}\put(-190,22){($f$)}   \put(-175,20){\fontsize{8pt}{\baselineskip}\selectfont {$\rm Mode^{LB}$, $(\sigma+$i$\omega)_{\rm BF}=1.6972\times10^{-2}+$i$0.1782$}}
 \vspace{0.3cm}
\qquad \qquad \centering\includegraphics[trim=0.0cm 0.0cm 4.15cm 0.0cm,clip,width=0.47\textwidth]{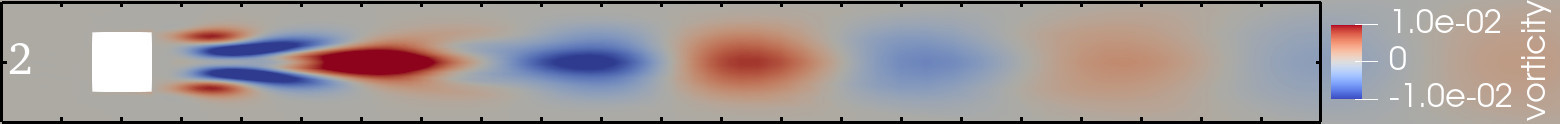} \put(-191,22){($g$)}   \put(-175,20){\fontsize{8pt}{\baselineskip}\selectfont {$\rm Mode^{LA}$, $(\sigma+$i$\omega)_{\rm MF} =8.4211\times10^{-3}+$i$0.1431$}} \ \
\quad \centering\includegraphics[trim=0.0cm 0.0cm 4.15cm 0cm,clip,width=0.47\textwidth]{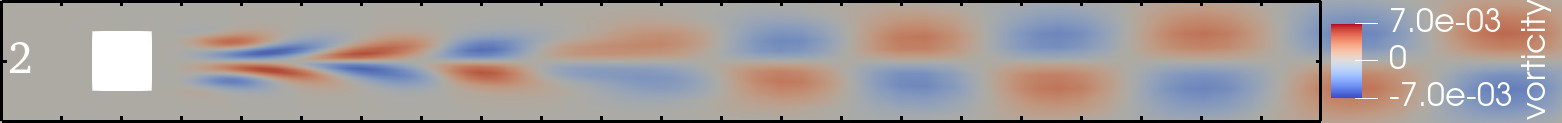} \put(-191,22){($h$)}   \put(-175,20){\fontsize{8pt}{\baselineskip}\selectfont {$\rm Mode^{LB}$, $(\sigma+$i$\omega)_{\rm MF}= -4.3530\times10^{-2}+$i$0.1915$}}
	\caption{Comparison of the DMD modes with the global modes at the point LP2 ($\alpha=0.05,Re=305$). Four DMD modes and the corresponding DMD eigenvalues are shown in the panels ($a-d$). The global modes LA ($e,g$) and LB ($f,h$) in the global stability analysis based on the SFD base flow ($e,f$) and mean flow($g,h$).
	All figures colored by streamwise vorticity. 
	The eigenfrequencies of the DMD modes are in good agreement with those obtained from FFT method (figure \ref{fig:PSD} $a$), with a relative error of $0.48\%$. Moreover, the eigenfrequencies of the DMD modes match well with the eigenfrequencies of the mean flow global modes (panel $g$ and $f$), with a relative error of $0.83\%$. However, the difference between the characteristic frequencies of the DMD modes and the eigenfrequencies of the SFD base flow global mode is slightly larger, with a relative error of $2.22\%$. Additionally, the topological structure of the DMD modes is almost identical to that of the mean flow global modes.}
	\label{fig:dmdmode_p2}
\end{figure}

\begin{figure}
	\centering\includegraphics[trim=0.0cm 0cm 0.0cm 0cm,clip,width=0.485\textwidth]{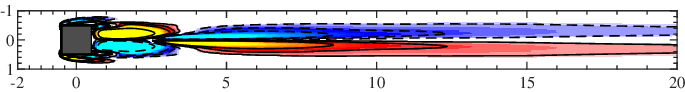} \put(-196,27){($a$)} \put(-180,25){\fontsize{8pt}{\baselineskip}\selectfont {Mode 0, $\sigma+$i$\omega =3.7369\times10^{-8}+$i$0.0$}}  \quad 
	\centering\includegraphics[trim=0.0cm 0cm 0.0cm 0cm,clip,width=0.485\textwidth]{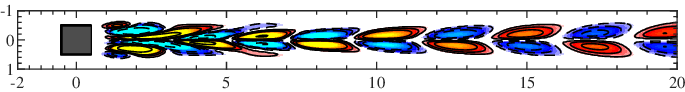} \put(-196,27){($b$)}   \put(-180,25){\fontsize{8pt}{\baselineskip}\selectfont {Mode 1, $\sigma+$i$\omega =-5.2874\times10^{-6}+$i$0.1976$}}	\vspace{0.2cm}
	\centering\includegraphics[trim=0.0cm 0cm 0.0cm 0cm,clip,width=0.485\textwidth]{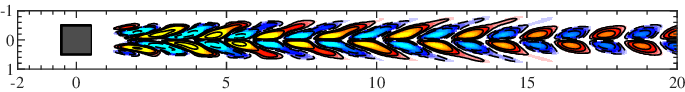} \put(-196,27){($c$)}  \put(-180,25){\fontsize{8pt}{\baselineskip}\selectfont {Mode 2, $\sigma+$i$\omega= -1.9908\times10^{-5}+$i$0.3953$}} \quad 
	\centering\includegraphics[trim=0.0cm 0cm 0.0cm 0cm,clip,width=0.485\textwidth]{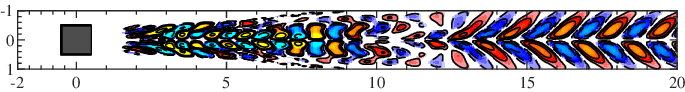} \put(-196,27){($d$)}   \put(-180,25){\fontsize{8pt}{\baselineskip}\selectfont {Mode 3, $\sigma+$i$\omega =-3.8395\times10^{-5}+$i$0.5929$}}	\vspace{0.2cm}
	\qquad \qquad \centering\includegraphics[trim=0.0cm 0.0cm 3.98cm 0.0cm,clip,width=0.47\textwidth]{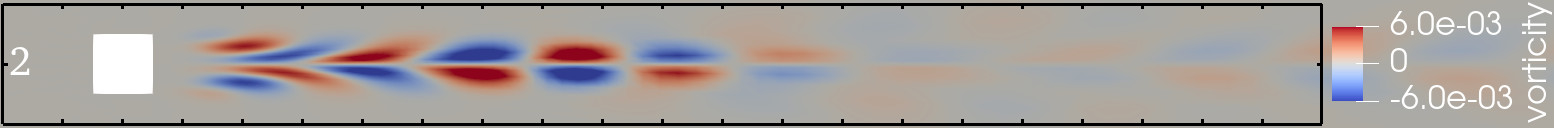} \put(-191,22){($e$)}   \put(-175,20){\fontsize{8pt}{\baselineskip}\selectfont {$\rm Mode^{LB}$, $(\sigma+$i$\omega)_{\rm BF} =9.3466\times10^{-2}+$i$0.2077$}} \ \ 
   \quad \centering\includegraphics[trim=0.0cm 0.0cm 4.15cm 0cm,clip,width=0.47\textwidth]{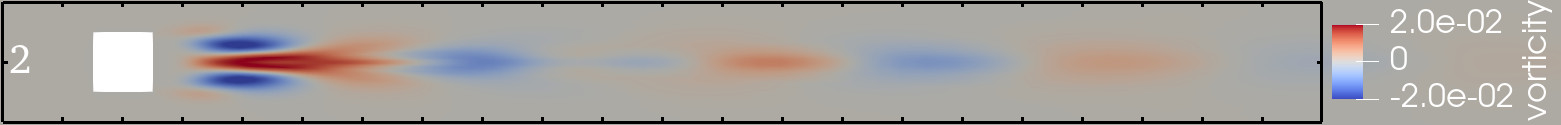} \put(-191,22){($f$)}   \put(-175,20){\fontsize{8pt}{\baselineskip}\selectfont {$\rm Mode^{LA}$, $(\sigma+$i$\omega)_{\rm BF}= 4.6550\times10^{-2}+$i$0.1451$}} \vspace{0.3cm}
   	\qquad \qquad \centering\includegraphics[trim=0.0cm 0.0cm 0cm 0.0cm,clip,width=0.47\textwidth]{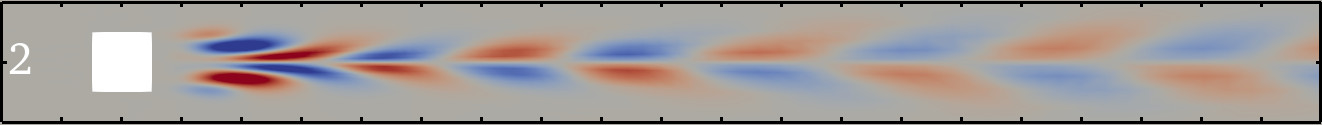} \put(-191,22){($g$)}   \put(-175,20){\fontsize{8pt}{\baselineskip}\selectfont {$\rm Mode^{LB}$, $(\sigma+$i$\omega)_{\rm MF} =5.7977\times10^{-3}+$i$0.1917$}} \ \
   \quad \centering\includegraphics[trim=0.0cm 0.0cm 4.15cm 0cm,clip,width=0.47\textwidth]{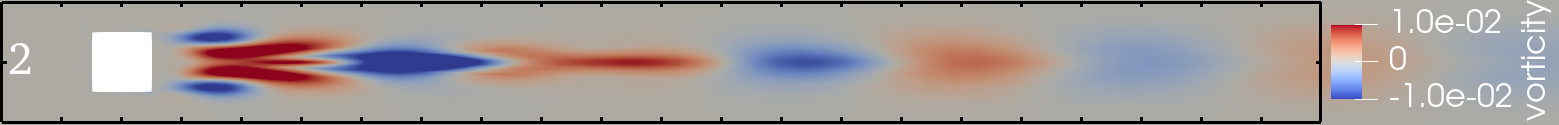} \put(-191,22){($h$)}   \put(-175,20){\fontsize{8pt}{\baselineskip}\selectfont {$\rm Mode^{LA}$, $(\sigma+$i$\omega)_{\rm MF}= 6.8022\times10^{-3}+$i$0.1481$}}
	\caption{Comparison of the DMD modes with the global modes at the point LP4 ($\alpha=0.1,Re=330$). Four DMD modes and the corresponding DMD eigenvalues are shown in the panels ($a-d$). The global modes LA ($f,h$) and  LB ($e,g$) in the global stability analysis based on the SFD base flow ($e,f$) and mean flow($g,h$). All figures colored by streamwise vorticity. 
	 The eigenfrequencies of the DMD modes are in good agreement with those obtained from FFT analysis (figure \ref{fig:PSD} $a$), with a relative error of $0.05\%$. However, the characteristic frequencies of DMD modes are not in complete agreement with those of mean flow and SFD base flow, with errors of $3.03\%$ and $4.81\%$, respectively. Additionally, the topological structure of DMD modes is nearly identical to that of mean flow global modes, as compared to the SFD base flow. }
	\label{fig:dmdmode_p4}
\end{figure}
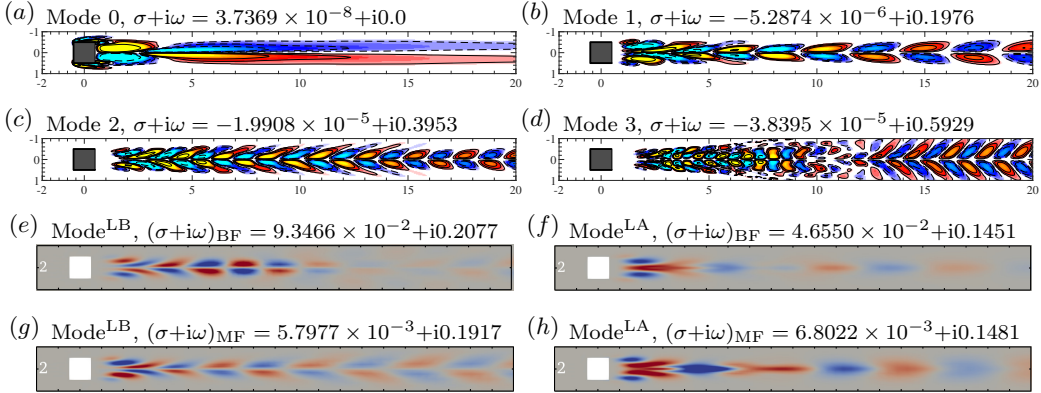

Figures \ref{fig:dmdmode_p2} and \ref{fig:dmdmode_p4} show the comparison of the DMD modes with the global modes based on the SFD base flow and time-mean base flow at selected points LP2 and LP4, respectively.
Figure \ref{fig:dmdmode_p2} shows that DMD mode 0 is the time-averaged flow filed. The DMD mode 1 presents alternated flow structures downstream. According to the imaginary part of the DMD eigenvalues, DMD modes 2, 3 are the second and third harmonics of mode 1, respectively. At point LP2, the frequencies of the leading DMD mode (panel $b$) and the leading global eigenmode based on the SFD base flow (panel $e$) are 0.1436 and 0.1411, respectively, with a difference of $1.7\%$.
But there is no DMD mode with a frequency close to $\omega=0.1782$, which is also a linearly unstable mode (figure \ref{fig:dmdmode_p2}$f$). 
In the global stability analysis, the (unstable) base flow is solved by the SFD method, whereas our DMD method analyses the saturated flow regime, which hints that the difference in the frequencies in the two methods may be reduced if we use the time-mean flow of the saturated regime in the global stability analysis. This has been carried out and the results are shown in the last row of figure \ref{fig:dmdmode_p2} with the subscript $_{\rm MF}$. Now we can see that the frequency of the leading global mode based on the time-mean flow is indeed closer to the leading DMD mode, comparing panel $g$ with panel $b$.

Similarly, the DMD and global LSA results for the point LP4 are shown in figure \ref{fig:dmdmode_p4}. One can see that the frequencies of the leading DMD mode (panel $b$) and the global mode based on the SFD base flow (panel $e$) are 0.1976 and 0.2077, respectively, with a difference of $4.9\%$. Also, there is no DMD mode with a frequency closed to $\omega=0.1451$, which is a linearly unstable mode (figure \ref{fig:dmdmode_p4} $f$). The leading global mode based on the time-mean flow in panel $g$ has a closer frequency (0.1917) compared with the one based on the SFD base flow to the DMD mode. The global eigenfunction based on the time-mean flow also looks more similar to the DMD mode.

 \begin{figure}
	\centering\includegraphics[trim=0.0cm 0cm 0.0cm 0cm,clip,width=0.24\textwidth]{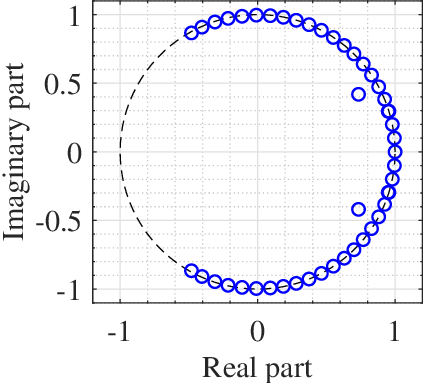} \put(-95,90){($a$)}	\put(-55,85){Point HP2} \ 
	\centering\includegraphics[trim=0.0cm 0cm 0.0cm 0cm,clip,width=0.24\textwidth]{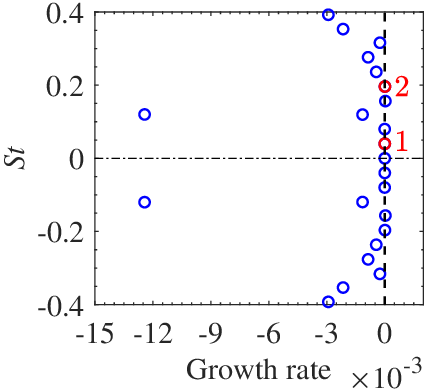} \put(-95,90){($b$)}	\put(-55,85){Point HP2}
	\centering\includegraphics[trim=0.0cm 0cm 0.0cm 0cm,clip,width=0.24\textwidth]{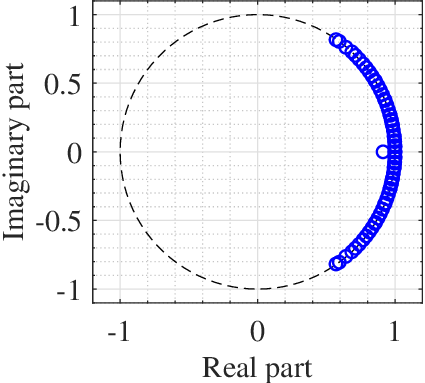} \put(-95,90){($c$)} \put(-55,85){Point HP3} \ 
	\centering\includegraphics[trim=0.0cm 0cm 0.0cm 0cm,clip,width=0.24\textwidth]{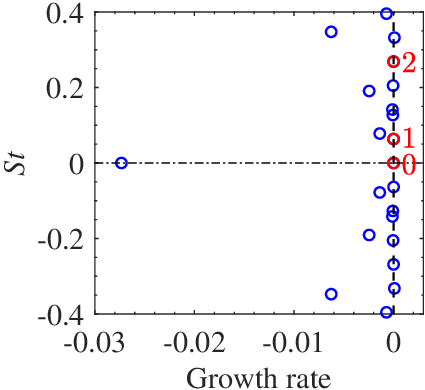} \put(-95,90){($d$)} \put(-55,85){Point HP3} \\
	\caption{The DMD eigenvalues spectrum of points HP2 ($a$, $b$) and HP3 ($c$, $d$) on the unit circle and on the growth-rate-$St$ plane.
}
	\label{fig:dmdeigenvalue_pointP4P5}
\end{figure}

\begin{figure}
	\centering
	\includegraphics[trim=0cm 0cm 0cm 0cm,clip,width=0.485\textwidth]{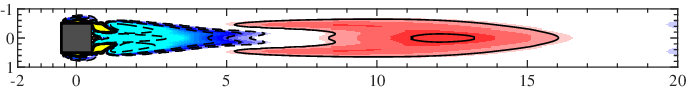} \put(-195,27){($a$)} \put(-182,25){\fontsize{8pt}{\baselineskip}\selectfont {Mode 1, $\sigma+$i$\omega =-9.660\times10^{-7}+$i$0.03994$}}  \quad 
	\centering
	\includegraphics[trim=0cm 0cm 0cm 0cm,clip,width=0.485\textwidth]{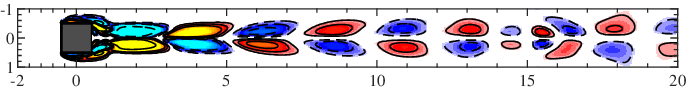} \put(-195,27){($b$)}  \put(-182,25){\fontsize{8pt}{\baselineskip}\selectfont {Mode 2, $\sigma+$i$\omega= 7.360\times10^{-6}+$i$0.1963$}} 
	\vspace{0.1cm}
	\centering
	\qquad\ \includegraphics[trim=0cm 0cm 4.05cm 0cm,clip,width=0.468\textwidth]{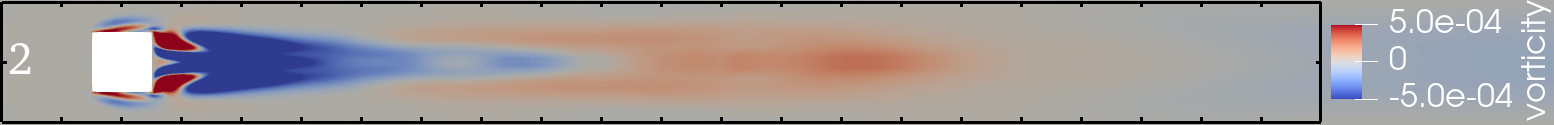} \put(-193,27){($c$)}   \put(-180,20){\fontsize{8pt}{\baselineskip}\selectfont {$\rm Mode^{HA}$, $(\sigma+$i$\omega)_{\rm BF} =-7.499\times10^{-5}+$i$0.04026$}} \qquad 
	\centering
	\includegraphics[trim=0cm 0cm 4.05cm 0cm,clip,width=0.468\textwidth]{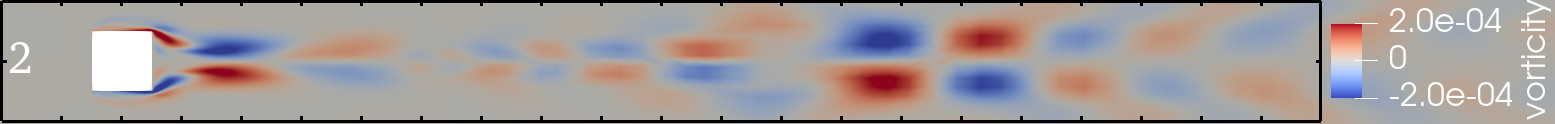} \put(-193,27){($d$)}   \put(-180,20){\fontsize{8pt}{\baselineskip}\selectfont {$\rm Mode^{HB}$, $(\sigma+$i$\omega)_{\rm BF}= 1.905\times10^{-3}+$i$0.1968$}} 
	\vspace{0.2cm}
	\centering
	\qquad \includegraphics[trim=0cm 0cm 4.25cm 0cm,clip,width=0.468\textwidth]{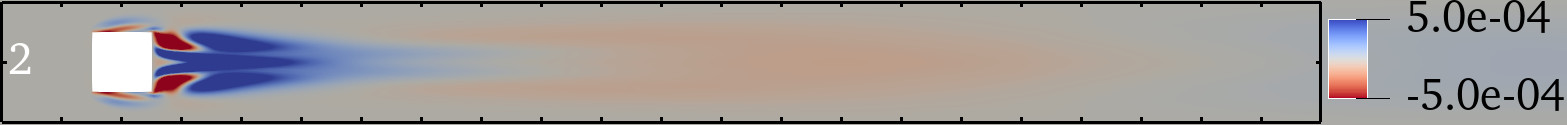} \put(-193,27){($e$)}   \put(-180,20){\fontsize{8pt}{\baselineskip}\selectfont {$\rm Mode^{HA}$, $(\sigma+$i$\omega)_{\rm MF} =7.740\times10^{-4}+$i$0.04004$}} \qquad 
	\centering
	\includegraphics[trim=0cm 0cm 4.2cm 0cm,clip,width=0.468\textwidth]{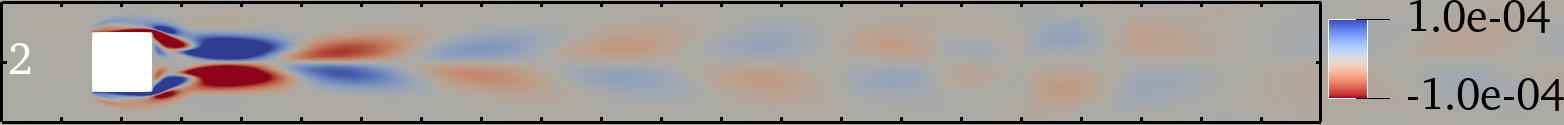} \put(-193,27){($f$)}   \put(-180,20){\fontsize{8pt}{\baselineskip}\selectfont {$\rm Mode^{HB}$, $(\sigma+$i$\omega)_{\rm MF}= -3.019\times10^{-3}+$i$0.1964$}} 
	\caption{Comparison of the DMD modes ($a, b$) with the global SFD base flow modes ($c, d$) and mean flow modes ($e, f$) at point HP2. The corresponding eigenvalues are shown in the panels ($a-f$). All figures colored by streamwise vorticity. 
	The eigenfrequencies of the DMD modes are in good agreement with those obtained from FFT analysis (figure \ref{fig:PSD} $d$), with a relative error of $0.05\%$. Because the point HP2 is very close to the neutral curve, the eigenvalues of both mean flow and SFD base flow can predict the true frequency well, with errors of $0.2\%$ and $0.8\%$, respectively. }
	\label{fig:dmdmode_hp2}
\end{figure}

\begin{figure}
	\centering
	\includegraphics[trim=0cm 0cm 0cm 0cm,clip,width=0.485\textwidth]{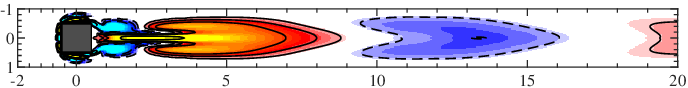} \put(-197,27){($a$)} \put(-182,25){\fontsize{8pt}{\baselineskip}\selectfont {Mode 1, $\sigma+$i$\omega =3.307\times10^{-6}+$i$0.06350$}}  \quad 
	\centering
	\includegraphics[trim=0cm 0cm 0cm 0cm,clip,width=0.485\textwidth]{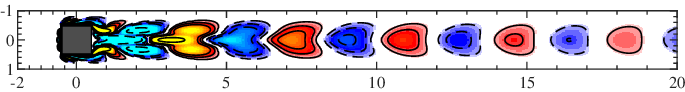} \put(-197,27){($b$)}  \put(-182,25){\fontsize{8pt}{\baselineskip}\selectfont {Mode 2, $\sigma+$i$\omega= -1.165\times10^{-6}+$i$0.2686$}} 
    \vspace{0.1cm}
   	\centering
    \qquad\ \includegraphics[trim=0cm 0cm 4.1cm 0cm,clip,width=0.468\textwidth]{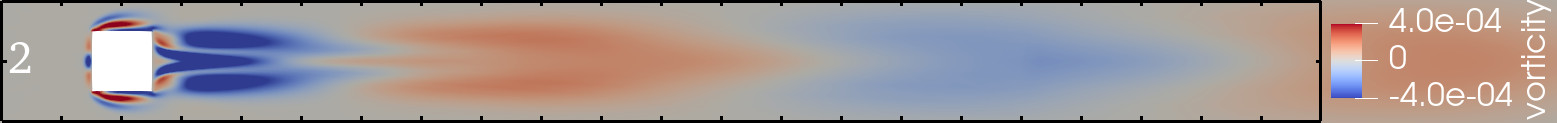} \put(-193,27){($c$)}   \put(-180,20){\fontsize{8pt}{\baselineskip}\selectfont {$\rm Mode^{HA}$, $(\sigma+$i$\omega)_{\rm BF} =-2.261\times10^{-2}+$i$0.06755$}} \qquad 
	\centering
    \includegraphics[trim=0cm 0cm 4cm 0cm,clip,width=0.468\textwidth]{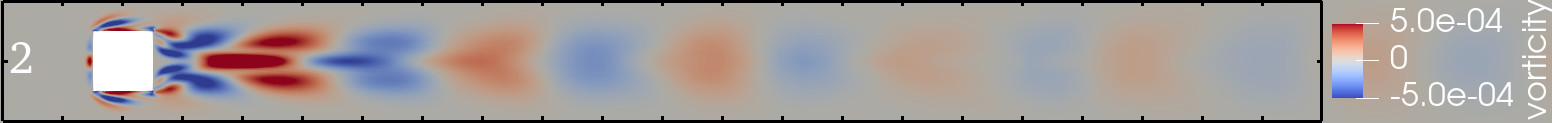} \put(-193,27){($d$)}   \put(-180,20){\fontsize{8pt}{\baselineskip}\selectfont {$\rm Mode^{HC}$, $(\sigma+$i$\omega)_{\rm BF}= 3.050\times10^{-2}+$i$0.2646$}} 
    \vspace{0.2cm}	
   	\centering
    \qquad \includegraphics[trim=0cm 0cm 4.05cm 0cm,clip,width=0.468\textwidth]{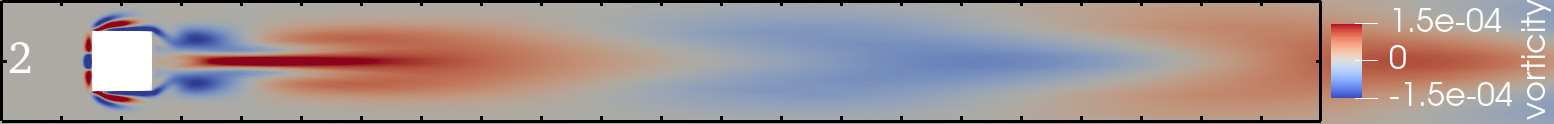} \put(-193,27){($e$)}   \put(-180,20){\fontsize{8pt}{\baselineskip}\selectfont {$\rm Mode^{HA}$, $(\sigma+$i$\omega)_{\rm MF} =-5.764\times10^{-2}+$i$0.06698$}} \qquad 
    \centering
    \includegraphics[trim=0cm 0cm 4.05cm 0cm,clip,width=0.468\textwidth]{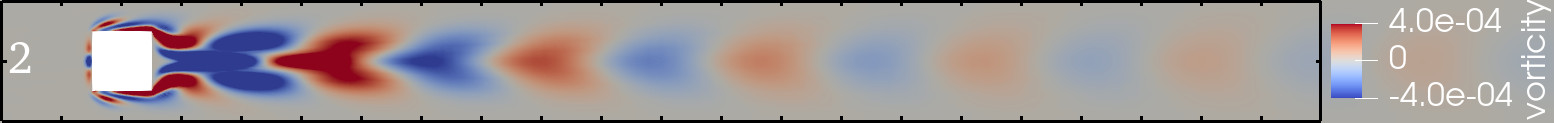} \put(-193,27){($f$)}   \put(-180,20){\fontsize{8pt}{\baselineskip}\selectfont {$\rm Mode^{HC}$, $(\sigma+$i$\omega)_{\rm MF}= 1.673\times10^{-2}+$i$0.2707$}} 
	\caption{Comparison of the DMD modes ($a, b$) with the global SFD base flow modes ($c, d$) and mean flow modes ($e, f$) at point HP3. The corresponding eigenvalues are shown in the panels ($a-f$). All figures colored by streamwise vorticity. 
	The eigenfrequencies of the DMD modes are in good agreement with those obtained from FFT analysis (figure \ref{fig:PSD} $e$), with a relative error of $0.7\%$. However, the frequencies of DMD modes are not in complete agreement with those of mean flow and SFD base flow, with errors of $5.8\%$ and $6.6\%$, respectively. Additionally, the topological structure of DMD modes is nearly identical to that of mean flow global modes, as compared to the SFD base flow. }
	\label{fig:dmdmode_hp3}
\end{figure}
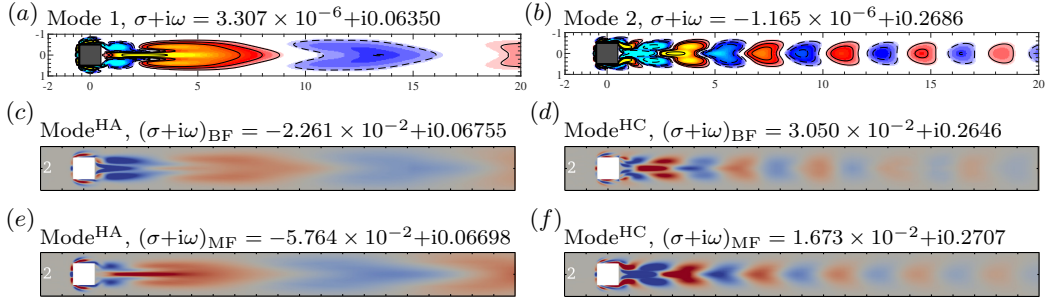

The above results pertain to the low-rotation-rate cases. Next, we compare the linear and nonlinear results for the high-rotation-rate flows. We plot the DMD eigenspectra in figure \ref{fig:dmdeigenvalue_pointP4P5} for the selected HP2 and HP3 cases. We discuss the DMD modes labelled in red in the figures. 
For HP2, the DMD modes are plotted in figure \ref{fig:dmdmode_hp2}($a,b$) along with the global modes based on the SFD base flow ($c,d$) and the time-mean flow ($e,f$). Compared with the case of a low rotation rate in figures \ref{fig:dmdmode_p2},\ref{fig:dmdmode_p4}, the similarity of the DMD modes with the global modes in the high-rotation-rate case is greater; for example, the first column in figure \ref{fig:dmdmode_hp2} shows that the three modes look similar, and the global mode based on the mean flow is again slightly better compared to the DMD mode. In the second column, we can also find DMD mode 2 that resembles the global modes in the global stability analyses, where the global mode based on the time-mean base flow looks closer to the DMD mode. The same conclusion can be drawn for the HP3 point in figure \ref{fig:dmdmode_hp3}. As the discussions are similar, we will not go into detail about them. To sum up, through the comparison of the frequencies and the shapes of the linear global modes and DMD modes, we can establish a connection between the linear and nonlinear systems.

\begin{figure}
	\hfil
	\centering\includegraphics[trim=0.0cm 0cm 0.0cm 0cm,clip,width=0.48\textwidth]{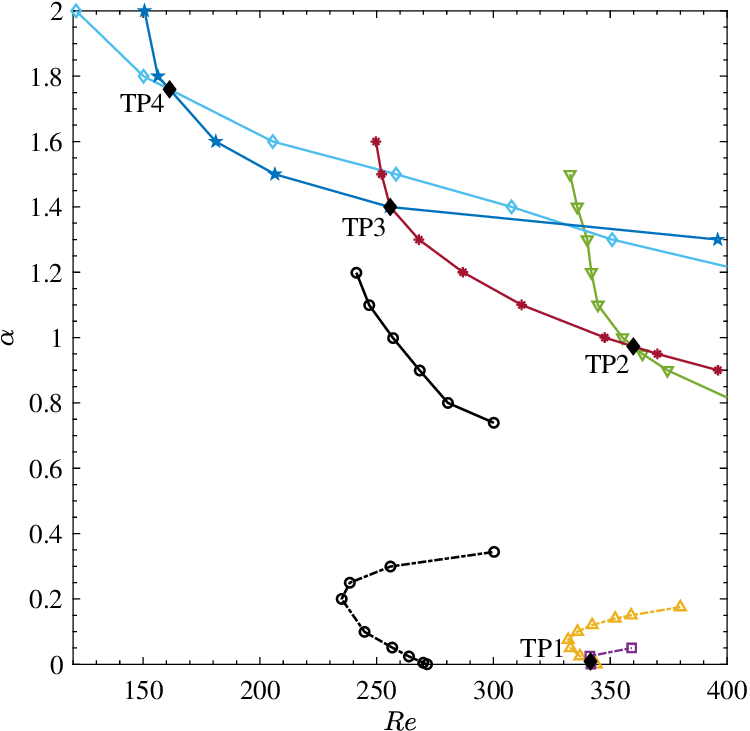}
	\put(-190,175){(a)} \ \ \quad
	\centering\includegraphics[trim=0.0cm 0cm 0.0cm 0cm,clip,width=0.48\textwidth]{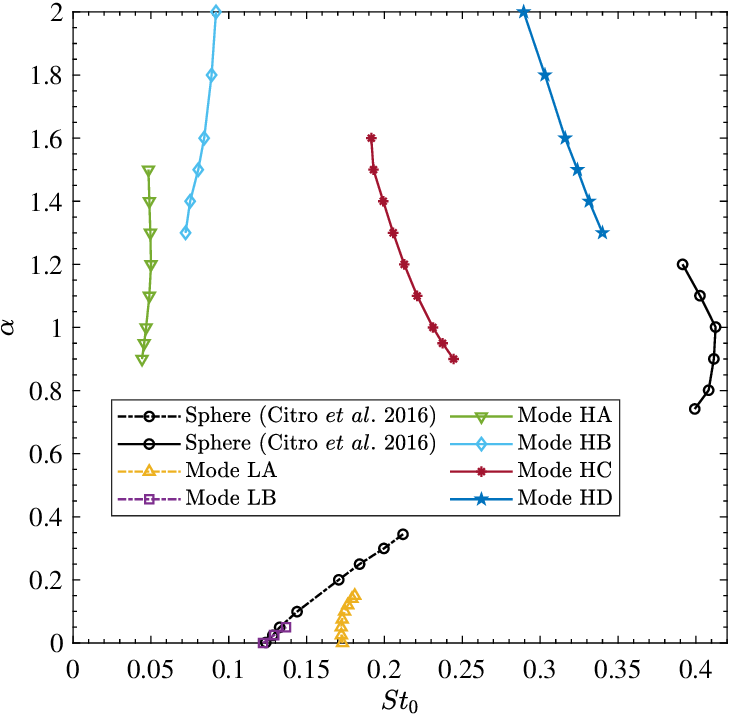}
	\put(-190,175){(b)} \ \ \quad
	\caption{Neutral stability curves and the corresponding frequencies for the flow past a short rotating cylinder at $\ar=0.75$,  compared with a vertically rotating sphere \citep{citro2016}.}
	\label{fig:ar075alpha_Re}
\end{figure}

\subsection{Effect of $\ar$} \label{subsec:EffectOfAR}
In this last section, we will discuss the effect of $\ar$ on the global modes in a short rotating cylinder wake flow, as this information seems to be scarce in the literature on the short (rotating) cylinder flows.

Figure \ref{fig:ar075alpha_Re} shows the neutral stability curves in panel ($a$) and the shedding frequency in panel ($b$) for the case of $\ar=0.75$. The results of the flow past a sphere \citep{citro2016} are also shown for a comparison. Similar to the $\ar=1$ results in figures \ref{fig:alpha_Re_AR1},\ref{fig:highalpha_Re_AR1}, the low and high rotation flows at $\ar=0.75$ present dissimilar behaviours. In the case of low rotation speeds, two modes LA, LB undergo Hopf bifurcation due to linear instability around $Re=330, 340$, respectively. The linear unstable region of the mode LB is very small as shown in panel \ref{fig:ar075alpha_Re}$(a)$, almost completely inside the unstable region of mode LA. In the case of high rotation speeds, four global unstable modes are identified up to $\alpha=2$, successively experiencing Hopf bifurcation, resulting in three turning points (TP2 to TP4). All these modes are characterised by different frequencies as shown in panel $(b)$. The frequencies in the low-rotation-rate flows are close to the frequencies of the slowly-rotating sphere, whereas the frequencies of the high-rotation-rate modes are smaller than those of the corresponding sphere. The structures of the global modes at low and high rotation rates are shown in figures \ref{fig:eigenmode_ar075low},\ref{fig:eigenmode_ar075high} and discussed in Appendix \ref{app_ar2}.

\begin{figure}
	\hfil
	\centering\includegraphics[trim=0.0cm 0cm 0.0cm 0cm,clip,width=0.48\textwidth]{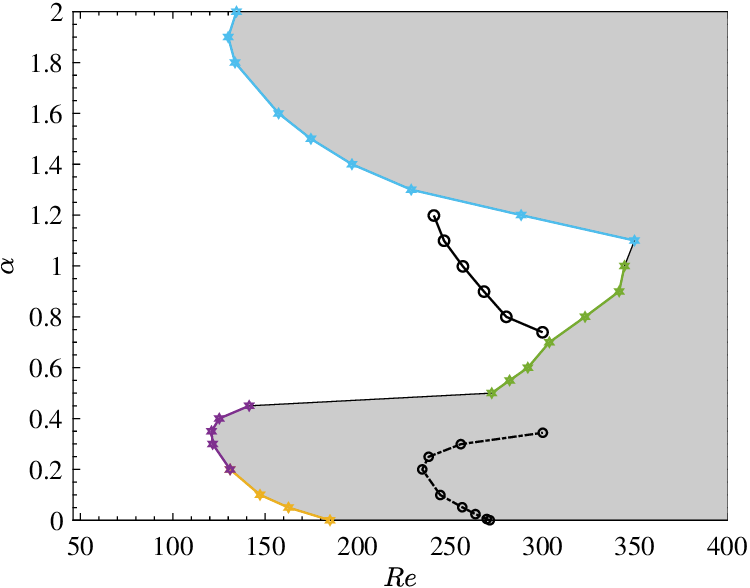}
	\put(-190,145){(a)} \ \ \quad
	\centering\includegraphics[trim=0.0cm 0cm 0.0cm 0cm,clip,width=0.48\textwidth]{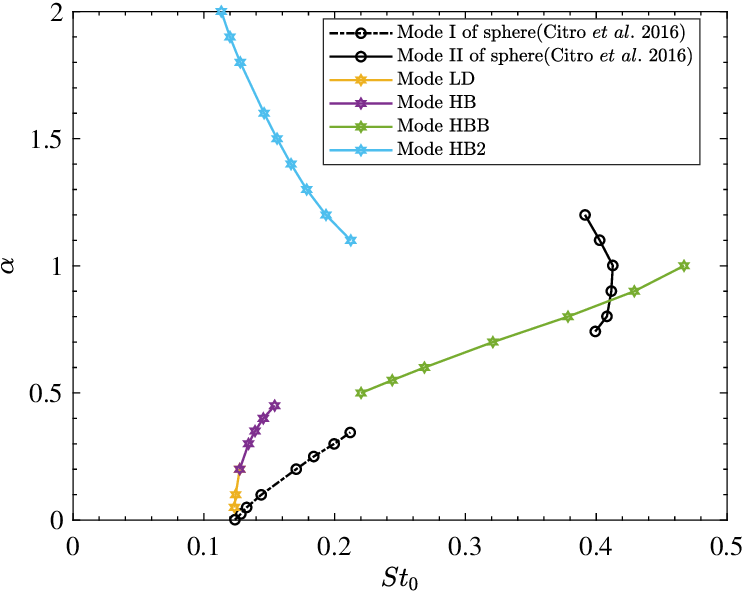}
	\put(-190,145){(b)} \ \ \quad
	\caption{Neutral stability curves and the corresponding frequencies for the flow past a short rotating cylinder at $\ar=2$, compared to a vertically rotating sphere \citep{citro2016}.}
	\label{fig:ar2alpha_Re}
\end{figure}

The results of $\ar=2$ are shown in figure \ref{fig:ar2alpha_Re}. In this case, the calculation seems to be more difficult to converge. We will interpret the results with caution. The shapes of the corresponding global modes are shown in figure \ref{fig:eigenmode_ar2} in Appendix \ref{app_ar2}. When the rotation rate is small in this case, we can identify an unstable mode appearing similarly to the LD mode in $\ar=1$; this mode will be similarly called LD. Further increasing $\alpha$, the shape of the most unstable global mode changes, see the colour transition from orange to purple in panels \ref{fig:ar2alpha_Re}($a,b$) and the comparison between figure \ref{fig:eigenmode_ar2}($a$) and figure \ref{fig:eigenmode_ar2}($b$). In the higher $\alpha$ regime, the most unstable global mode changes abruptly, denoted by green and blue lines in figure \ref{fig:ar2alpha_Re}. We tried to converge as many unstable modes as we could, but some calculations were not converged. Thus, we will not go into details for the high-rotation-rate cases in the $\ar=2$. Such difficulty in converging the 3-D wake flow is not uncommon, reflecting the complex nature of these flows and highlighting more research efforts to decipher their dynamics. 

\begin{figure}
	\hfil
	\centering\includegraphics[trim=0.0cm 0cm 0.0cm 0cm,clip,width=0.99\textwidth]{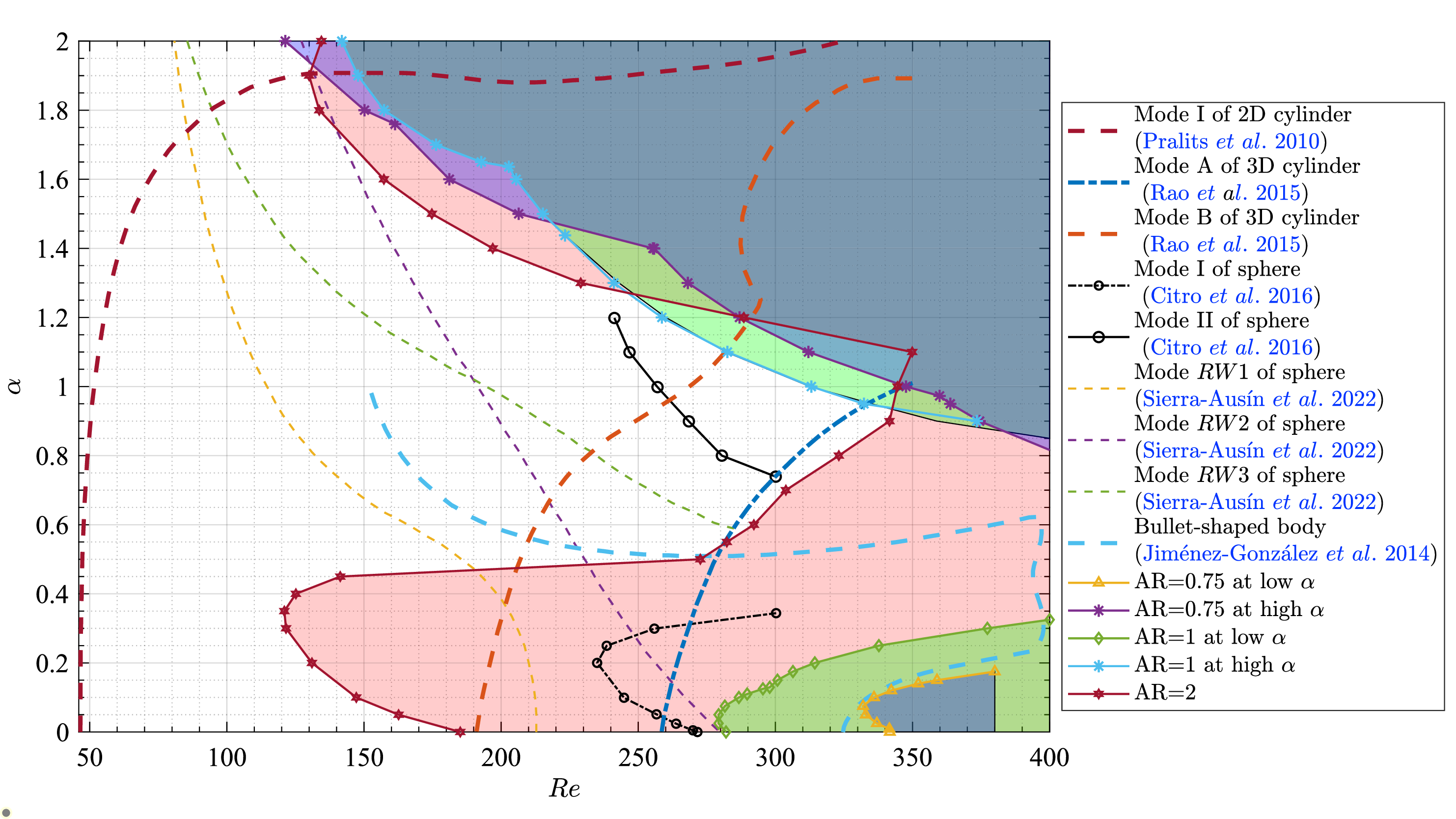}
	\caption{Neutral stability curves of Hopf bifurcation for the flow past a rotating finite cylinder at $\ar=0.75,1,2$. Comparison with streamwise rotating infinite cylinder \citep{pralits2010instability,rao2015review}, sphere \citep{citro2016}, and streamwise rotating sphere \citep{sierra2022unveiling} and bullet-like body \citep{spinningbullet_2014}.}
	\label{fig:alpha_Re}
\end{figure}

In the end, we consolidate and compare all the significant results in this work spanning a large parameter space with $Re\in[100,500], \alpha\in[0,2]$ for $\ar=0.75,1,2$ in figure \ref{fig:alpha_Re}. To place our results in a more general context, we also compare our results with other rotating bluff-body flows such as 2-D rotating cylinders \citep{pralits2010instability,rao2015review}, spheres \citep{citro2016,sierra2022unveiling} and bullet-like body \citep{spinningbullet_2014}. 
It can be seen from figure \ref{fig:alpha_Re} that in the low-rotation-rate regime, larger $\ar$ renders the flow more unstable as the critical $Re$ decreases from $Re_c=340$ for $\ar=0.75$ to $Re_c=50$ for $\ar=\infty$ (2-D case, \citealt{pralits2010instability}). In the high-rotation-rate end (with the maximum rotation rate being $\alpha=2$ in our work), it is difficult to summarise a trend of increasing $\ar$ from our finite-length cylinders to the infinitely long cylinder. As discussed above, the computations in this regime are more difficult, calling for more research efforts to elucidate the difference. The 3-D instability in the infinitely-long cylinder \citep{rao2015review} presents a smoother transition in the range of $\alpha\in[0,2]$, different from our 3-D results where distinct behaviours in the low- ($\alpha\lesssim0.6$) and high-rotation rate ($0.6\lesssim\alpha\lesssim2$) cases can be identified. This is because $\alpha=2$ is not a high rotation rate for infinitely-long cylinder flows; in Fig. 1 of \cite{rao2015review}, dissimilar wake behaviours in this flow are separated by $\alpha\approx2.5$. 
The dynamics of the rotating sphere wake flow along the transverse direction \citep{citro2016} is similar to that of the short rotating cylinder with $\ar$ slightly larger than 1, whereas the rotating sphere wake along the streamwise direction \citep{sierra2022unveiling} looks more dissimilar than ours. This is likely because our cylinder is also rotating along a transverse axis. In the end, the flow past a spinning bullet-shaped bluff body \citep{spinningbullet_2014} is also shown for a comparison and its low-rotation-rate behaviour appears similarly to our flow.
By studying the effect of $\ar$, we can qualitatively connect our results with those for the sphere, the cylinders of infinite length, etc. and explain the difference between these benchmark flows in a large parameter space.

\section{Conclusions} \label{sec:Conclusions}
In this work, a 3-D flow stability problem past a short rotating cylinder has been studied. The motivation for considering this flow configuration is due to its applications in various engineering settings and its relevance to flow control strategy by rotation. New flow modes have been identified in our direct numerical simulations and global stability analyses of this flow. The linear results have also been compared with the nonlinear results to find their traces in real flows. The wavemaker region responsible for the instability generation has been delimited. We have also studied the effect of aspect ratio $\ar$ to understand how the 3-D flows change with its geometry parameters. 

Firstly, for a cylinder with $\ar=1$, when the rotation rate $\alpha$ is slower than 0.3, the rotation effect only trivially affects the flow past a short cylinder. The two unstable global modes (see figure \ref{fig:alpha_Re_AR1}$a$) in the low-rotation-rate cases resemble those in the non-rotating flows, corresponding to the vortices shedding from the flat ends of cylinder and the vortices shedding from the curved surface, respectively. The rotating effect swings the recirculation region towards the rotating direction and in general decreases the separation bubble length, defined in our work. Besides, the rotation also casts its effect on the flow instability and bifurcation. For example, the rotation strengthens the symmetry breaking caused by the inherent wake mechanism observed in the regular bifurcation without rotation, leading to a similar asymmetric recirculation region in the non-rotating flows generated by the regular bifurcation. 
We have also investigated the lift and drag coefficients in the short rotating cylinder flow. Larger rotation rates both increase the absolute value of the drag coefficient and lift coefficient in the transverse direction. An interesting correspondence of the lift coefficients between the short rotating cylinder and the rotating sphere with a doubled rotation rate is observed when $Re$ is relatively large. 

The global stability analyses reveal that the parameter space $\alpha$ can be divided into low and high regimes when $Re$ is relatively low $<500$.
When the rotation rate is smaller than approximately $0.3$, two unstable global modes exist with non-zero frequencies undergoing Hopf bifurcation, whose interaction and competition giving rise to a codimension-two transition state where the two Hopf bifurcations can occur simultaneously. The critical $Re$ at a certain $\alpha$ slightly decreases and then obviously increase with increasing $\alpha$. When the rotate rate is large, more unstable modes are observed experiencing Hopf bifurcations. The critical $Re$ in this case decreases with increasing $\alpha$, highlighting the different effects of $Re$ in the two rotation regimes. The eigenvectors as well as their superposition with the corresponding adjoint modes have also been probed. Especially, we observed that the sensitivity region of the mode HB (with a high rotation rate) is distinguished from other modes close to the cylinder, indicating that its control may be achieved separately.

The comparison of the linear and nonlinear results aims to attach more physical significance to the linear analyses. The traces of the global modes are identified in the nonlinear simulations by comparing their frequencies (i.e., eigenfrequency in the linear analysis and the shedding frequency in the nonlinear DNS). The DMD method is employed to conduct the comparison with the global stability analysis based on the steady flow and the time-mean flow (averaged over several oscillating periods). In general, we can find better correspondence of the time-mean flow results with the DNS results, simply because both analyses were applied to the nonlinear saturated oscillation.
In term of the phase diagram, the low-rotation-rate cases characterise limit cycles whereas high-rotation-rate cases present an increasing degree of complexity, encompassing limit cycles, limit torus and chaos, reflecting more complex flow structures and dynamics when the cylinder rotates faster.

Then, the effect of the aspect ratio $\ar$ has also been investigated. The aim of this investigation is to compare our flow configuration with other bluff body wake dynamics in a large parameter space. Even though more data points are needed, we can find the trend of how the increasing $\ar$ renders the short rotating flow more unstable in the low-rotation-rate cases $\alpha<0.6$. When the rotation rate is large but less than $\alpha=2$, the critical $Re$ decreases with increasing $\alpha$. Compared with other bluff body dynamics, we found that the dynamics of the rotating sphere wake flow along the transverse direction is similar to that of the short rotating cylinder with $\ar$ slightly larger than 1, and the rotating sphere wake along the streamwise direction differs more significantly than our results due to the different flow configuration. 

The current work focuses on the first instability in the short rotating cylinder wake flow. The flow dynamics already presents a high degree of complexity. In order to further understand the underlying mechanism, as a future direction, the subsequent flow bifurcations in this flow can be studied in detail by employing the global linear stability and weakly nonlinear stability analyses.

\begin{acknowledgments}
\textbf{Acknowledgments.} The simulations were performed at National Supercomputing Centre, Singapore (NSCC). The authors would like to thank Ms. Xuerao He and Mr. Dongdong Wan for  insightful discussions.

\textbf{Funding.} We acknowledge the financial support of a Tier 1 grant from the Ministry of Education, Singapore (WBS No. A-8001172-00-00). Y.L. is supported by the National Natural Science Foundation of China (No. 12202200) and China Postdoctoral Science Foundation (No. 2022M711641).
\end{acknowledgments}
\textbf{Declaration of Interests.} The authors report no conflict of interest.

\begin{appendix}

\section{Landau model}\label{appA}

Section \ref{subsubsec:DMDmodes} mentions that the steady-state flow transitions supercritically. The calculation of the Landau coefficient is presented in this appendix. 
The growth rate ($c_1=1.4473\times10^{-2}$) obtained by nonlinear DNS is in good agreement with the growth rate ($1.4642\times10^{-2}$ in table \ref{tab:pointseigevalues}) of SFD base flow linear stability analysis. The coefficient $c_3=-3.673$ indicates that the Hopf bifurcation caused by mode LA is supercritical. This is the same as the bifurcation property of the non-rotating cylinder \citep{YANG2021}, also caused by mode LA.
	Now, based on Stuart--Landau equation $ \frac{\rm{d}\it A_m}{\rm{d}\it t}=\it c_{\rm1} A_m + c_{\rm3} A_m|A_m|^2$ (where $\it A_m$ can be viewed as the amplitude of $C_{lz}$) or, equivalently, $\rm d(ln|\it A_m\rm |)/d\it t=c_{\rm1}+ c_{\rm3}|A_m|^2$, the Landau coefficients $c_{\rm1}$ and $c_{\rm3}$ can be calculated by plotting $\rm d(ln|\it A_m\rm |)/d\rm t$ versus $|A_m|^2$ \citep{thompson2001kinematics,sheard2004spheres}, as shown in panel ($c$). Therefore, the transverse intercept point gives an estimation of $c_1\rm=1.4473\times10^{-2}$ and the gradient near this point is an approximation of $ c_3=-3.673$.
	
\begin{figure}
	\hfil
	\centering\includegraphics[trim=0.0cm 0.0cm 0.0cm 
	0.0cm,clip,width=0.30\textwidth]{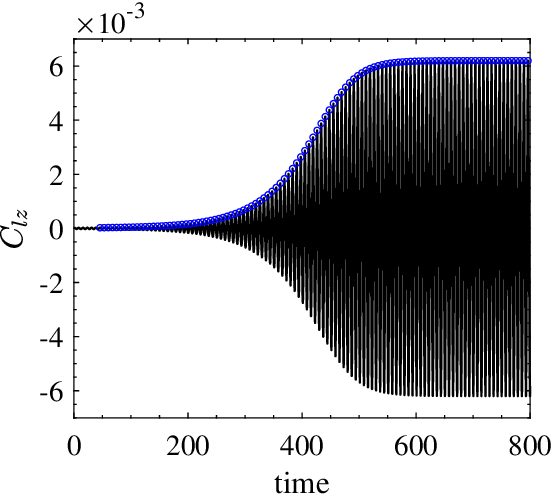} \put(-120,95){($a$)}
	\quad\ \centering\includegraphics[trim=0.0cm 0.0cm 0.0cm 
	0.0cm,clip,width=0.30\textwidth]{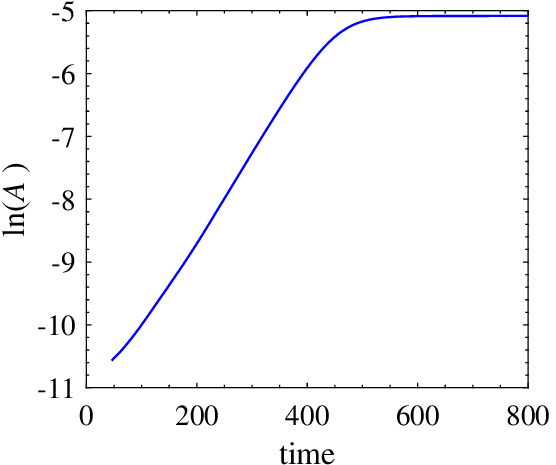} \put(-120,95){($b$)}
	\quad\ \centering\includegraphics[trim=0.0cm 0cm 0.0cm 
	0cm,clip,width=0.30\textwidth]{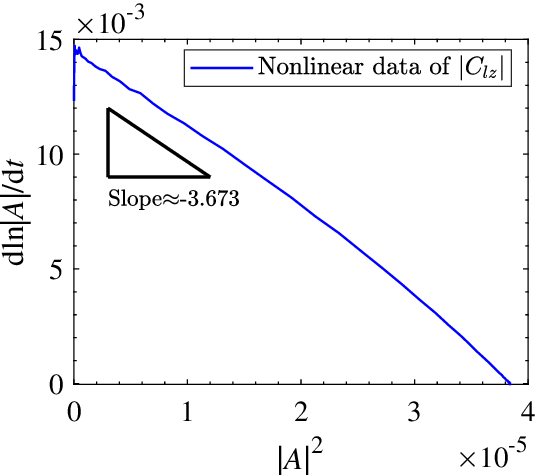} \put(-120,95){($c$)}
	\caption{Landau coefficients $c_1=1.4473\times10^{-2}$ and $c_3=-3.673$ computed by the nonlinear DNS at point LP1.}
	\label{fig:landauDNS}
\end{figure}

\section{Global modes for the cases $\ar=0.75$ and $\ar=2$}\label{app_ar2}

\begin{figure}
\centering\includegraphics[trim=0.0cm 0cm 0.0cm 0cm,clip,width=0.47\textwidth]{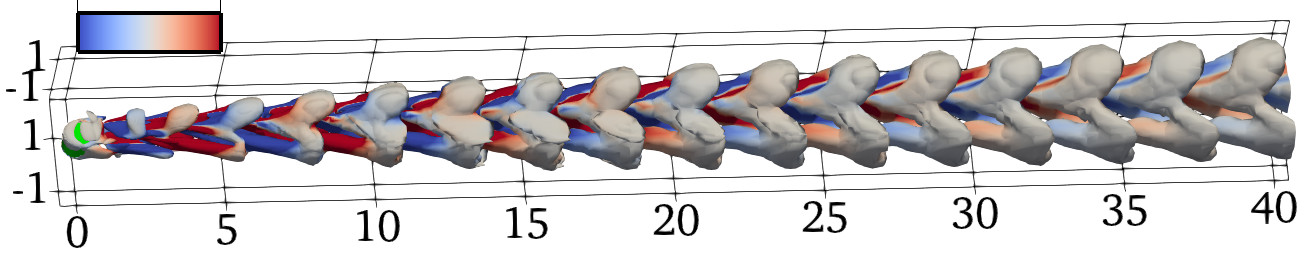}\put(-190,40){($a$)} \put(-172,40){LA mode $\lambda_{\rm LA} = 4.062\times10^{-2} + \rm{i}0.1749$}  \qquad  %f=1.099068
\centering\includegraphics[trim=0.0cm 0cm 0.0cm 0cm,clip,width=0.47\textwidth]{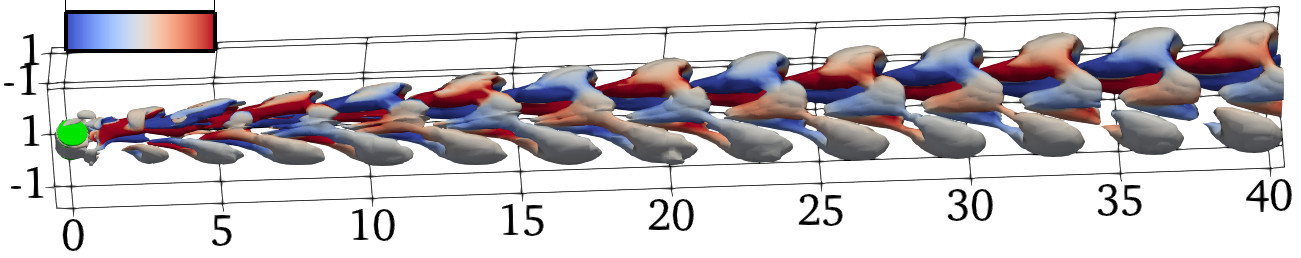}\put(-190,40){($b$)} \put(-172,40){LD mode $\lambda_{\rm LD} = -5.222\times10^{-3} + \rm{i}0.1431$}         %f=0.899065
\caption{The global mode LA ($a$) and LD ($b$) for the flow past a rotating finite cylinder at ($\ar=0.75, Re=365, \alpha=0.05$). 
	The $Q=1\times 10^{-6}$ isosurfaces are colored by the streamwise vorticity ranging from $-2\times 10^{-3}$ to $2\times 10^{-3}$.}
\label{fig:eigenmode_ar075low}
\end{figure}

\begin{figure}
\centering\includegraphics[trim=0.0cm 0cm 0.0cm 0cm,clip,width=0.47\textwidth]{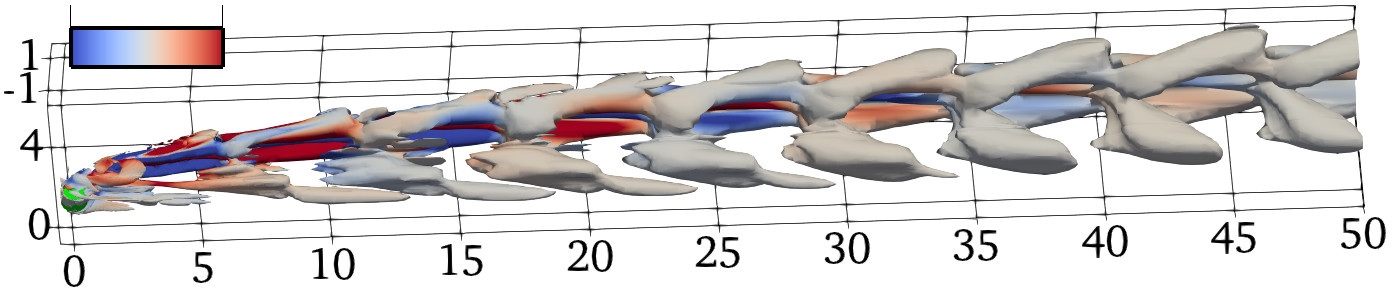} \put(-190,45){($a$)} \put(-172,43){\fontsize{8pt}{\baselineskip}\selectfont{HA1 mode $\lambda_{\rm HA1} = 6.177\times10^{-2} + \rm{i}0.07840$}} \qquad
\centering\includegraphics[trim=0.0cm 0cm 0.0cm 0cm,clip,width=0.47\textwidth]{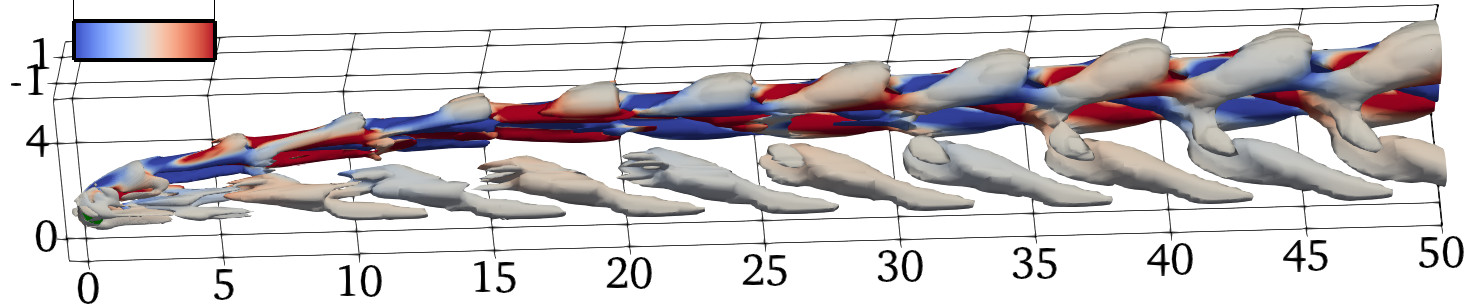} \put(-190,45){($b$)} \put(-172,43){\fontsize{8pt}{\baselineskip}\selectfont{HA2 mode $\lambda_{\rm HA2} = 9.240\times10^{-3} + \rm{i}0.08931$}} \\
\vspace{0.3cm}
\centering\includegraphics[trim=0.0cm 0cm 0.1cm 0cm,clip,width=0.475\textwidth]{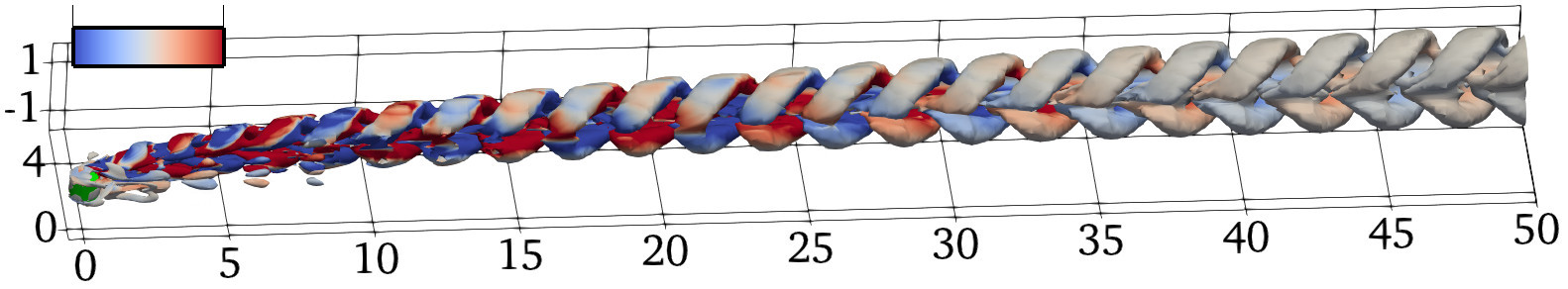} \put(-190,42){($c$)} \put(-172,40){\fontsize{8pt}{\baselineskip}\selectfont{HB mode $\lambda_{\rm HB} = 0.1078 + \rm{i}0.1950$}} \qquad 
\centering\includegraphics[trim=0.0cm 0cm 0.0cm 0cm,clip,width=0.47\textwidth]{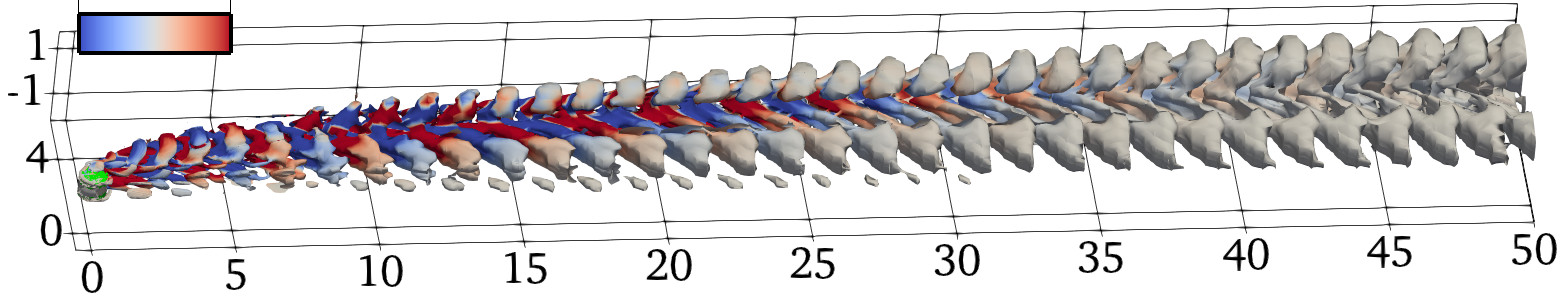} \put(-190,42){($d$)} \put(-172,40){\fontsize{8pt}{\baselineskip}\selectfont{HC mode $\lambda_{\rm HC} = 0.1146 + \rm{i}0.3012$}} \\
\caption{The global modes HA1 ($a$), HA2 ($b$), HB ($c$) and HC ($d$) for the flow past a rotating finite cylinder at ($\ar=0.75, Re=330, \alpha=1.5$). 
The $Q=1\times 10^{-6}$ isosurfaces are colored by the streamwise vorticity ranging from $-2\times 10^{-3}$ to $2\times 10^{-3}$.}
	\label{fig:eigenmode_ar075high}
\end{figure}
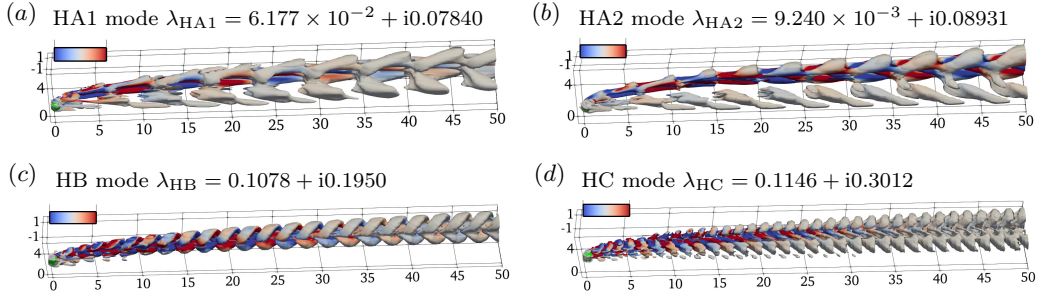

Figures \ref{fig:eigenmode_ar075low},\ref{fig:eigenmode_ar075high} display the structures of global linear modes for $\ar=0.75$ at low and high rotation rates, respectively. In figure \ref{fig:eigenmode_ar075low}, the modes LA and LD are presented. They are named as such because they resemble the LA and LD modes in the $\ar=1$ flow as shown in figure \ref{fig:eigenmodeABCD}$(a,d)$. Figure \ref{fig:eigenmode_ar075high} features the global modes HA1, HA2, HB and HC. Again, comparison can be made to the global modes for the $\ar=1$ flow in figure \ref{fig:eigenmodeEF} at a similar rotation rate.

\begin{figure}
\centering\includegraphics[trim=0.0cm 0cm 0.0cm 0cm,clip,width=0.48\textwidth]{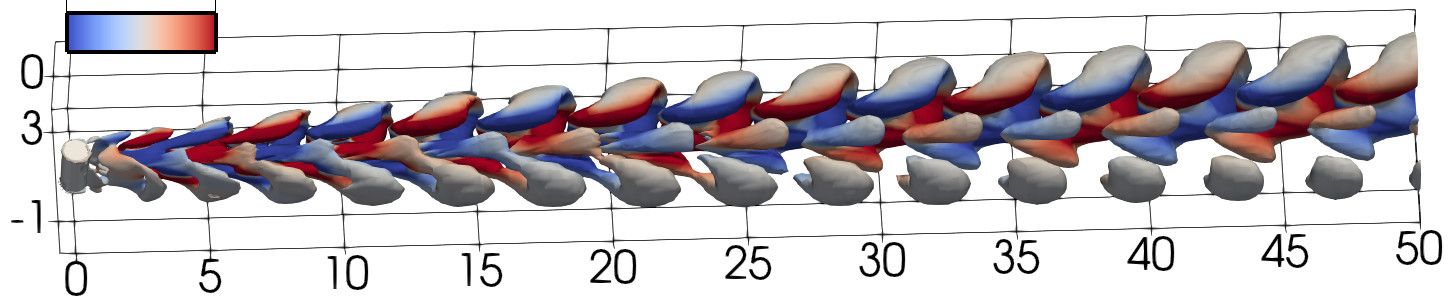}
\put(-185,45){($a$)} \ \put(-165,45){\fontsize{8pt}{\baselineskip}\selectfont {LD mode $\lambda_{\rm LD} =1.105\times10^{-2}+$i$0.1255$}}
\centering\includegraphics[trim=0.0cm 0cm 0.0cm 0cm,clip,width=0.48\textwidth]{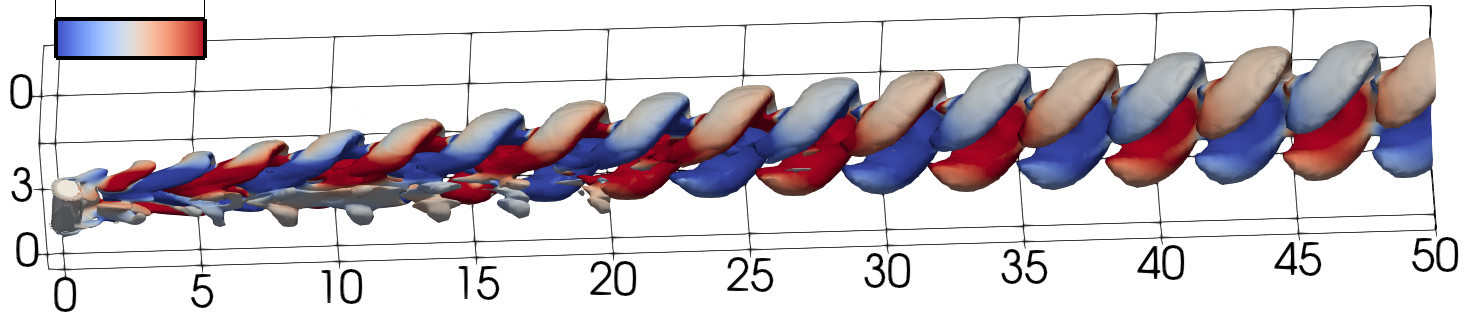} 
\put(-185,45){($b$)} \ \put(-165,45){\fontsize{8pt}{\baselineskip}\selectfont {LB mode $\lambda_{\rm LB} =9.314\times10^{-3}+$i$0.1470$}}\\
 \vspace{0.4cm}
\centering\includegraphics[trim=0.0cm 0cm 0.0cm 0cm,clip,width=0.48\textwidth]{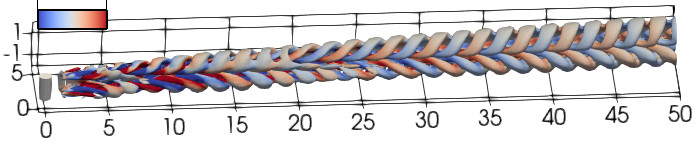}
\put(-185,50){($c$)} \ \put(-165,50){\fontsize{8pt}{\baselineskip}\selectfont {HB1 mode $\lambda_{\rm HB1} =2.899\times10^{-2}+$i$0.2720$}}
\centering\includegraphics[trim=0.0cm 0cm 0.0cm 0cm,clip,width=0.48\textwidth]{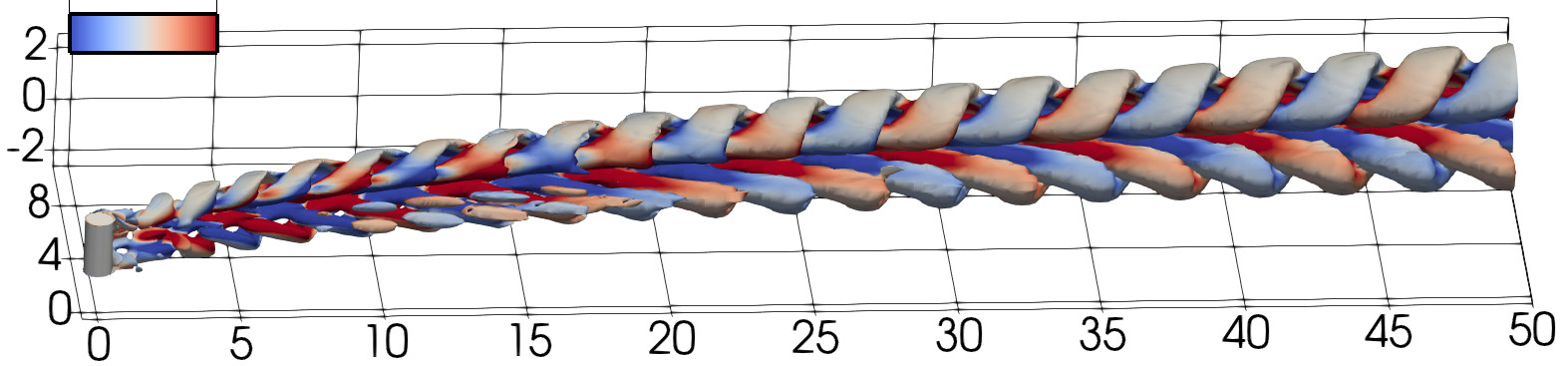}
\put(-185,50){($d$)} \ \put(-165,50){\fontsize{8pt}{\baselineskip}\selectfont {HB2 mode $\lambda_{\rm HB2} =2.028\times10^{-2}+$i$0.1562$}}
\caption{Selected global modes (SFD base flow) for $\ar=2$. ($a$) mode LD ($Re=150$, $\alpha=0.1$, which is similar to the mode LD in figure \ref{fig:eigenmodeABCD} $d$). ($b$) mode LB (panel $b$ at $Re=130$ and $\alpha=0.4$), mode HB1 (panel $c$ at $Re=310$ and $\alpha=0.6$), and mode HB2 (panel $d$ at $Re=190$ and $\alpha=1.5$). The $Q=1\times 10^{-7}$ isosurfaces are colored by streamwise vorticity ranging from $-0.02$ to $0.02$.
}
	\label{fig:eigenmode_ar2}
\end{figure}

The structures of some selected global linear modes for $\ar=2$ are displayed in figure \ref{fig:eigenmode_ar2}. Panel $a$ shows an unstable mode at a low rotation rate $\alpha=0.1$. It is called LD mode because the wake structure looks very similar to the LD mode in figure \ref{fig:eigenmodeABCD} for $\ar=1, Re=290, \alpha=0.1$. 
Under the action of enhanced speed ratio, the mode LB in panel $b$ can be regarded as the result of mode LD losing the vortex shedding from the cylinder's arc-surface that rotates along the streamwise direction. It closely resembles  mode HB in figure \ref{fig:eigenmodeEF} for $\ar=1, Re=170, \alpha=1.8$ at a high rotation rate. 
Mode HB1 in panel $c$ exhibits similar vortex structures to the mode LB, but with smaller vortices, resulting in higher oscillation frequencies.
Mode HB2 in panel $d$ can be regarded as the result of stronger lateral deflection of mode LB due to the stronger rotation.

\section{Validation of numerical codes}\label{app_validation}
\begin{figure}
%	\hfil
	\centering\includegraphics[trim=0.0cm 0cm 0.0cm 0cm,clip,width=0.49\textwidth]{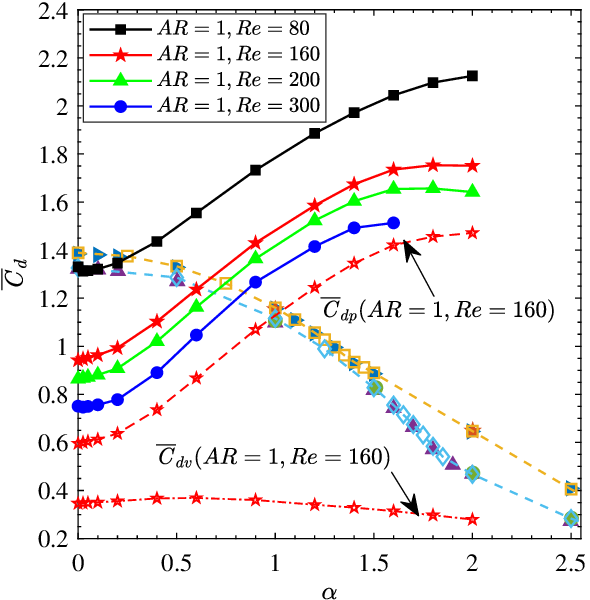}	
	\put(-190,190){($a$)}\quad 
	\centering\includegraphics[trim=0.0cm 0cm 0.0cm 0cm,clip,width=0.475\textwidth]{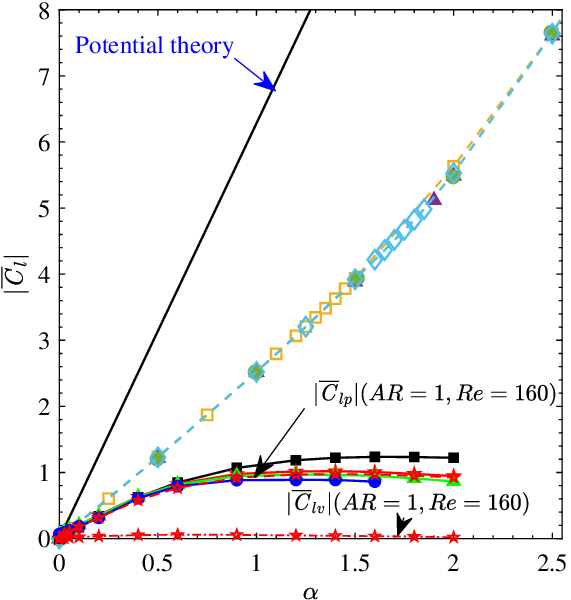}
	\put(-185,190){($b$)} \\ 
	\vspace{0.3cm}
	\centering\includegraphics[trim=0.0cm 0cm 0.0cm 0cm,clip,width=0.32\textwidth]{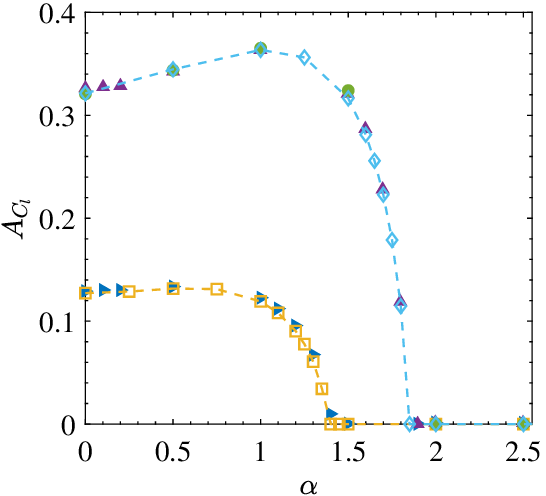} \put(-128,115){($c$)}\ \ 
	\centering\includegraphics[trim=0.0cm 0cm 0.0cm 0cm,clip,width=0.32\textwidth]{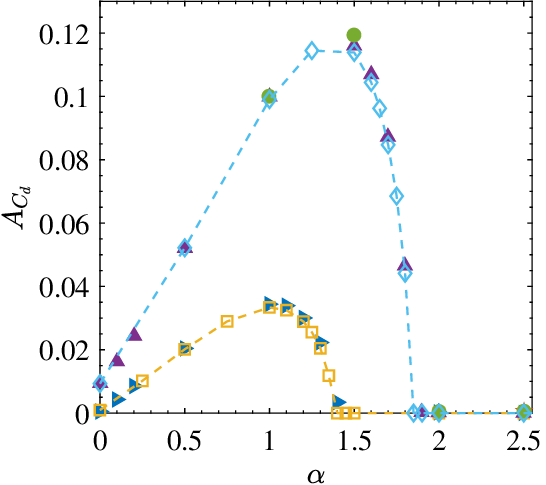} \put(-125,115){($d$)}\quad
	\centering\includegraphics[trim=0.0cm 0cm 0.0cm 0cm,clip,width=0.32\textwidth]{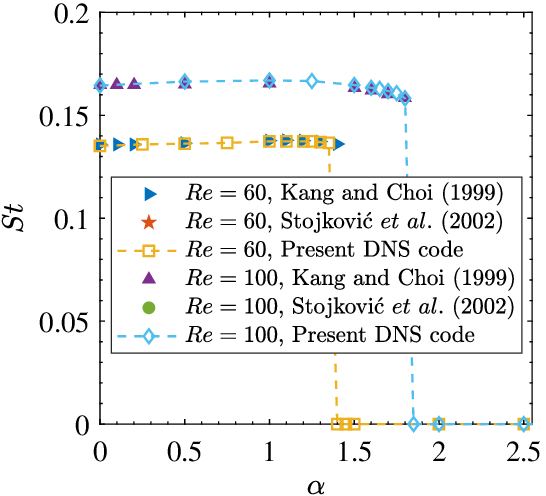} \put(-125,115){($e$)} \quad
	\caption{Validations by comparing the lift and drag coefficients of 2-D rotating cylinder flow between the present DNS code results with those in \cite{kang1999laminar,stojkovic2002effect}. ($a$) time-averaged drag coefficient; ($b$) time-averaged lift coefficient; ($c$) amplitude of $C_l$; ($d$) amplitude of $C_d$; ($e$) Strouhal number calculated using the $C_l$ signal. Besides, the drag and lift coefficients of steady SFD base flow are added in panels ($a$) and ($b$) for comparative analysis, and decompose them into $\bar C_l=\bar C_{lp}+\bar C_{lv}$, $\bar C_d=\bar C_{dp}+\bar C_{dv}$ for the case $(\ar=1, Re=160)$.}
	\label{fig:alpha_Coefficient}
\end{figure}

\begin{figure}
	\hfil
	\centering\includegraphics[trim=1.0cm 0cm 1.1cm 0cm,clip,width=0.45\textwidth]{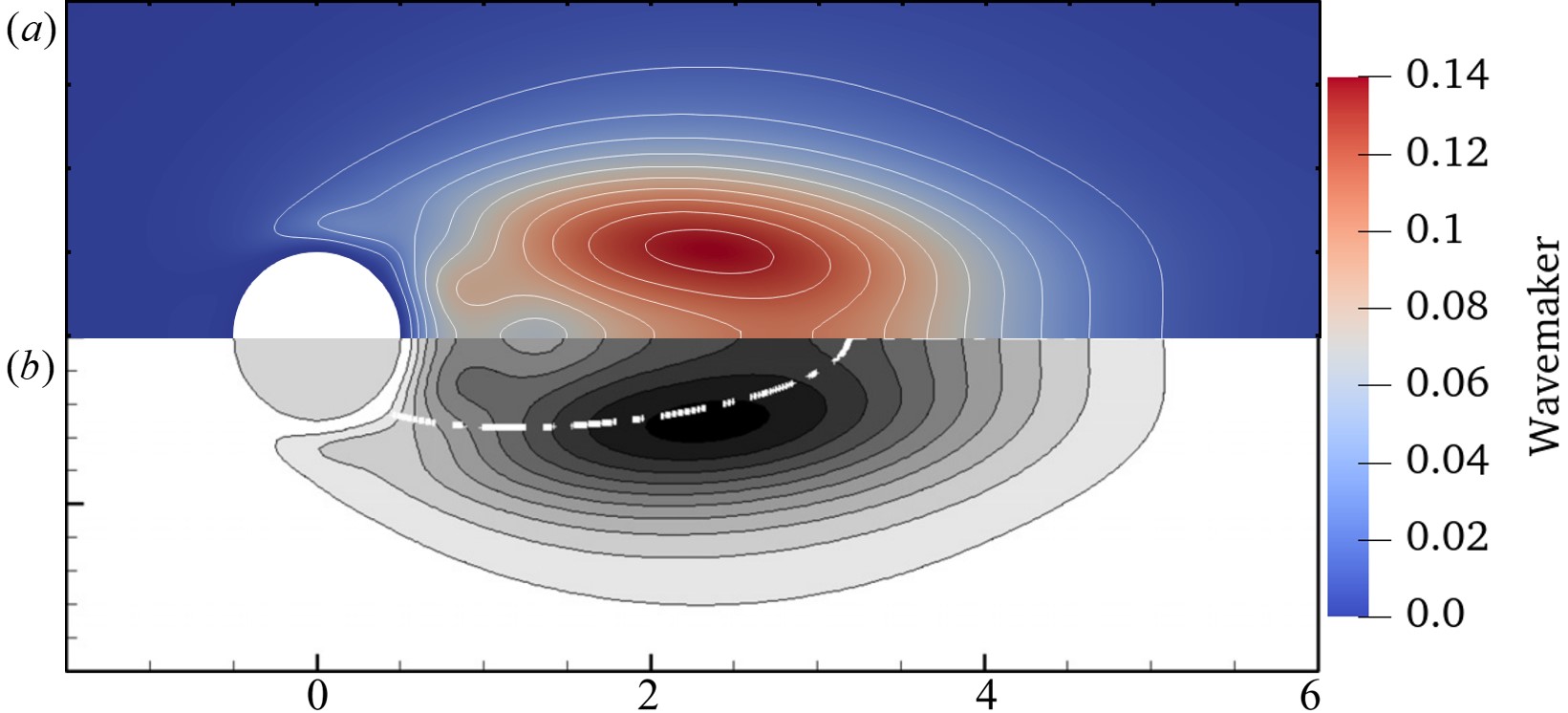}
	\put(-185,77){($a$)} 
	\put(-185,36){($b$)}
	\quad \  \centering\includegraphics[trim=0.0cm 0.0cm 1.0cm 0cm,clip,width=0.475\textwidth]{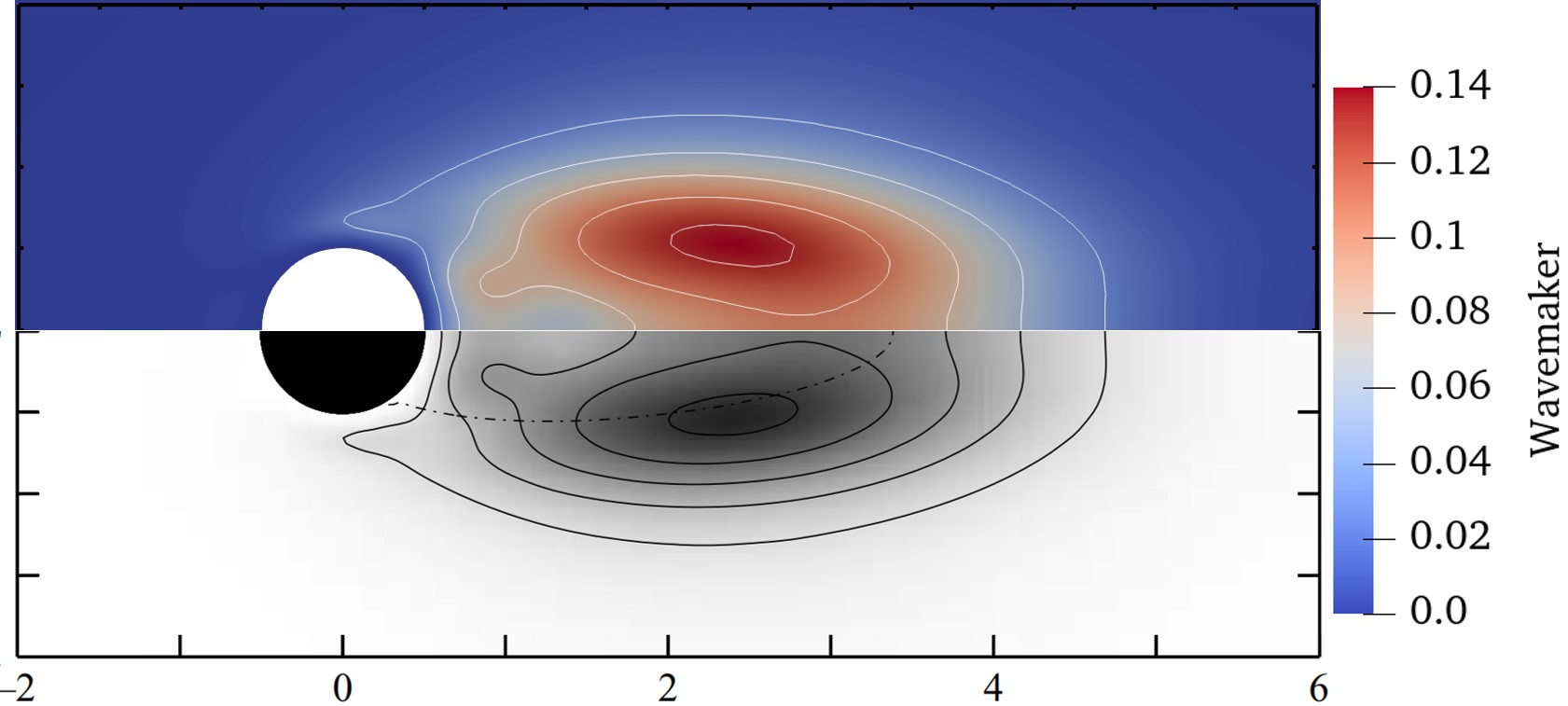}
	\put(-193,77){($c$)} 
	\put(-193,36){($d$)}
	\caption{Comparison of the wavemaker region $\boldsymbol{\zeta}$ between the results generated by the present code (panels $a, c$) and those in \cite{marquet2008}(panel $b$), \cite{giannetti2007}(panel $d$) for 2-D non-rotating cylinder flows at $Re=46.8$ (panels $a, b$) and $Re=50$ (panels $c, d$).}
	\label{fig:wavemaker2D}
\end{figure}

This appendix demonstrates the validation of the numerical codes used in the present work. The results of $\alpha=0$ connects smoothly with the non-zero-$\alpha$ results in the main text as qualitative proof of the accuracy of our codes. To further test the accuracy in a quantitative manner, we compare our simulated results of the lift and drag coefficients in the 2-D rotating cylinder flow with those in \cite{kang1999laminar,stojkovic2002effect}. Figure \ref{fig:alpha_Coefficient} displays a good comparison of our results with theirs at $Re=60$ and 100, indicating the accuracy of the used nonlinear numerical code in the current work. For the verification of the linear code, figure \ref{fig:wavemaker2D} presents the wavemaker region in the classical 2-D cylindrical wake flow in \cite{giannetti2007,marquet2008}. We can see a very good comparison is achieved between our results (top) and theirs (bottom). This good comparison entails the linear code solving correctly both the global modes and the adjoint modes. 

In the end, we furnish a test study on the size of the computational domain. In our previous work on the 3-D non-rotating short cylinder \citep{YANG2021}, we converged a suitable computational domain balancing the computational efficiency and accuracy. With the rotation effect, we found that the size of the computational domain should increase to accommodate the swinging effect brought by the rotation. 
The result is shown in table \ref{tab:meshtest} for different values of $L_a$ and $L_o$, see their definitions in figure \ref{fig:CD_BC}. The symmetry boundary condition is imposed on the surfaces $S_{xy,t}$, $S_{xy,b}$, $S_{xz,f}$ and $S_{xz,b}$, where the position is set to a range of $\pm 10D$ to $\pm 15D$, that is $L_a=10-15$.

As shown in table \ref{tab:meshtest}, unlike the non-rotating cylinder \citep{CADIEUX2017}, the drag coefficient is more sensitive to the domain size than $St$ number. The relative errors of the drag, lift coefficients and the frequencies from mesh M2 to M4 are less than $1\%$ compared with the reference results of mesh M5. Thus, in order to balance the efficiency and accuracy, the mesh M2 and the order $N_{ord.}=7$ are adopted in the present work to compute all  computational instances.
\begin{table}\centering	
\begin{tabular}{p{0.8cm} p{1.2cm} p{1.8cm} p{1.0cm} p{1.2cm} p{1.2cm} p{1.2cm} p{2.1cm} p{1.2cm}}
 Mesh & $L_{a}\times L_{o}$ & $N_{tot.}(N_{ord.})$ & $\overline{C}_d$ & $\overline{C}_{ly}$ & $A_{{C}_{d}}$ & $A_{{C}_{lz}}$ & $St_{C_{ly}}$ & $St_{C_{lz}}$\\ 
 M1& $10\times 40$ & 6350  (7) & 1.412  & -0.8991 & 0.01537 & 0.01618 & 0.1848(0.1456) & 0.04049\\%& Nb=5
 M2& $15\times 50$ & 10928 (7) & 1.409  & -0.8964 & 0.01525 & 0.01604 & 0.1876(0.1433) & 0.04054\\%& Nb=6
 M3& $15\times 50$ & 10928 (9) & 1.408  & -0.8961 & 0.01528 & 0.01606 & 0.1876(0.1431) & 0.04049\\%& Nb=6
 M4& $12\times 50$ & 15804 (7) & 1.410  & -0.8974 & 0.01529 & 0.01609 & 0.1876(0.1428) & 0.04055\\%& Nb=6
 M5& $15\times 60$ & 18648 (7) & 1.409  & -0.8963 & 0.01527 & 0.01605 & 0.1874(0.1432) & 0.04053\\%& Nb=8
\end{tabular}
\caption{A grid sensitivity test for the nonlinear DNS case $Re=290, \alpha=1.2$, $\textsc{ar}=1$, $\Delta t=10^{-3}$. As shown in figure \ref{fig:CD_BC}, $L_{a}$ is length from surfaces $S_{in}$, $S_{xz}$ and $S_{xy}$ to cylinder centre; $L_{o}$ is length from surface $S_{out}$ to cylinder centre. $N_{tot.}$ is the total number of hexahedral elements inside the computational domain. $N_{ord.}$ is the polynomial order of each hexahedral element.}\label{tab:meshtest}
\end{table}

\end{appendix}

\bibliography{ref.bib}
\end{document}